\documentclass[a4paper,11pt]{article}
\pdfoutput=1
\usepackage{mymacros}
\usepackage{nccmath}

\title{Replica analysis of entanglement properties}
\author{Arvind Shekar${}^{\diamondsuit}$ and Marika Taylor${}^{\spadesuit}$}
\affiliation{${}^{\diamondsuit}$ STAG Research Centre, Highfield, University of Southampton, SO17 1BJ Southampton, UK \\
and School of Mathematical Sciences, Highfield, University of Southampton, SO17 1BJ Southampton, UK \\
${}^{\spadesuit}$ College of Engineering and Physical Sciences, University of Birmingham, B15 2TT Birmingham, UK
}

\emailAdd{A.Shekar@soton.ac.uk, M.M.Taylor@bham.ac.uk}

\begin{document}	
\abstract{In this paper we develop a systematic analysis of the properties of entanglement entropy in curved backgrounds using the replica approach. We explore the analytic $(q-1)$ expansion of R\' enyi entropy $S_q$ and its variations; our setup applies to generic variations, from symmetry transformations to variations of the background metric or entangling region. Our methodology elegantly reproduces and generalises results from the literature on entanglement entropy in different dimensions, backgrounds and states. We use our analytic expansions to explore the behaviour of entanglement entropy in static black hole backgrounds under specific scaling transformations, and we explain why this behaviour is key to determining whether there are islands of entanglement.}	

\maketitle
\newpage

\section{Introduction}
Entanglement entropy has been increasingly studied over the last twenty years \cite{Calabrese:2004eu, Rangamani:2016dms, Headrick:2019eth, Nishioka:2018khk, Casini:2022rlv, Casini:2009sr}. On the one hand, entanglement entropy provides a robust measure of quantum entanglement in the context of quantum information and quantum computation\cite{Calabrese:2004eu, Nielsen, RevModPhys.74.197, Horodecki:2009zz, Eisert:2006kue}: for example, it is preserved under local unitary transformations i.e. quantum gates. Entanglement entropy can also be used to characterise several features of quantum systems such as thermalization, many-body localization \cite{PhysRevB.82.174411}, different phases of a system \cite{RevModPhys.80.517, Calabrese:2009kka, Vidal:2002rm, Skinner:2018tjl}, confinement \cite{Klebanov:2007ws}, quantum chaos \cite{Bertini:2018fbz}. On the other hand, entanglement entropy plays a key role in gravity/gauge theory dualities. The famous proposal of Ryu-Takayanagi \cite{Ryu:2006bv} relates entanglement entropy in a holographic quantum field theory to the area of a minimal surface in the holographic dual geometry. Entanglement entropy in holographic theories thus leads to deep insights into the holographic reconstruction of dual spacetimes \cite{VanRaamsdonk:2009ar, VanRaamsdonk:2010pw, Faulkner:2013ica, Bianchi:2012ev,Myers:2013lva,Pourhasan:2014fba,Faulkner:2017tkh,Haehl:2015rza,Swingle:2014uza}. 

Entanglement entropy also plays an important role in the formulation of the black hole information loss paradox by Don Page \cite{PhysRevLett.44.301}. Page formulates information loss in terms of entanglement entropy of quantum fields in the far exterior of the black hole. In Hawking's calculation \cite{Hawking:1975vcx}, these quantum fields are in a thermal state and the entanglement entropy increases monotonically through the evaporation. Page expresses the recovery of information during the black hole evaporation in terms of the entanglement entropy increasing to a maximum, and then decreasing to zero as the black hole evaporates, signalling recovery of all information contained in the black hole.

In recent years the Page formulation of the information loss paradox, together with quantum extremal surfaces\cite{Engelhardt:2014gca}, has stimulated new understanding of how information is recovered. In the islands of entanglement paradigm, one considers the entanglement entropy of quantum fields in a black hole background and then uses the quantum extremal surface approach (combining black hole and quantum field entropy) to determine the quantum extremal surface. The latter has been found to have a component behind the black hole horizon, a so-called island of entanglement\cite{Almheiri:2019psf, Almheiri:2019yqk, Penington:2019npb}, which is associated with recovery of information from the black hole during the evaporation.  

An essential part of the analysis of entanglement islands is the computation of entanglement entropy of quantum fields within black hole backgrounds. Most of the literature explores two-dimensional black holes within JT gravity, with the quantum fields described by a two-dimensional conformal field theory. Accordingly, the entanglement of the quantum fields is obtained from the well-known results for finite temperature two-dimensional conformal field theory, with the details of the background captured by suitable conformal transformations. 

Generalisation of the islands approach to higher dimensional black holes requires computation of the entanglement of quantum fields within such backgrounds. This is computationally challenging, even for static black holes and conformal fields. Most of the quantum field theory computations in the literature relate to quantum entanglement in vacuum states, in either non-interacting quantum theories or conformal theories, in flat backgrounds \cite{Callan:1994py, Calabrese:2004eu, Calabrese:2005zw, Calabrese:2009qy, Lewkowycz:2012qr, Calabrese:2011eb, Rosenhaus:2014woa, Casini:2011kv, Hertzberg:2010uv, Fursaev:1995ef, Solodukhin:2008dh, Fursaev:2013fta, Landgren:2024ccz}. Holographic approaches to entanglement entropy allow access to a wider range of states in strongly interacting quantum field theories, but these computations still primarily cover quantum theories in conformally flat backgrounds i.e. the boundary of the holographic geometry is conformally flat. 

\begin{figure}[H]
\centering
\includegraphics[scale=1.3]{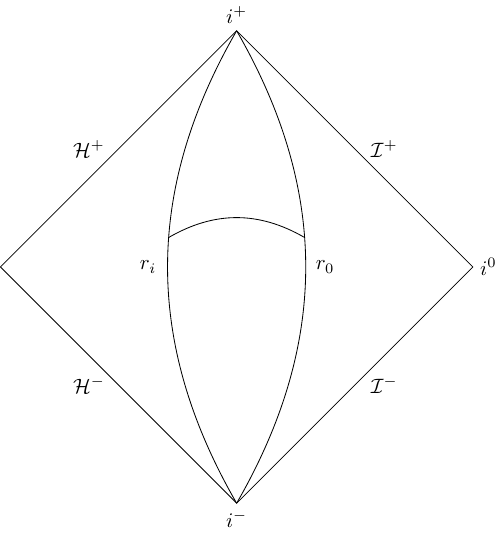}
\caption{Entangling region on asymptotically flat black hole background bounded by inner radius $r_i$ and outer radius $r_o$}
\label{region}
\end{figure}
\bigskip

The relevant setup for exploring islands of entanglement is quantum field theory within a black hole background. Here it is important that the background is viewed as representing a typical state in the thermal ensemble, rather than the thermal state itself. In practice this means that the geometry is well represented by the black hole geometry outside the horizon but that we are agnostic about the details of the black hole interior, which capture the specifics of the microstate. The entanglement entropy region is a slab outside the horizon i.e. the region between specified inner and outer radii, see Figure~\ref{region}. The quantum fields are implicitly assumed to be in quasi-equilibrium with the background, and their state is thus approximated to leading order by the thermal state. Again, however, one should keep in mind that the actual state of this radiation is pure, rather than thermal. 

Thus, entanglement islands require analysis of entanglement entropy at finite temperature in black hole backgrounds. Interesting examples of backgrounds include black strings: near extremal black strings have AdS$_3$/BTZ near horizon geometries, and thus one is led to consider entanglement entropy within slab regions outside the horizon in such geometries. Given the underlying symmetry of such backgrounds, the most natural representation of the quantum fields is as conformal field theories i.e. there are no characteristic scales in the quantum field theory. Analysis of this system would allow interpretation of islands within the holographic duals to the asymptotically AdS$_3$ near horizon regions\cite{Landgren:2024ccz}. 

More generally, to determine the existence and features of islands, one would need to analyse entanglement entropy in generic black hole backgrounds - including the asymptotic regions, not just near horizon regions. In this paper we initiate a systematic analysis of the properties of entanglement in such backgrounds, using the replica trick \cite{callan1994geometric, holzhey1994geometric,Calabrese:2009qy}. 

\bigskip

The entanglement entropy corresponding to a reduced density matrix $\rho_A$, associated with a region $A$ takes the form:
\begin{equation}
S_{\rm QFT} = - {\rm Tr} (\rho_A \log \rho_A ).
\end{equation}
The premise of the replica approach is to rewrite this as a limit of the R\'{e}nyi entropy $S_q$:
\begin{equation}
S_{\rm QFT} = \underset{q \rightarrow 1}{\lim} \left ( S_q \right ) \equiv \underset{q \rightarrow 1}{\lim} \left ( - \frac{\partial}{\partial q} ( \log ({\rm Tr} \rho_A^q)) \right ). 
\end{equation}
This implicitly assumes that $S_q$ is analytic in $q$, and there are well-known examples where this assumption is not valid e.g. spin glass systems. However, known replica anomalies are in discrete systems, rather than in continuum quantum field theories. Moreover, we are agnostic about the details of the quantum field theory describing the fields outside the black hole; our entanglement analysis would be valid for any quantum field theory in which there is no replica anomaly. 

In this paper we explore the analytic $(q-1)$ expansion of the R\'{e}nyi entropy $S_q$ and its variations $\delta S_q$, and use these expansions to compute entanglement and its variations. Our setup applies to generic variations, from symmetry transformations through to variations of the background metric or the shape of the entangling region. Here we focus on the example of variation under a specific scaling transformation: as we discuss in the conclusions, this quantity is key to determining whether there are islands of entanglement.

Analytic $(q-1)$ expansions have been used in previous literature on entanglement entropy, particularly for computing the universal (conformal anomaly driven) contributions to entanglement entropy in vacua of conformal field theories. In this context systematic studies of $(q-1)$ expansions of curvature invariants were carried out in \cite{Fursaev:1995ef, Fursaev:2013fta, Solodukhin:2008dh}.  

In this paper we extend such analysis to generic quantum states on static backgrounds, exemplifying our approach with conformal field theories. Let $\langle T_{\mu \nu} \rangle^{[0]}$ represent the stress tensor of the quantum state on the base manifold. The stress tensor on the replica manifold $\langle T_{\mu \nu} \rangle_q$ is then expressed analytically as 
\begin{equation}
\langle T_{\mu \nu} \rangle_q = \langle T_{\mu \nu} \rangle^{[0]} + (q-1) \langle T_{\mu \nu} \rangle^{[1]} + \cdots   
\end{equation}
The leading correction $\langle T_{\mu \nu} \rangle^{[1]}$ is determined from the stress tensor on the base manifold. We derive explicitly the equations for this correction term for states in a conformal field theory, from the conservation and trace equations of the stress tensor. We then use the analytic expansion of the stress tensor to compute properties of entanglement entropy in such states, focussing on variations of entanglement entropy.  

In the case of vacuum states of conformal field theories, the stress tensor on the base manifold is expressed in terms of curvatures, with the trace of the stress tensor corresponding to the conformal anomaly. Our methodology reproduces results from the literature on entanglement in such states, and its variations under scaling transformations. 

Entanglement in quasithermal states has not been studied extensively in the literature. In general quantum entanglement is primarily studied for vacuum and near vacuum states; in finite temperature states, quantum effects are entwined with thermal behaviour. As discussed above, entanglement in quasithermal states is however relevant to the study of islands of entanglement in black holes. Accordingly, we use our methodology to analyse the replica stress tensor for quasithermal states, and the corresponding properties of entanglement in such states. Our results make contact with earlier literature on stress tensors of finite temperature states in backgrounds with conical deficits, obtained in the context of cosmology and cosmic strings.

\bigskip

The plan of this paper is as follows. In section \ref{sec:two} we review the replica trick for entanglement entropy and in section \ref{sec:three} we discuss the geometric properties of replica manifolds. In section \ref{sec:four} we consider the parametrisations of variations of the entanglement entropy. In section \ref{section6} we set up the analytic expansion in $(q-1)$ of the entanglement entropy and its variations. These quantities involve the $(q-1)$ expansion of the stress tensor and in section \ref{sec:six} we show how the replica stress tensor can be determined from the stress tensor on the base manifold, illustrating with conformal field theory examples. In section \ref{sec:seven} we solve the equations for the replica stress tensor explicitly in the context of vacuum states of conformal field theories on static backgrounds. Finally, in sections \ref{sec:eight} and \ref{sec:nine} we give results for the entanglement entropy variations in vacuum and finite temperature states in such backgrounds. We conclude in section \ref{sec:ten}, discussing how our results relate to entanglement islands. 

\section{Entanglement entropy of a QFT and the replica trick}\label{sec:two}

Let us consider a quantum field theory living on a $d-$dimensional manifold $M_1$ with a general non-dynamical metric. Consider a state $\rvert\psi\rangle$ on the manifold $M_1$ with the corresponding density matrix $\rho= \rvert\psi\rangle \langle \psi \rvert$. If the Hilbert space $\mathcal{H}$ on $M_1$ can be factorised as $\mathcal{H} = \mathcal{H}_A \otimes \mathcal{H}_{\bar A}$ where $\mathcal{H}_A$ is the Hilbert space corresponding to a subregion $A$ on $M_1$, then we can define a reduced density matrix $\rho_A = \text{Tr}_{\bar A} \rho$ corresponding to the subregion $A$. A discussion on how to define states and density matrices in the path integral formalism \cite{TomHartman} is in appendix \ref{A}. 

The entanglement entropy over a region $A$ on the manifold $M_1$ is given by the von Neumann entropy corresponding to the density matrix $\rho_A$ associated with the region $A$
\begin{equation}\label{EE}
    \Sqft(A) = -\text{Tr} (\rho_A \log \rho_A). 
\end{equation}
When $\rho$ corresponds to a pure quantum state, we have $S(A) = S(\overline{A}
)$. 
$\Sqft$ is difficult to calculate owing to the requirement to calculate the logarithm of the density matrix. We can instead work with polynomials of the density matrix. The $q^{th}$ R\'{e}nyi entropy $S_q$ is given by,
\begin{equation}
    S_q=\frac{1}{1-q}\log(\Tr_A\rho_A^q).
\end{equation}
The von Neumann entropy is then given by
\begin{equation} \label{sqftreplica}
    \Sqft=\underset{q \to 1}{\lim}S_q = \underset{q \to 1}{\lim}-\frac{\partial}{\partial q}\log(\Tr_A\rho_A^q)=\underset{q \to 1}{\lim}-\frac{\Tr_A(\rho_A^q \log \rho_A)}{\Tr\rho_A^q}=-\Tr(\rho_A \log\rho_A).
\end{equation}
This is called the replica trick  \cite{callan1994geometric, holzhey1994geometric,Calabrese:2009qy}. We first compute $\Tr_A\rho_A^q$ for positive integers and then we analytically continue $q$ to general complex values in order to take the limit $q\to 1$. The existence and uniqueness of a proper analytic continuation is hence necessary for the replica trick to work (which is not well understood in certain systems e.g. spin glass).

In a QFT computing $\Tr_A\rho_A^q$ amounts to calculating the Euclidean path integral over a $q-$fold replicated manifold $M_q$ (henceforth referred to as replica manifold) obtained by taking $q$ copies of the original manifold $M_1$ and gluing them with specific boundary conditions along the couple of identical open cuts at the $\tau=\tau_0$ slice where the entangling region $x\in A$ is defined on each manifold. If $\phi_i(x,\tau)$ is the field on the $i^{th}$ copy of the manifold $M_1$,
\begin{equation}\label{replicadensitytrace}
    \Tr_A\rho_A^q = \frac{1}{Z_1^q}\int_{BC} D\phi_1 D\phi_2...D\phi_q \text{exp}\left( -\int_\mathbb{C}d\tau d^{d-1}x \sum_{i=1}^q\mathcal{L}_E[{\phi_i(x, \tau)}]\right) \equiv \frac{Z_q}{Z_1^q}.
\end{equation}
Along the couple of identical open cuts in each manifold the above path integral is restricted by the boundary conditions(BC) given by
\begin{align}\label{BC}
    &\phi_i(x,\tau_0^+)=\phi_{i}(x,\tau_0^-)=\phi_{i}(x \notin A);\text{ if $x \notin A$ },\nonumber\\
    &\phi_i(x,\tau_0^+)=\phi_{i+1}(x,\tau_0^-);\text{ if $x \in A$ },
\end{align}
where $i \in \{1,2,...,q\}$ and $q+i\equiv i$. In the replica path integral above, on each one-manifold $M_1$, the fields along the points $x \notin A$ are identified and summed over all such possible field configurations $\phi_{i}(x \notin A)$, and the points $x \in A$ are cyclically identified between one-manifolds by the above boundary conditions.

\begin{figure}[H]
\centering
\includegraphics[scale=1.3]{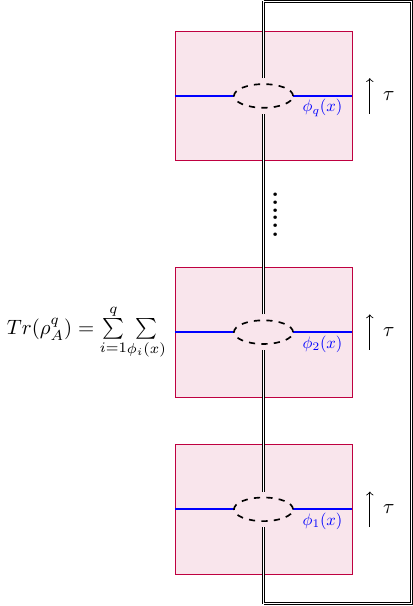}
\caption{$q$ copies of manifold $M_1$ with replica boundary conditions}
\label{Reptr}
\end{figure}

It now follows that
\begin{equation}\label{Sqft}
    \Sqft=\underset{q \to 1}{\lim}\frac{1}{1-q}\left(W_q- q W_1\right),
\end{equation}
where $W_q = \log Z_q$ is the connected generating functional on the replica manifold $M_q$.

From now on we will refer to terms that depend on every point in spacetime as densities and terms that take a value over the entire manifold (integrated value of densities over the entire manifold) as global terms. Expanding $W_q$, we have
\begin{equation}\label{Wqexp}
    W_q=q W_1+(q-1) W^{[1]}+\mathcal{O}((q-1)^2).
\end{equation}
If we ignore the subtleties due to the boundary condition \eqref{BC} (which would contribute in powers of $(q-1)$ for any quantity calculated on $M_q$), we simply have $q$ copies of $M_1$ contributing one $W_1$ on each copy, giving us the $\mathcal{O}((q-1)^0)$ term in the expression above. Such an expansion is true for any global quantity (as opposed to densities) calculated on $M_q$. Densities on $M_q$ (say $F_q$) have an expansion of the form,
\begin{equation}
    F_q = F_1 + (q-1) F^{[1]}+\mathcal{O}((q-1)^2).
\end{equation}

In other words, any global quantity on $M_q$ (say $G_q$) can be written as 
\begin{align}
   G_q= \int_{M_q} F_q &= \int_{M_q}(F_1 + (q-1) F^{[1]}+\mathcal{O}((q-1)^2))\nonumber\\
    &= q\int_{M_1} F_1 +(q-1)\underset{q\to_1}{lim}\int_{M_q}F^{[1]}+\mathcal{O}((q-1)^2)\nonumber\\
    &= q G_1+ (q-1) G^{[1]}+\mathcal{O}((q-1)^2).
\end{align}

Using \eqref{Wqexp} in \eqref{Sqft} we have the following expressions for the R\'{e}nyi and entanglement entropy,
\begin{equation}\label{Sq}
    S_q=- (W^{[1]}+ (q-1)W^{[2]}+...),
\end{equation}
\begin{equation}\label{Sqft1}
    \Sqft=- W^{[1]}.
\end{equation}
One can think of the entanglement structure between two subregions as being encoded in the replica corrections to $W_q$ arising due to the replica boundary conditions \eqref{BC}. $\Sqft$ is clearly dependent on only the correction $W^{[1]}$, but $S_q$ is dependent on all the replica corrections (terms with $(q-1)^{i>0}$) to $W_q$. 

\section{Geometry of the replica manifold} \label{sec:three}
The metric and entangling region on $M_1$ along with the replica boundary conditions \eqref{BC} completely determines the metric on the replica manifold $M_q$. We can impose the cyclic boundary condition \eqref{BC} on the fields on the replica manifold by constructing a manifold respecting the boundary conditions and then defining a field on it. Hence, the replica manifold $M_q$ is made up of $q-$copies of the base manifold $M_1$ glued by the replica boundary conditions \eqref{BC}. Therefore, the metric on the replica manifold depends on, the metric of $M_1$ as well as the shape of the entangling region since that determines how the $q-$copies of $M_1$ are glued respecting \eqref{BC}. In the section that follows we will see how the replica metric $g_{\mu\nu}^{(q)}$ depends on both these details. We will keep $g_{\mu\nu}$ on $M_1$ general and keep the entangling region $A$ to be of an arbitrary shape.

\subsection{Replica manifold \texorpdfstring{$M_q$ and Replica metric $g_{\mu\nu}^{(q)}$}{}}
For the sake of highlighting each aspect of the replica metric $g_{\mu\nu}^{(q)}$, we will restrict $g_{\mu\nu}$ and the shape of the entangling region $A$ at varying levels and later put all aspects together for arbitrary metric and entangling region shape. 
\subsubsection{Entangling regions with no extrinsic curvature}
Let us first consider metric $g_{\mu\nu}$ and entangling region $A$ on $M_1$ such that the entangling boundary $\Sigma$ has no extrinsic curvature \cite{Fursaev:1995ef}. Any entangling region in 2-dimensions trivially satisfies this. In $d>2$ an example would be planar entangling region on a flat metric 
\begin{equation}\label{flatpl}
    ds^2_{M_1}=d\tau^2+dx^2+\delta_{ij}dz^i dz^j,
\end{equation}
with entangling region $x\in[x_1,x_2]$ on a $\tau=$constant slice, spanning $z^i{}'s$.
 
\begin{figure}[H]
\centering
\includegraphics[angle=0, width=0.9\linewidth]{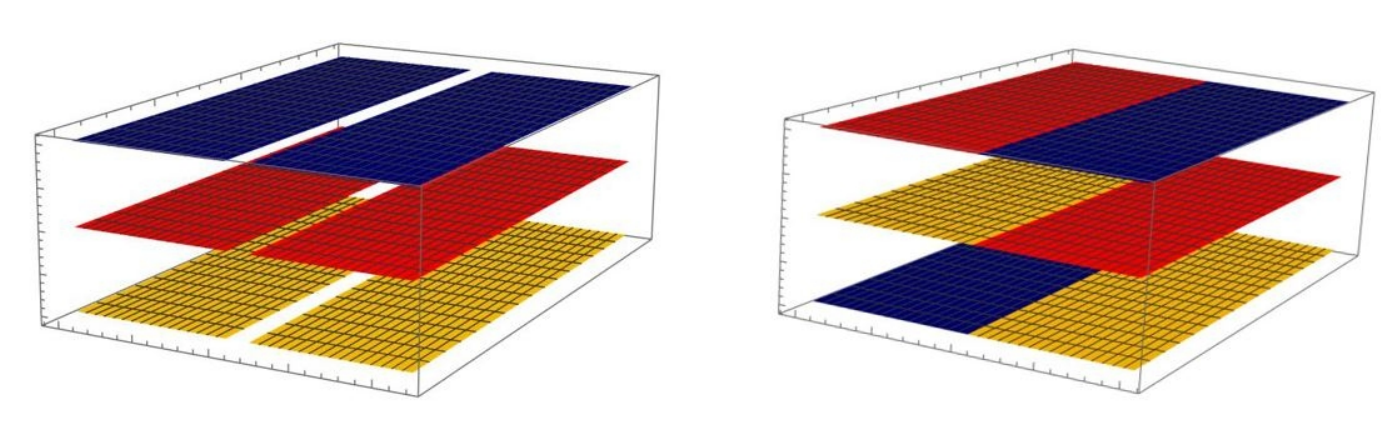}
\caption{$q=3$ copies of manifold after replica boundary conditions - in the bulk of the entangling region}
\label{bulkrep}
\end{figure}

 For sake of simplicity let us suppress the boundary directions $\Sigma$ in $M_1$ and consider only the resulting manifold $\mathbb{R}^2$. We do this since along these directions the metric remains unaffected going from $M_1$ to $M_q$. Away from the boundary of the entangling region, the replica boundary conditions imply that the replica manifold is simply $q$ copies of the base manifold $M_1$, see figure \eqref{bulkrep}. Therefore in the bulk of the entangling region (i.e., sufficiently away from endpoints/ entangling boundaries) the replica manifold inherits the same topology and metric as the base manifold $M_1$.

Locally around the boundary of the entangling region the metric on $M_1$ is flat
\begin{equation}
    ds^2_{M_1}\text{ around boundary}=d\tau^2+dx^2.
\end{equation}

\begin{figure}[H]
\centering
\includegraphics[angle=0, width=1.1\linewidth]{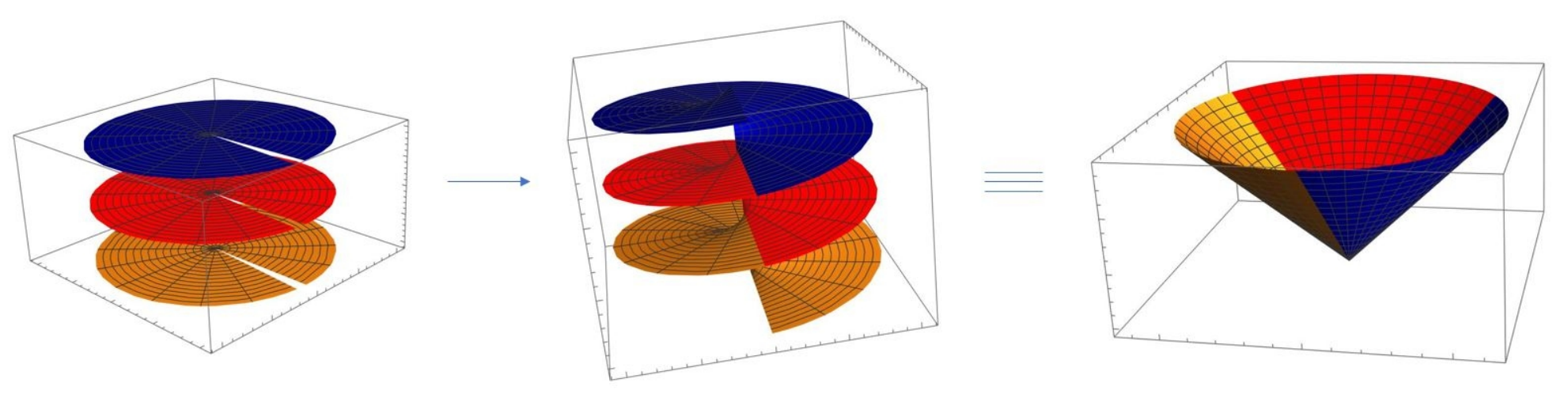}
\caption{$q=3$ copies of manifold after replica boundary conditions - locally around the boundary of the entangling region (the free cut in the first manifold is identified with the free cut in the last manifold although it isn't shown in the figure). This is diffeomorphic to segments of a cone}
\label{bdyrep}
\end{figure}

Around the entangling boundary, the replica boundary conditions imply that the replica manifold locally takes the topology of a $q-$fold spiral sheet with two ends attached as shown in figure \eqref{bdyrep}. One can choose a chart $\{\rho,\psi\}$ where $\rho$ is the distance from the boundary with $\rho\in [0,\infty)$ and $\psi$ is the angle from the cut in the first manifold with $\psi \in [0,2\pi q)$. If $\{x_0,\tau_0\}$ is the endpoint, we have,
\begin{align}\label{conversion}
    &x-x_0=\rho \cos(\psi),\nonumber\\
    &\tau-\tau_0=\rho \sin(\psi).
\end{align}
Every $2\pi$ range of $\psi$ starting from 0 with $\rho \in [0,\infty)$ on the replica manifold covers one complete range of $\{x,\tau\}$ i.e., one copy of $M_1$. There are $q$ copies of $M_1$ in $M_q$. 
One can consider a smooth diffeomorphism that maps the replica space around the boundary, to a cone as shown in figure \eqref{bdyrep} i.e., every copy of $M_1$ in $M_q$ around the boundary is mapped to $1/q^{th}$ segment of a cone. Therefore the replica manifold around the boundary has the topology $C_q \times \Sigma$ where $C_q$ is the conical space with the corresponding range of the angle $[0,2\pi q)$. When $q\to 1$, $C_q \to \mathbb{R}^2$ and we recover the base manifold $M_1$ around the boundary of the entangling region i.e., $\mathbb{R}^2 \times \Sigma$  

The cone can be seen as embedded in a 3-dimensional (pseudo) Euclidean manifold,
\begin{equation}
    X=q \rho'\cos(\psi'/q),\nonumber\\
    Y=q \rho'\sin(\psi'/q),\nonumber\\
    Z=\sqrt{|1-q^2|}\rho',\nonumber\\
    i.e.,  Z^2=\frac{|1-q^2|}{q^2}(X^2+Y^2),
\end{equation}
where $\rho' \in [0,\infty)$ is the distance from the Z-axis and $\psi' \in [0,2\pi q)$ is the angle from X-axis. 
We have the metric on the cone,
\begin{equation}
ds^2_{\text{cone}}=dX^2+dY^2+dZ^2=d\rho'^2+\rho'^2 d\psi'^2.
\end{equation}
The conical singularity can be regulated by redefining $Z=\sqrt{|1-q^2|}f(\rho',a)$ with $\d_{\rho'}f(\rho',a)|_{\rho'=0}=0$ and $f(\rho',a)|_{a\to0}=\rho'$, where $a$ is a regulating parameter. We are essentially introducing a smooth transition in derivative around the tip in order to regulate the singularity. After regulating the metric on the cone goes to,
\begin{equation}
    ds^2_{\text{reg cone}}=U(\rho',a)d\rho'^2+\rho'^2 d\psi'^2,
\end{equation}
where $U(\rho',a)=q^2+(1-q^2)\d_{\rho'}f(\rho',a)$.
\begin{figure}[H]
\centering
\includegraphics[angle=0, width=0.5\linewidth]{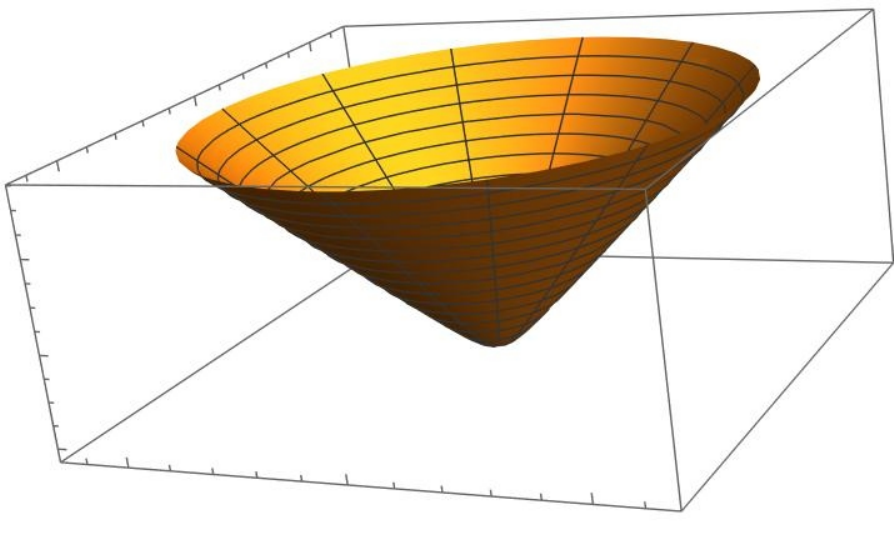}
\caption{Regulated cone}
\label{regcone}
\end{figure}
Using the properties of $f(\rho,a)$, it can be shown that $U(\rho,a)$ satisfies
\begin{equation}\label{limits}
  U(\rho,a)\big|_{\rho=0} \to q^2;\quad U(\rho,a)\big|_{\rho>>a} \to 1 ,
\end{equation}
i.e., in the $\rho\to 0$ limit, the metric becomes $q^2d\rho^2 + \rho^2 d\psi^2 = d\Tilde{\rho}^2 + \Tilde{\rho}^2 d\Tilde{\psi}^2$ where $\{\Tilde{\rho}=q \rho,\Tilde{\psi}=\psi/q\}$ in which there is no conical singularity as desired. Outside the local neighborhood around $\rho=0$, the metric expansion simply yields the flat metric as expected. 

 A particular $U(\rho,a)$ satisfying the limits (\ref{limits}) is 
\begin{equation}\label{Uexplicit}
    U(\rho,a)  = \frac{\rho^2 + a^2 q^2}{ \rho^2+a^2 };\quad f(\rho,a)=\rho-a \tan^{-1}(\rho/a).
\end{equation}
Every such choice of the function $U(\rho,a)$
satisfying the limits in (\ref{limits}) corresponds to choosing different surfaces to roll off the tip. Our calculations must be insensitive to these choices and must only depend on the defining properties of $ U(\rho,a)$ given by the limits (\ref{limits}). Hence, we will keep $U(\rho,a)$ general in our discussions along with the conditions given by \eqref{limits}. The first condition in \eqref{limits} fixes
\begin{equation}
    U(\rho,a) = 1 + (q-1) (2+ O(\rho))+ \mathcal{O}((q-1)^2),
\end{equation}
i.e.,
\begin{equation}\label{Ufixed}
    U\rvert_{q=1}=1; \quad\d_q U\rvert_{q=1,\rho=0}=2.
\end{equation}
We could also express these limits in the coordinate $x=\rho/a$ in which $U$ has no $a$ dependence
\begin{equation}\label{limits2}
  U(x)\big|_{x=0} \to q^2;\quad U(x)\big|_{x>>1} \to 1. 
\end{equation}
This metric on the (regulated) cone is also the same metric on the $q-$fold spiral sheeted replica manifold around the entangling boundary in $\{\rho,\psi\}$ coordinates with the diffeomorphism $\rho' \to \rho, \psi'\to \psi$. We will also reinstate the $z^i$ directions.  The metric along the $z^i$ directions on $M_q$ remains same as that on $M_1$\footnote{If we however did not have symmetries along the $y^i$ directions (Ex: stationary case) and chose a finite entangling region along $z^i$, then $g^{(q)}_{ij}$ on $M_q$ would have conical singularities as well, and would be different from $g_{ij}$ on $M_1$.}. Locally around the boundary, we have,
\begin{equation}\label{repmet1}
    ds_{q\textcolor{blue}{, reg}}^2 =g^{(q)}_{\mu \nu}dx^\mu dx^\nu|_{bdy} =\textcolor{blue}{U(\rho,a)} d\rho^2 + \rho^2 d\psi^2 +\delta_{ij}dz^i dz^j,
\end{equation}
where metric on $\Sigma$ has no extrinsic curvature and is same as the metric on $M_1$. We will refer to \eqref{repmet1} as the unsquashed conical metric (to differentiate it from the squashed conical metric that follows).

In the bulk of the entangling region i.e., sufficiently away from boundary, 
\begin{equation}
    g_{\mu\nu}^{(q)}=g_{\mu\nu},
\end{equation}
as discussed before.

Therefore, the metric on the replica manifold $g^{(q)}_{\mu \nu}$ (replica metric) varies from the metric on $M_1$ i.e., $g_{\mu \nu}$ only in a small region around the end-points/ boundary of the entangling region (characterised by $r=r_1$ or $r=r_2$, $\tau=\tau_0$ and spanning $M^{d-2}$), where it has the topology $C_q \times \Sigma$.

\subsubsection{Entangling regions with extrinsic curvature(s): squashed Cone}
Let us now consider entangling regions and metrics
where the entangling boundary $\Sigma$ has an extrinsic curvature \cite{Fursaev:2013fta}. An example would be, non planar entangling regions (spherical, cylindrical etc...) in flat space:
\begin{equation}
     ds^2_{M_1}=d\tau^2+dr^2+d\Omega^2_{d-2}.
\end{equation}
Consider the entangling region with boundary $\Sigma$ given by $r=r_0$, $\tau=$constant, spanning all other coordinates. In suitable polar coordinates defined by 
\begin{equation}\label{conversion1}
    r-r_0=\rho \cos(\psi),\nonumber\\
    \tau-\tau_0=\rho\sin(\psi),
\end{equation}
the flat space metric can be written in the form,
\begin{equation}
    ds^2_{M_1}=d\rho^2 +\rho^2 d\psi^2 +(r_0+\rho \cos\psi)^2h_{ij}(y)dy^i dy^j,
\end{equation}
for a spherical entangling region with $h_{ij}(y)dy^i dy^j=(d\theta^2+\sin^2\theta d\phi^2)$ in $d=4$, and
\begin{equation}
    ds^2_{M_1}=d\rho^2 +\rho^2 d\psi^2 +(r_0+\rho \cos\psi)^2d\phi^2 +dz^2,
\end{equation}
for a cylindrical entangling region.

 Under replica boundary conditions the replica manifold $M_q$ and the replica metric $g_{\mu\nu}^{(q)}$ has similar features as discussed before - remains same as $M_1$ and $g_{\mu\nu}$ in the bulk of the entangling region, has the topology $C_q\times \Sigma$ locally around the boundary of the entangling region. We could hence try a similar regularization procedure modifying $g_{\rho\rho}^{(q)}|_{bdy}$ as before to smoothen the conical tip. Locally around an entangling boundary, we have
\begin{align}\label{repmetsq}
   & ds_q^2 = g^{(q)}_{\mu \nu}dx^\mu dx^\nu |_{bdy}\nonumber\\
    &=U(\rho,a) d\rho^2 + \rho^2 d\psi^2 + (r_0+\rho \cos(\psi))^2h_{ij}(y)dy^i dy^j; \text{ spherical}\nonumber
    \\&=U(\rho,a) d\rho^2 + \rho^2 d\psi^2 + (r_0+\rho \cos(\psi))^2d\phi^2 +dz^2; \text{ cylindrical}.
\end{align}
However, the resulting metric in this case has a curvature singularity (for example in the Ricci scalar) at $\rho=0$. Unlike the previous case (i.e., the case where $\Sigma$ has no extrinsic curvature), in the metric on $M_q$ above, $\d_\psi$ is not a killing vector (due to explicit $\psi$ dependence in $g_{ij}^{(q)}$). Hence the $O(2)$ symmetry in the $\{\rho,\psi\}$ directions present in the previous case is now absent. Instead, we now have a discrete symmetry $\psi\to\psi+2\pi k$ for $k\in \mathbb{Z}$, hence this is termed a squashed conical singularity. The geometry near $\Sigma$ is a warped product of two spaces. This is regularised by introducing a new regularization parameter $p$, modifying the warp factor $(r_0+\rho \cos(\psi))\to(r_0+\rho^p c^{1-p}\cos(\psi))$. Locally around an entangling boundary, we have 
\begin{align}\label{sqcone}
    ds_q^2 &= g^{(q)}_{\mu \nu}dx^\mu dx^\nu |_{bdy}\nonumber\\
    &=U(\rho,a) d\rho^2 + \rho^2 d\psi^2 + (r_0+\rho^p c^{1-p} \cos(\psi))^2 h_{ij}(y)dy^i dy^j; \text{ spherical}
    \\&=U(\rho,a) d\rho^2 + \rho^2 d\psi^2 + (r_0+\rho^p c^{1-p} \cos(\psi))^2d\phi^2 +dz^2; \text{ cylindrical}.\label{sqconecyl}
\end{align}
The curvature singularity vanishes for $p\geq 2$. Since we will be dealing with integrals quadratic in curvature, therefore it is enough if $p\geq 1$. We choose $p(q)=q$. We will refer to \eqref{sqcone}, \eqref{sqconecyl} as squashed conical metrics. Different regularisation schemes than the ones described here have been considered in \cite{Camps:2013zua, Dong:2013qoa, Cooperman:2013iqr}.

In $\{r,\tau,y^i\}$ coordinates the replica metric locally around an entangling boundary takes the form,
\begin{align}
\begin{split}
    ds_q^2&= g^{(q)}_{\mu \nu}dx^\mu dx^\nu|_{bdy}\\
    &=\frac{(r-r_0)^2+(\tau-\tau_0)^2U(\rho,a)}{(r-r_0)^2+(\tau-\tau_0)^2}d\tau^2+\frac{(r-r_0)^2U(\rho,a)+(\tau-\tau_0)^2}{(r-r_0)^2+(\tau-\tau_0)^2}dr^2 
    \\
    &+\frac{(r-r_0)(\tau-\tau_0)(U(\rho,a)-1)}{(r-r_0)^2+(\tau-\tau_0)^2}dr d\tau\\
    &+\Bigg((r_0+(r-r_0)[(r-r_0)^2+(\tau-\tau_0)^2]^{\frac{p-1}{2}} c^{1-p})^2 h_{ij}(y)dy^i dy^j; \text{ spherical}\\
    &\text{or } (r_0+(r-r_0)[(r-r_0)^2+(\tau-\tau_0)^2]^{\frac{p-1}{2}} c^{1-p})^2 d\phi^2 +dz^2; \text{ cylindrical}\Bigg).
    \end{split}
\end{align}
Notice that one requires $q$ copies of $\{r,\tau,y^i\}$ to cover the entire replica manifold.

\paragraph{Static metric:}More generally let us consider static metric in coordinates $\{\tau,r,\tilde{y}^i\}$,
\begin{equation}
   ds^2_{M_1} = g_{\tau\tau}(r)d\tau^2 +g_{rr}(r)dr^2+r^2 h_{ij}(y)d\tilde{y}^id\tilde{y}^j.
\end{equation}  
 Consider the entangling region with boundary $\Sigma$ characterised by $\tau=$constant and $f(r,\tilde{y}^i)=$constant. Around the codimension-2 entangling boundary, we expand the metric in suitable coordinates $\{x^a\}_{a=1}^2$, $\{y^i\}_{i=1}^{d-2}$ as follows \cite{Fursaev:2013fta},
 \begin{align}\label{static fol }
     ds^2_{M_1} &= dx_1^2+dx_2^2+(\gamma_{ij}(y)+2 x_2 K^{x_1}_{ij}+\dots)dy^i dy^j+\dots\nonumber\\
     &= d\rho^2+\rho^2d\psi^2+(\gamma_{ij}(y)+2 \rho\cos(\psi) K^{x_1}_{ij}+\dots)dy^i dy^j+\dots
 \end{align}
where we have used 
\begin{equation}\label{conversion3}
    x_1=\rho\cos(\psi), \quad x_2=\rho\sin(\psi).
\end{equation}
${x_1}$ and $x_2$ span the transverse space to $\Sigma$. The entangling boundary $\Sigma$ is at $x_1=0=x_2$, spanning $y^i{}'s$. $\gamma_{ij}$ is the induced metric. In the static case, the extrinsic curvature corresponding to the normal vector directed along $\d_\tau$ is zero. Therefore we have only one non-zero extrinsic curvature $K_{ij}^{x_1}$ for the normal vector along $\d_{x_1}$ 

If the entangling region corresponds to $r=r_0$, $\tau=\tau_0$ spanning $y^i{}'s$, we then have 
\begin{equation}
    x_1=r-r_0;\quad x_2=\tau-\tau_0,
\end{equation}
and the extrinsic curvature
\begin{equation}
    K^r_{ij}=r_0 g^{rr}(r_0) h_{ij}; \quad K^\tau_{ij}=0.
\end{equation}

The corresponding metric on the replica manifold has a conical singularity at $\Sigma$ and a curvature singularity as discussed before. We have the regulated replica metric given by,
\begin{multline}\label{repmetstat}
     ds_q^2 \text{ expanded around endpoints }  = g^{(q)}_{\mu \nu}dx^\mu dx^\nu|_{bdy}\\= U(\rho,a)d\rho^2+\rho^2d\psi^2+(\gamma_{ij}(y)+2 \rho^p c^{1-p}\cos(\psi) K^{x_1}_{ij}+\dots)dy^i dy^j+\dots
\end{multline}

\subsubsection{General metric and arbitrary entangling region shape}
Since the replica metric $g_{\mu\nu}^{(q)}$ differs from the metric on $M_1$ only at $\Sigma$ where it picks up a conical singularity and an additional curvature singularity (in the case of a non zero extrinsic curvature at $\Sigma$), it is useful to consider the expansion of a general metric around the boundary $\Sigma$ of an entangling region of an arbitrary shape. Such a foliation perturbatively in distance from the entangling boundary was considered in \cite{Rosenhaus:2014woa}
\begin{align}\label{general fol}
d s_{M_1}^2 & =\left(\delta_{a b}-\left.\frac{1}{3} \mathcal{R}_{a c b d}\right|_{\Sigma} x^c x^d\right) d x^a d x^b+\left(A_i+\left.\frac{1}{3} x^b \varepsilon^{d e} \mathcal{R}_{i b d e}\right|_{\Sigma}\right) \varepsilon_{a c} x^a d x^c d y^i \nonumber\\
& +\left(\gamma_{i j}+2 K_{a i j} x^a+x^a x^c\left(\delta_{a c} A_i A_j+\left.\mathcal{R}_{i a c j}\right|_{\Sigma}+K_{c i l} K_{a j}{ }^l\right)\right) d y^i d y^j+\mathcal{O}\left(x^3\right),
\end{align}
where $\{y^i\}^{d−2}_{i=1}$ parametrize $\Sigma$ and $\{x^a\}^2_{a=1}$ the 2-dimensional transverse space. The entangling surface $\Sigma$ is located at $x^a = 0$, $\gamma_{ij}$ is the corresponding induced metric, $\epsilon_{ac}$ is the volume form of the transverse space, $R_{\mu\nu\alpha\beta}$ and $K^a_{ij}$ are the background and extrinsic curvatures, respectively, and $A_i$ is the analog of the Kaluza-Klein vector field associated with dimensional reduction over the transverse space, that comes from $g_{ic}$. Note that by construction the structure of $\Sigma$ is built into the above ansatz. Upto linear order in distance from the entangling boundary, the metric reduces to
\begin{align}
d s_{M_1}^2 &=\delta_{a b} d x^a d x^b+A_i \varepsilon_{a c} x^a d x^c d y^i +\left(\gamma_{i j}+2 K_{a i j} x^a\right) d y^i d y^j+\mathcal{O}\left(x^2\right)\nonumber\\
&=d\rho^2+\rho^2 d\psi^2+A_i \varepsilon_{a c} x^a d x^c dy^i +(\gamma_{ij}+2\rho\cos(\psi)K_{1ij}+2\rho\sin(\psi)K_{2ij})d y^i d y^j+\mathcal{O}\left(x^2\right),
\end{align}
which is a generalization of \eqref{static fol } to non-static metrics. We have used \eqref{conversion3} to express in $\{\rho,\psi,y^i\}$ coordinates. 

On $M_q$ we can regularise the conical and curvature singularity in $g_{\mu\nu}^{(q)}$ as before,
\begin{align}\label{repgen}
& d s_{M_q}^2\text{ around boundary}=g_{\mu\nu}^{q} dx^\mu dx^\nu\rvert_\text{around $\Sigma$}\nonumber\\
=&U(\rho,a)d\rho^2+\rho^2 d\psi^2+(\gamma_{ij}+2\rho^p c^{1-p}\cos(\psi)K_{1ij}+2\rho^p c^{1-p}\sin(\psi)K_{2ij})dy^idy^j
\\\nonumber&+A_i \varepsilon_{a c} x^a d x^c dy^i +\mathcal{O}\left(x^2\right).
\end{align}
Notice that sufficiently far away from the entangling boundary caustics might be encountered and the coordinate system will break down. Hence, the metric expansions \eqref{general fol} and \eqref{repgen} are only valid locally around $\Sigma$. This is sufficient since the replica metric differs from the metric on $M_1$ only around $\Sigma$, which \eqref{repgen} captures.  

 In general, the replica metric is simply a local expansion around the entangling boundary. This can be noticed by taking the $q\to1$ limit in the replica metric and noticing that we only recover a flat metric. Therefore, replica metric and coordinates are only well defined in a small region around the entangling boundary specified by the conical regulator $'a'$. Away from the entangling boundary the metric is same as that on $M_1$. Specifically, in the case of flat metric on $M_1$, the corresponding replica metric and coordinates are well defined throughout $M_q$.
     
\section{QFT entanglement entropy (EE) and its variations}\label{sec:four}

We are ultimately interested in understanding the conditions on the spectrum of a QFT modeling the Hawking radiation on a black hole background in the semiclassical regime (i.e., metric is not quantised) if unitarity has to be preserved. We follow the recent island rule \cite{Almheiri:2019yqk, Penington:2019npb} which uses the Quantum extremal surface prescription \cite{Engelhardt:2014gca}. See the discussion in section \ref{sec:ten} following \eqref{QESpres} for a more detailed discussion. For this purpose, it is enough to understand how the EE of the QFT i.e., $\Sqft$ varies under the scaling of the entangling region \eqref{island extremum}. Since energy is inversely related to length, the scaling of the entangling region, including the scaling of all other length parameters in the theory, is related to the RG flow up to a negative sign.

\subsection{General QFTs and RG flows} 
For a $d$-dimensional QFT on a Riemannian manifold $M_1$, we consider the behavior of the action under the local renormalization group \cite{osborn1991local,Nakayama:2013wda} where the cut-off and the energy scale depends on the spatial coordinates say $x$. Since we need well-defined local operator equations, it is necessary to be able to define general finite local operators $\O_i(x)$ with corresponding sets of renormalizable couplings $\lambda^i$ which form a closed set under operator mixing. This is ensured by promoting the coupling to be a function of $x$, $\lambda^i(x)$, so that they act as sources for the corresponding local operators. We have the QFT action $I_{QFT}$ given by                                      
\begin{equation}
 I=\int_Md^dx\sqrt{g} \lambda^i(x) \mathcal{O}_i(x) \subset     I_{\textnormal{QFT}}.
\end{equation}

$I$ remains invariant under the following local classical  dilatation
 \begin{equation}\label{cl dil}
     g_{\mu\nu}(x) \to  e^{2\sigma(x)}g_{\mu\nu}(x), \quad \lambda^i(x) \to e^{\beta^i(\lambda) \sigma(x)}\lambda^i(x),\quad \mathcal{O}_i(x) \to e^{-(d+\beta^i(\lambda)) \sigma(x)} \mathcal{O}_i(x),
 \end{equation}
where the Weyl transformation, capturing local rescaling of the metric by an arbitrary continuous function $\sigma(x)$, is an extension of the usual renormalization group to a local renormalization group. $\beta^i(\lambda)=\Lambda(x)\frac{d\lambda^i(x)}{d\Lambda(x)}$  is the beta function, a vector field on the space of couplings and $\Lambda(x)$ is the $x$ dependent energy scale. 

For a strictly renormalizable theory the operators are marginal (\ie $\O_i$ has dimension $d$ and in $\beta^i(\lambda)=k_i\lambda^i+O(\lambda^2)$ we have $k_i=0$). For now, we will restrict our attention to marginal operators. 
 
We have the connected generating functional,
 \begin{equation}\label{genfn}
     e^{W[\lambda^i(x)]}=\int D[\phi] e^{-S_0[\phi]-\int_Md^dx\sqrt{g} \lambda^i(x) \mathcal{O}_i(x)},
 \end{equation}
 with 
\begin{align}
    \left<T_{\mu\nu}(x)\right>&=\frac{-2}{\sqrt{g}}\frac{\delta W}{\delta{g^{\mu\nu}(x)}},\label{tuv}\\
    \left<\O_i(x)\right> &= \frac{-1}{\sqrt{g}}\frac{\delta W}{\delta \lambda^i(x)}.\label{O}
\end{align}

We have the following operator that captures the variation w.r.t. the energy scale $\Lambda(x)$

\begin{align}\label{rgop}
    \Delta_{\Lambda}\equiv\Lambda(x)\frac{\d}{\d\Lambda(x)}=-\delta L^i \frac{\delta}{\delta L^i}&=\int d^d x  \left(\delta_\Lambda g_{\mu\nu}\frac{\delta}{\delta g_{\mu \nu}}+ \delta \lambda^i(x) \frac{\delta}{\delta \lambda^i(x)} \right)\\
    &=\int d^d x  \left(\delta g_{\mu\nu}\frac{\delta}{\delta g_{\mu \nu}}+ \sigma(x)\beta^i(\lambda) \frac{\delta}{\delta \lambda^i(x)} \right),
\end{align}
where in the last line we have used $\sigma(x)\beta^i(\lambda)=\Lambda(x)\frac{d\lambda^i(x)}{d\Lambda(x)}$. The variation wrt the energy scale is inversely related to varying the length scales $L^i$. We will have a corresponding local renormalization group operator on the replica manifold with the metric and theory now on the replica manifold $M_q$, and the integral now promoted to be on the replica manifold. 

\paragraph{Dilatation Ward Identity and anomaly:}Considering the operation of $\Delta_\Lambda$ on the generating functional $W[\lambda^i(x)]$ from (\ref{genfn}), the classical dilatation symmetry ($\ref{cl dil}$) manifests as the trace Ward identity. To keep it general, we will consider the operation of the replica local renormalization group operator on the replica connected generating functional $W_q$
\begin{align}  \label{QFTtrward}
    &\int_{M_{q}} d^d x  \left(2\sigma(x)g^{(q)}_{\mu\nu}(x)\frac{\delta}{\delta g^{(q)}_{\mu \nu}(x)}+ \sigma(x)\beta^i(\lambda) \frac{\delta}{\delta \lambda^i(x)} \right) W_q[\lambda^i(x)] = \int_{M_{q}} d^dx \sqrt{g_q} (-\mathcal{A}_{M_q}),\nonumber\\
    &\mathcal{A}_{M_{q}} = \sigma(x)\left<T^\mu_\mu (x)\right>_{M_{q}}+\sigma(x)\beta^i(\lambda) \left<\mathcal{O}_i\right>_{M_{q}}.
\end{align}
 In the expression above we have used that the dilatation symmetry is under $\delta g^{(q)}_{\mu\nu}=2\sigma(x)g^{(q)}_{\mu\nu}(x)$ and $\lambda^i(x) \to e^{\beta^i(\lambda) \sigma(x)}\lambda^i(x)$ and expressions (\ref{tuv}), (\ref{O}). $\Delta_{\Lambda}$ with this transformation is also the local renormalization group operator, under which the generating functional is dilatation invariant upto a QFT dilatation anomaly $\mathcal{A}$. If we consider a CFT, $\mathcal{A}$ would simply be the well-known Weyl anomaly. The expression (\ref{QFTtrward}) is essentially the local renormalization group equation. 

\subsection{Scale transformations of entanglement}
Let us now look at how the R\'{e}nyi $q-$entropy, $S_q$ and the EE, $\Sqft$ varies under the scaling of the coordinates $x^\mu \to x^\mu{}'= x^\mu + \xi^\mu(x^\mu)$ (this could include scaling of the entangling region boundary, symmetry transformations, and/or deformations of the background metric). With the definition $\delta_{\xi^i}F=F(x^i+\xi^i)-F(x^i)$, we have

\begin{align}\label{genrepentscaling}
 &\delta_L S_q = \Big(\sum_i \delta_{\xi^i}+\sum_j\delta_{l^j}+\delta_g\Big)S_q =-\Delta_{\Lambda}S_q\nonumber\\
 =& -\int_M d^dx \left(\delta_{\Lambda} g_{\mu\nu}\frac{\delta}{\delta g_{\mu\nu}}+\sigma(x)\beta^i(\lambda)\frac{\delta}{\delta \lambda^i(x)}\right) \left(\frac{1}{1-q} (W_q-q W_1)\right)\nonumber\\
    =& \int_M d^dx \Bigg\{\left( \left[\delta_L g_{\mu\nu}=\Big(\sum_i \delta_{\xi^i}+\sum_j\delta_{l^j}+\delta_g\Big) g_{\mu\nu}\right]\frac{\delta}{\delta g_{\mu\nu}}-\sigma(x)\beta^i(\lambda)\frac{\delta}{\delta \lambda^i(x)}\right) \nonumber\\
    &\quad\quad\quad\quad\left(\frac{1}{1-q} (W_q-q W_1)\right)\Bigg\}\nonumber\\
    =&  \frac{1}{q-1} \Bigg\{ \frac{1}{2} \int_{M_q} d^dx \sqrt{g_q}  \Big(\sum_i (\mathcal{L}_{\xi^i}+B_i)+\sum_j\delta_{l^j}+\delta_g\Big) g^{(q)}_{\mu \nu} \left< T^{\mu \nu} \right>_q   \nonumber\\
    &\hspace{0.6cm}- \frac{1}{2}q\int_{M_1} d^dx \sqrt{g} \Big(\sum_i (\mathcal{L}_{\xi^i}+B_i)+\sum_j\delta_{l^j}+\delta_g\Big) g_{\mu \nu} \left< T^{\mu \nu} \right>  \nonumber\\ 
     &\hspace{0.6cm}-\int_{M_q} d^dx\sqrt{g_q}\sigma(x)\beta^i(\lambda)\left<\mathcal{O}_i(x)\right>_q+q\int_{M_1} d^dx\sqrt{g}\sigma(x)\beta^i(\lambda)\left<\mathcal{O}_i(x)\right>_{q=1} \Bigg\};\nonumber\\
     &\delta_L \Sqft =\underset{q \to 1}{\lim}\delta_L S_q.
\end{align}
here $l^i$ are other variable length scales. We use $\delta_{l^j}$ to denote the variation of these length scales and $\delta_g$ the deformations of the metric. We have used,
\begin{equation} \label{metvar}   \delta_{\xi^i}g_{\mu\nu}^{(q)}=\mathcal{L}_{\xi^i}g_{\mu\nu}^{(q)}+B_ig_{\mu\nu}^{(q)}.
\end{equation}
$\mathcal{L}_{\xi^i}g_{\mu\nu}^{(q)}$ is the Lie derivative along $\xi^i$, $B_i g_{\mu\nu}^{(q)}$ captures the contribution from $dx^\mu dx^\nu$ in the line element i.e., under one of the coordinates $x^a \to x^a{}'= x^a + \xi^a(x^\mu)$ we have 
\begin{align}
    &g_{\mu\nu}^{(q)}(\textbf{x})dx^\mu dx^\nu \to g_{\mu\nu}^{(q)}(x^a+\xi^a(\textbf{x}))d(x^\mu+\delta^{\mu}_a\xi^a(\textbf{x})) d(x^\nu+\delta^{\nu}_a\xi^a(\textbf{x}))=g_{\mu\nu}^{(q)}{}'(\textbf{x})dx^\mu dx^\nu; \nonumber\\
   &\delta_{\xi^a}g_{\mu\nu}^{(q)}=g_{\mu\nu}^{(q)}{}'-g_{\mu\nu}^{(q)}.
\end{align}

The variations of $S_q$ and $\Sqft$ \eqref{genrepentscaling} captures $S_q$ and $\Sqft$ up to an overall constant. When one of the directions $x^i$ corresponds to an isometry on $M_1$ i.e., $\mathcal{L}_{\xi^i}g_{\mu\nu}=0$, and the entangling region does not have a boundary along this direction then $\mathcal{L}_{\xi^i}g_{\mu\nu}^{(q)}=0$. In the presence of entangling boundaries along $x^i$, $M_q$ and $g_{\mu\nu}^{(q)}$ develops a conical singularity along the boundary which breaks the isometry along $x^i$. 

Notice that \eqref{genrepentscaling} reduces to expressions in \cite{Rosenhaus:2014woa}, where perturbation in the reduced density matrix $\rho_A$ in $M_1$ under deformations of $g_{\mu\nu}$ and entangling surface $\Sigma$ is calculated, to calculate the perturbation of $\Sqft$ given by \eqref{EE}.

\paragraph{Calculation of variation of entropy:}
For the most general metric and entangling region with boundary $\Sigma$ characterised by $\{x^a\}_{a=1}^2=0$, the replica metric $g_{\mu\nu}^{(q)}$ locally around $\Sigma$ is given by $\eqref{repgen}$, and sufficiently away from the boundary we have $g_{\mu\nu}^{(q)}=g_{\mu\nu}$. The replica stress tensor $\langle T_{\mu\nu}\rangle_q$ can now be obtained by solving the replica conservation equations on $M_q$,
\begin{equation}\label{genrepstresscons}
    \nabla^{(q)}_\mu \langle T_{\mu\nu}\rangle_q=0.
\end{equation}
In the case of a CFT ($\beta^i(\lambda)=0$) we additionally have
\begin{equation}
   \langle T^\mu_{\ \mu}\rangle_q=\mathcal{A}[g_{\mu\nu}^{(q)}].
\end{equation}
These are enough to evaluate the scaling of the CFT R\'{e}nyi and entanglement entropy given by \eqref{genrepentscaling} with $\beta^i(\lambda)=0$.

For calculating the scaling of the entanglement entropy of a QFT we can further simplify the problem by taylor expanding $g_{\mu\nu}^{(q)}$ and $\langle T_{\mu\nu}\rangle_q$ around $q=1$, 
\begin{align}\label{repmetexp}
    g^{(q)}_{\mu \nu} &= g_{\mu \nu} + (q-1)\Big(g_{\mu\nu}^{[1]}=(\partial_q g^{(q)}_{\mu \nu}) \rvert_{q=1}\Big) + \mathcal{O}\left( (q-1)^2 \right),\nonumber\\
    \sqrt{g_q} &= \sqrt{g} +\frac{q-1}{2}\sqrt{g}g^{ \sigma\kappa} (\d_q g^{(q)}_{\sigma\kappa} ) \rvert_{q=1}  + \mathcal{O}\left( (q-1)^2 \right),\nonumber\\
    \left< T^{\mu \nu} \right>_q &= \left< T^{\mu \nu} \right>+(q-1)\left< T^{\mu \nu} \right>^{[1]}+ \mathcal{O}\left( (q-1)^2 \right).
\end{align}
Using this in (\ref{genrepentscaling}) and keeping only $\mathcal{O}((q-1)^0)$ terms in the scaling of the $q^{th}$ R\'{e}nyi entropy (since we take the limit $q\to1$ of this to get the scaling of $\Sqft$ with the scaling of the entangling region), we get 
\begin{multline}\label{gensqftvar}
      \delta_L\Sqft =  \lim_{q \to 1}  \Bigg\{\Bigg(I_1 \equiv\frac{1}{2} \int_{M_q} d^dx \sqrt{g} \delta_{L}( ( \d_q g^{(q)}_{\mu \nu})\rvert_{q=1}) \left< T^{\mu \nu} \right> \\+\frac{1}{4} \int_{M_q} d^dx\sqrt{g} g^{\sigma\kappa} (\d_q g^{(q)}_{\sigma\kappa} ) \rvert_{q=1}  \delta_{L} g_{\mu \nu}  \left< T^{\mu \nu} \right> \Bigg) \\
      +\left(I_2 \equiv\frac{1}{2}\int_{M_q} d^dx \sqrt{g}  \left( \delta_{L} g_{\mu \nu}\right) \left< T^{\mu \nu} \right>^{[1]}\right) \\
        -\frac{1}{q-1}\Bigg(\int_{M_q} d^dx\sqrt{g_q}\sigma(x)\beta^i(\lambda)\left<\mathcal{O}_i(x)\right>_q-q\int_{M_1} d^dx\sqrt{g}\sigma(x)\beta^i(\lambda)\left<\mathcal{O}_i(x)\right>_{q=1}\Bigg) \Bigg\}, 
\end{multline}
where $\delta_L=\sum_i \delta_{\xi^i}+\sum_j\delta_{l^j}+\delta_g$ and $\delta_{\xi^i}g_{\mu\nu}^{(q)}$ is given by \eqref{metvar}.
We have also used,
\begin{equation}
        \int_{M_q} d^dx \sqrt{g} \left( \delta_{L} g_{\mu \nu} \left< T^{\mu \nu} \right>  \right) = q\int_{M_1} d^dx \sqrt{g} \left( \delta_{L} g_{\mu \nu} \left< T^{\mu \nu} \right>  \right).
\end{equation}
Note that there are implicit $q$ dependencies coming from the $M_q$ integration limits which we have not series expanded in the above expression \eqref{gensqftvar}. Specifically, the $q$ dependence comes from the $\psi$ integration on $M_q$ with integration limits $[0,2\pi q)$. We expect these to give terms with factors of $q^{i\geq 0}$ since we are integrating over $q$ copies of $M_1$ (if the $\psi$ integration had resulted in factors of $(q-1)^{i<0}$ then we had to retain $O((q-1)^2)$ terms in the series expansion of replica metric and replica stress tensor). We hence retain the limit $q\to 1$ for such unaccounted factors of $q^{i>0}$.

If
\begin{align}\label{zeros}
    \int_{M_q}\sqrt{g} \d_q{g_{\mu\nu}^{(q)}}\rvert_{q=1} F^{\mu\nu}=0, \text{ and }
    \int_{M_q}\sqrt{g} \delta_L\left(\d_q{g_{\mu\nu}^{(q)}}\rvert_{q=1}\right) F^{\mu\nu}=0,
\end{align}
where $F^{\mu\nu}$ is on $M_1$, then
\begin{multline}\label{gensqftvar2}
      \delta_L\Sqft =  \lim_{q \to 1}  \Bigg\{\left(I_2 \equiv\frac{1}{2}\int_{M_q} d^dx \sqrt{g}  \left( \delta_{L} g_{\mu \nu}\right) \left< T^{\mu \nu} \right>^{[1]}\right) \\
        -\frac{1}{q-1}\Bigg(\int_{M_q} d^dx\sqrt{g_q}\sigma(x)\beta^i(\lambda)\left<\mathcal{O}_i(x)\right>_q-q\int_{M_1} d^dx\sqrt{g}\sigma(x)\beta^i(\lambda)\left<\mathcal{O}_i(x)\right>_{q=1}\Bigg) \Bigg\}. 
\end{multline}
For the conformal vacuum of CFT, $\beta^i(\lambda)=0$ and the second line in the above expression is trivial. 
\begin{equation}\label{scft}
      \delta_L\Sqft =  \lim_{q \to 1} \frac{1}{2}\int_{M_q} d^dx \sqrt{g}  \left( \delta_{L} g_{\mu \nu}\right) \left< T^{\mu \nu} \right>^{[1]}.
\end{equation}
The $O(q-1)$ correction to the replica stress tensor is the only quantity on $M_q$ required to compute the variation of $S_{CFT}$ with the scaling of the entangling region. When there are enough symmetries, $\left< T^{\mu \nu} \right>^{[1]}$ can be determined in terms of $\left< T^{\mu \nu} \right>$ on $M_1$ using the replica conservation equations \eqref{genrepstresscons}. We will show later that this is possible in the static case. \eqref{zeros} is reasonable, since $\d_q{g_{\mu\nu}^{(q)}}\rvert_{q=1}$ is non zero only at the entangling boundary $\Sigma$ where there is a conical singularity contribution. We will also explicitly show this in the general case and specifically in the static case in later sections.

For a general QFT, given the action and connected generating functional on $M_1$, one can obtain the corresponding action and connected generating functional on $M_q$ by replacing $g_{\mu\nu}$ by $g_{\mu\nu}^{(q)}$. In addition to solving $\langle T_{\mu\nu}\rangle_q$ explicitly using \eqref{tuv}, one can then also solve for the one-point functions on $M_q$ defined by \eqref{O} which can then be used to evaluate \eqref{genrepentscaling}. 

For the specific case, when $g_{\mu\nu}$ is flat and the entangling surface is planar and we consider a vacuum state, 
\begin{itemize}
    \item \cite{Rosenhaus:2014woa}, \cite{Smolkin:2014hba} gives the $N$ point vacuum correlation function on $M_q$ in terms of specific $N+1$ point vacuum correlation functions on $M_1$ given by,
\begin{equation}\label{smolkin}  -\underset{q\to1}{lim}\d_q\langle\mathcal{O}_1\left(x_1\right) \ldots \mathcal{O}_N\left(x_N\right)\rangle_q=\langle\mathcal{O}_1\left(x_1\right) \ldots \mathcal{O}_N\left(x_N\right) K_0\rangle_c,
\end{equation}
where $K_0=2\pi H_R$, and $H_R$ is the Rindler (modular) Hamiltonian.
\begin{equation}
    K_0=-2 \pi \int d^{d-2} y \int_0^{\infty} d x_1 x_1 T_{22}\left(x_1, x_2=0, y\right),
\end{equation}
here $\{x_1,x_2\}$ span the transverse space to $\Sigma$ (located at $x_1=x_2=0$).

\item Replica stress tensor one-point functions on corresponding replica space given by \eqref{repmet1} (unsquashed cone) has also been considered in the literature on cosmic string backgrounds which have a similar conical singularity structure - massless scalar \cite{PhysRevD.35.3779, Moretti:1997dx, PhysRevD.36.3095}, massive scalar \cite{Iellici:1997ud}, fermion \cite{BezerradeMello:2012ajq}, photon \cite{Moretti:1996wd,PhysRevD.45.4486}, graviton \cite{PhysRevD.45.4486}. We will discuss more on this in later sections.
\end{itemize}

\section{Entanglement entropy on static backgrounds}\label{section6}
We will now restrict to static backgrounds on $M_1$ where we will carry out completely explicit computations.

\subsection{Metric on \texorpdfstring{$M_1$}{}} 
We restrict to static metrics on $M_1$. We can always choose coordinates/ gauge $\{\tau,r,y^i\}$ in which the metric can be written as, 
\begin{equation}\label{statmetric}
   ds^2_{M_1} = g_{\tau\tau}(r)d\tau^2 +g_{rr}(r)dr^2+r^2 h_{ij}(y)dy^idy^j,
\end{equation}
here $\tau \in \mathbb{R}$ is the euclidean time coordinate, $r$ $ \in [0,\infty)$ is a spatial radial coordinate, $y^i$'s are $(d-2)$ spatial coordinates that span a compact orientable submanifold $M^{d-2}$ with $Vol(M^{d-2})=\int d^{d-2}x\sqrt{h}$. 

Our analysis is valid for any asymptotics. It is however worth noting that in AdS we can have non-trivial codimension-2 $h_{ij}$ topologies,
\begin{equation}\label{hortopo}
     R_{ijkl}(h)=k(h_{ik}h_{jl}-h_{il}h_{jk}),
\end{equation}
i.e., $h_{ij}$ is a constant curvature space. We have three possible codimension-2 topologies depending on the value of $k$ - elliptic horizons $(k=1)$, flat horizons $(k=0)$, hyperbolic horizons $(k=-1)$ \cite{mann1997topological}.

\paragraph{Global flat space:}
We will also consider flat space with metric
\begin{equation}\label{flatmet1}
     ds^2_{M_1} =d\tau^2 +dx^2+\phi_0^2\, h_{ij}(y)dy^idy^j,
\end{equation}
with constant $\phi_0$ and $y^i$ spans a compact submanifold. Here 
\begin{equation}
    \int d^{d-2}x\sqrt{det(g_{ij})}=\phi_{0}^{d-2}\int d^{d-2}x\sqrt{h},
\end{equation}
is regulated.

We will also consider the flat spacetime in planar slicing, in which case the metric takes the form,
\begin{equation}\label{flatmet2}
   ds^2_{M_1} =d\tau^2 +dx^2+\delta_{ij}dz^idz^j,
\end{equation}
where now $x$ and $z^i$ are spatial cartesian coordinates with $x,z^i \in (-\infty,\infty)$.

\subsection{Entangling region \label{entreg}} 
We are interested in computing the entanglement entropy of a region given by: $\tau$ constant, $r_1\leq r \leq r_2$ (or $x_1\leq x \leq x_2$) spanning the range of $y^i$'s (or $z^i$'s) i.e., $M^{d-2}$. Due to symmetries along the $y^i$ (or  $z^i$) directions (spherical or toroidal or translation symmetries), we think such an entangling region is suitable to capture all the interesting dynamics. So our entangling region is a $(d-1)$ dimensional manifold, $A=M^{d-2}\times I$ ($I \in [r_1,r_2]\text{ or } [x_1,x_2]$). Notice that the entangling region has two disconnected boundaries at $r=r_1$ and $r=r_2$.

If we do not have complete isometry along the directions of $M^{d-2}$ (for example, perpendicular to the axis of rotation in the stationary metric), we would have to consider our entangling region to also span a finite length in the corresponding $y^i$ direction(s) with no isometry, in order to capture all interesting dynamics. 

\subsection{Scaling of entanglement entropy}
As mentioned before we are ultimately interested in calculating how the EE of a black hole starting in a pure state varies during evaporation, using the Quantum extremal surface prescription. We assume that the metric remains static throughout evaporation. In doing so we have ignored backreaction or are considering that this evaporation is a quasi-equilibrium process. For this purpose, it is enough to understand how the EE of the QFT modeling the Hawking radiation i.e., $\Sqft$ varies under the scaling of the entangling region \eqref{entreg} in the $r$ direction \eqref{QESpres}. On static spacetimes, this is closely related to the action of the renormalization group operator on $\Sqft$. The variation wrt the energy scale is inversely related to varying the length $\Delta r$ (the only length scale in our setup).\footnote{In non-static cases where we have to take finite lengths in other directions for the entangling region, their variation would also add to the variation of the energy scale.}
\begin{equation}
     \Delta_{\Lambda}\equiv\Lambda(x)\frac{\d}{\d\Lambda(x)}=-\delta \Delta r \frac{\delta}{\delta \Delta r}.
\end{equation}

Let us now look at how the EE $\Sqft$ varies under the scaling of the entangling region in the $r$ direction i.e., under the action of $\delta \Delta r\frac{\delta}{\delta \Delta r}$. Using (\ref{rgop}) on the expression for $\Sqft$ given by (\ref{Sqft}), we have
\begin{align}\label{repentscaling}
   & \delta (\Delta r) \frac{\delta}{\delta (\Delta r)}S_{QFT} =-\Delta_{\Lambda}S_{QFT}\nonumber\\&= -\int d^dx \left(\delta_{\Lambda} g_{\mu\nu}\frac{\delta}{\delta g_{\mu\nu}}+\sigma(x)\beta^i(\lambda)\frac{\delta}{\delta \lambda^i(x)}\right) \left(\underset{q \to 1}{\lim}\frac{1}{(1-q)} (W_q-q W_1)\right)\\
    &= \int d^dx \left(\delta_{\Delta r} g_{\mu\nu}\frac{\delta}{\delta g_{\mu\nu}}-\sigma(x)\beta^i(\lambda)\frac{\delta}{\delta \lambda^i(x)}\right) \left(\underset{q \to 1}{\lim}\frac{1}{(1-q)} (W_q-q W_1)\right).\nonumber
\end{align}

Using (\ref{tuv}) and (\ref{O}), we have
\begin{align}\label{delr}
      &\delta (\Delta r) \frac{\delta}{\delta (\Delta r)} \Sqft \nonumber\\
    =&  \lim_{q \to 1} \frac{1}{(q-1)} \left( \frac{1}{2} \int_{M_q} d^dx \sqrt{g_q}\Bigg( \delta_{\Delta r} g^{(q)}_{\mu \nu} \left< T^{\mu \nu} \right>_q  \right) - \frac{1}{2}q\int_{M_1} d^dx \sqrt{g} \left( \delta_{\Delta r} g_{\mu \nu} \left< T^{\mu \nu} \right>  \right)\nonumber\\
    &-\int_{M_q} d^dx\sqrt{g_q}\sigma(x)\beta^i(\lambda)\left<\mathcal{O}_i(x)\right>_q+q\int_{M_1} d^dx\sqrt{g}\sigma(x)\beta^i(\lambda)\left<\mathcal{O}_i(x)\right>_{q=1} \Bigg).
\end{align}

 Changing the length $\Delta r$ of the entangling region can also be achieved by scaling the coordinates (measuring scale) in the opposite sense which gets reflected in the change of the metric above ($\delta _{\Delta r}g^{(q)}_{\mu\nu}$).

 \begin{figure}[H]
\centering
\includegraphics[scale=0.7]{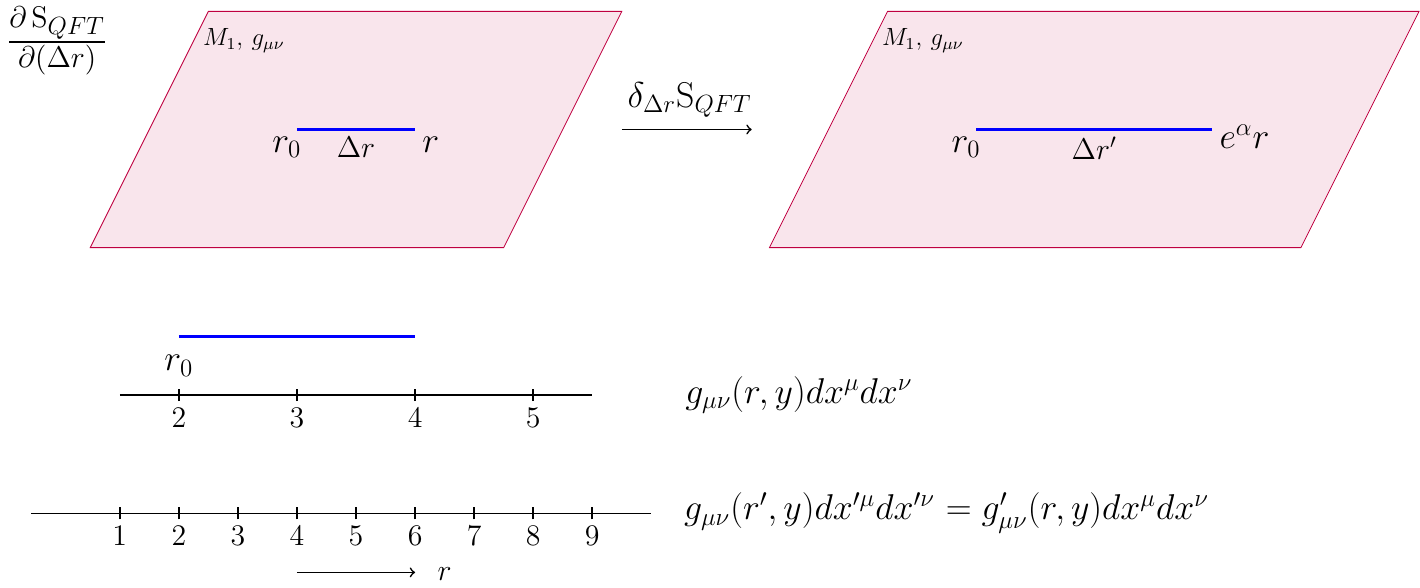}
\caption{Scaling coordinates and metric}
\label{ESBH5}
\end{figure}

\paragraph{$\Sqft$ only receives entangling region and its localised endpoint contributions:} 
From (\ref{delr}) we can see that under $r-$scaling, $\Sqft$ depends on the difference between integrals on the replica manifold and the base manifold. Specifically, the (replica)integrals involve the (replica) stress tensor expectation value wrt the quantum state under consideration and the $r-$scaling of the (replica)metric on the (replica)manifold. As we have already noticed, the metric on the replica manifold differs from the base metric only in a localised region around the boundary of the entangling region where it has a conical singularity. The corresponding replica stress tensor expectation value on the replica manifold $\left< T^{\mu \nu} \right>_q$ differs from the base manifold stress tensor expectation value $\left< T^{\mu \nu} \right>$ via purely metric dependent terms which are localised around the boundary of the entangling region, and non metric dependent (regulator independent) terms which are non trivial over the entire entangling region (we will discuss the regulator independent terms in detail in the following sections). Hence, calculating the $r-$scaling of $\Sqft$ amounts to calculating the difference between integrals on the replica and base manifold or in other words extracting these localised entangling boundary and entangling region contributions (here, coming from the stress tensor and the metric).

\begin{figure}[H]
\centering
\includegraphics[scale=1]{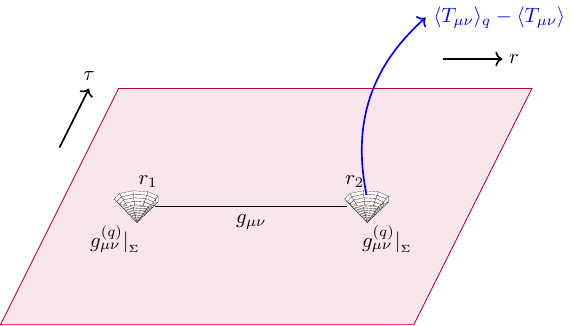}
\caption{Localised endpoint contribution of metric and stress tensor expectation on replica manifold }
\label{ESBH6}
\end{figure}

On $M_q$, we will expand the metric and the stress tensor expectation value around $q=1$ up to $\mathcal{O}(q-1)$ as before using \eqref{repmetexp}. Using this in (\ref{delr}), we get 
\begin{multline}\label{sqftvar}
      \delta(\Delta r) \frac{\delta \Sqft}{\delta(\Delta r)} =  \lim_{q \to 1}  \Bigg\{\Bigg(I_1 \equiv\frac{1}{2} \int_{M_q} d^dx \sqrt{g} \delta_{\Delta r}( ( \d_q g^{(q)}_{\mu \nu})\rvert_{q=1}) \left< T^{\mu \nu} \right> \\+\frac{1}{4} \int_{M_q} d^dx\sqrt{g} g^{\sigma\kappa} (\d_q g^{(q)}_{\sigma\kappa} ) \rvert_{q=1}  \delta_{(\Delta r)} g_{\mu \nu}  \left< T^{\mu \nu} \right> \Bigg) \\
      +\left(I_2 \equiv\frac{1}{2}\int_{M_q} d^dx \sqrt{g}  \left( \delta_{\Delta r} g_{\mu \nu}\right) \left< T^{\mu \nu} \right>^{[1]}\right) \\
        -\frac{1}{q-1}\Bigg(\int_{M_q} d^dx\sqrt{g_q}\sigma(x)\beta^i(\lambda)\left<\mathcal{O}_i(x)\right>_q-q\int_{M_1} d^dx\sqrt{g}\sigma(x)\beta^i(\lambda)\left<\mathcal{O}_i(x)\right>_{q=1}\Bigg) \Bigg\}. 
\end{multline}

Alternatively, we can choose to not expand the determinant of the metric on the replica manifold $g_q$, since we will be evaluating these integrals on the replica manifold. 
\begin{multline}\label{sqftvar0}
      \delta(\Delta r) \frac{\delta \Sqft}{\delta(\Delta r)} =  \lim_{q \to 1}  \Bigg\{\Bigg(I_1'=\frac{1}{2} \int_{M_q} d^dx \sqrt{g_q} \delta_{\Delta r}( ( \d_q g^{(q)}_{\mu \nu})\rvert_{q=1}) \left< T^{\mu \nu} \right>\Bigg) \\
      +\left(I_2'=\frac{1}{2}\int_{M_q} d^dx \sqrt{g_q}  \left( \delta_{\Delta r} g_{\mu \nu}\right) \left< T^{\mu \nu} \right>^{[1]}\right) \\
        -\frac{1}{q-1}\Bigg(\int_{M_q} d^dx\sqrt{g_q}\sigma(x)\beta^i(\lambda)\left<\mathcal{O}_i(x)\right>_q-q\int_{M_1} d^dx\sqrt{g}\sigma(x)\beta^i(\lambda)\left<\mathcal{O}_i(x)\right>_{q=1}\Bigg) \Bigg\}. 
\end{multline}
It however does not make a difference to the final result.\\

In (\ref{sqftvar}) and \eqref{sqftvar0} the integral $I_1$ or $I_1'$ captures the replica contribution coming from the difference between replica and base metric, and $I_2$ or $I_2'$ captures the replica contribution coming from the difference between replica and base stress tensor expectation value. One must note that it is the difference between terms on $M_q$ and $M_1$ that matters for $\Sqft$ and not the actual terms on the replica manifold $M_q$.  

\subsubsection{Local representation of regulator dependent replica terms:} In this section we will briefly review how to deal with integrals over the conical singular manifold $M_q$ of quantities defined on $M_q$, following \cite{Fursaev:1995ef}. In expression (\ref{sqftvar}) and \eqref{sqftvar0}, we have integrals of the kind
\begin{equation}
   I= \underset{q\to1}{lim}\frac{1}{(q-1)}\int_{M_q} d^d x \sqrt{g_q}(F_{\mu\nu}^{(q)}-F^{(1)}_{\mu\nu})G^{\mu\nu(1)},
\end{equation}
where $F_{\mu\nu}^{(q)}$, $G_{\mu\nu}^{(q)}$ are tensors on $M_q$, and $F_{\mu\nu}^{(1)}$, $G^{\mu\nu(1)}$ are the corresponding $O((q-1)^0)$ terms whose values reduce to corresponding tensors on $M_1$. In the integral above, we can expand $\sqrt{g_q}$ in powers of $(q-1)$ and keep terms up to order $(q-1)$. After this expansion, we also have integrals with $\sqrt{g}$ in the volume element. If $F_{\mu\nu}^{(q)}$ depends only on the replica metric then $(F_{\mu\nu}^{(q)}-F_{\mu\nu}^{(1)})$ is localised around the conical singular boundary $r=r_0$ of the entangling region. For the purpose of evaluating these integrals we could always write a local representation of $F_{\mu\nu}^{(q)}$ on $M_q$
\begin{equation}\label{localrep}
    F_{\mu\nu}^{(q)}=F_{\mu\nu}^{(1)}+ N_{\mu\nu} (q-1) \delta_{\Sigma}(\rho)+ \mathcal{O}\left( (q-1)^2 \right),
\end{equation}
where $N_{\mu\nu}$ is given by,
\begin{equation}
   N_{\mu\nu} = \underset{q\to1}{lim}\frac{1}{(q-1)Vol(M_{d-2})}\int_{M_q} d^d x \sqrt{g_q}  (F_{\mu\nu}^{(q)}-F^{(1)}_{\mu\nu}),
\end{equation}
where we have defined (normalised) the delta function $\delta_{\Sigma}(\rho)$ as,
\begin{equation}
    \int_{M_q} d^dx \sqrt{g_q} f\delta_{\Sigma}(\rho)=\int_{\Sigma}  d^{d-2}x \sqrt{h} f,
\end{equation}
and $Vol(M_{d-2})=\int_{\Sigma}d^{d-2} x\sqrt{h}$. With this we have,
\begin{equation}
    I= N_{\mu\nu}\int_{\Sigma}d^{d-2}x \sqrt{h} G^{\mu\nu(1)}.  
\end{equation}

When working with integrals of the kind $I$(i.e., involving the regulating function $U(\rho,a)$), it is convenient to work in $x=\rho/a$ coordinate since the regulating function $U(x)$ is independent of concial singularity regulator $(a)$ in these coordinates (see \eqref{limits2}) which makes it easy to read off the $a$ dependence in the integral.

\paragraph{Volume element:}The volume element in $x=\rho/a$ coordinate is given by
\begin{align}\label{volele}
  \sqrt{g_q} d^d x 
  &= a^2 x \sqrt{U(x)}  \sqrt{det(g_{ij})} dx d\psi d^{d-2}y^i;
  \end{align}
    where 
    \begin{equation}
    g_{ij}=\phi_0^2 h_{ij},\,\delta_{ij};  
    \end{equation}
     for $g_{\mu\nu}^{(q)}$ given by \eqref{repmet1}, and
  \begin{equation}
      g_{ij}=\gamma_{ij}(y)+2 \rho^p c^{1-p}\cos(\psi) K^{x_1}_{ij}+\dots
  \end{equation}  
  for the static case \eqref{repmetstat}. For $g_{\mu\nu}^{(q)}$ given by \eqref{sqcone}, we have
  \begin{align}
  \sqrt{g_q} d^d x 
   &= a^2 x \sqrt{U(x)}     \left(r_0 + (a x)^p c^{1-p} \cos (\psi)\right)^{(d-2)} \sqrt{h} dx d\psi d^{d-2}y^i.
\end{align}
We have not written the explicit form for a general non-static replica metric \eqref{repgen} since it also involves non-diagonal components and is easier to write for the specific case required.\\

We expand the volume element corresponding to \eqref{sqcone} in powers of $a$,
\begin{align}\label{volelea}
    \sqrt{g_q} d^d x =& a^2 x \sqrt{U(x)} \sqrt{h} dx d\psi d^{d-2}y^i \nonumber\\
    &\Bigg(\frac{1}{24} a^{4p} x^{4p} c^{4(1-p)} (d-5) (d-4) (d-3) (d-2) r_0^{d-6} \cos ^4(\psi )
    \nonumber\\
    &+\frac{1}{6} a^{3p} x^{3p} c^{3(1-p)} (d-4) (d-3) (d-2)  r_0^{d-5} \cos ^3(\psi )\nonumber\\
    &+\frac{1}{2}
   a^{2p} x^{2p} c^{2(1-p)} (d-3) (d-2)  r_0^{d-4} \cos ^2(\psi )\nonumber\\
    &+a^p x^p c^{1-p} (d-2)  r_0^{d-3} \cos (\psi )+r_0^{d-2}+O(a^{5p})\Bigg).
\end{align}
We need the limit in which the regulator $a \to 0$ (since this amounts to setting $U\to 1$). In $d-$dimensions, the most subleading term in $a$ in the volume element has $a$ dependence $a^{i=2+(d-2)p}$(flat squashed cones) and $a^{i=2}$(flat, zero extrinsic curvature). Therefore, in a $d-$dimensional spacetime, we need to consider all terms in $(F_{\mu\nu}^{(q)}-F^{(1)}_{\mu\nu})$ with coefficient $1/a^{r\geq i}$.

A similar expression and result for $\sqrt{g}d^d x$ (appearing in replica integrals) can be obtained by simply setting $q\to1$ ($U(x)=1$) in \eqref{volele} and \eqref{volelea}.

\paragraph{Example:}Consider the following regulated conical singular metric
\begin{equation}
    ds_q^2 = e^\sigma(U(\rho,a)d\rho^2+d\psi^2)=e^\sigma(ds_{\tilde{C}}^2); \quad \sigma=\sigma_1\rho^2+\sigma_2 \rho^4+...
\end{equation}
We have the Ricci scalar given by,
\begin{equation}
    R = e^{-\sigma} R_{\tilde{C}}-e^{-\sigma}\Box_{\tilde{C}}\sigma.
\end{equation}
$ R_{\tilde{C}}$ is the Ricci on $ds_{\tilde{C}}^2$ given by $\frac{U'(\rho)}{\rho U^2}$. We have 
\begin{equation}
\begin{split}
    \int_{M_{q(reg)}} \sqrt{g_q} R &=  2\pi q\int_{\rho=0}^{\rho=\infty}d\rho  \frac{U'(\rho)}{U^{3/2}}-\int_{\rho=0,\psi=0}^{\rho=\infty,\psi=2\pi q}\sqrt{U}\rho d\rho d\psi \Box_{\tilde{C}}\sigma\\
    &=4\pi(1-q)-\int_{\rho=0,\psi=0}^{\rho=\infty,\psi=2\pi q}\sqrt{U}\rho d\rho d\psi \Box_{\tilde{C}}\sigma.
\end{split}
\end{equation}
The differences between the replica and base space tensors that depend on the regulator $(a)$ are zero for $\rho>>a$. Therefore, for integrals involving the difference it is enough to do the integrals between $\rho \in [0,\rho_0]$ for $\rho_0>>a$. This must give the same result as taking the limit $\rho_0$ to $\infty$. For example:  

\begin{equation}
\begin{split}
    \int_{M_{q(reg)}} \sqrt{g_q} R_q -   \int_{M} \sqrt{g} R &=  \left(2\pi q\int_{\rho=0}^{\rho_0}d\rho  \frac{U'(\rho)}{U^{3/2}}\right)\Bigg\rvert_{\rho_0>>a}\\
    &=4\pi q\left(\frac{1}{q}-\frac{1}{U(\rho_0,a)}\right)\Bigg\rvert_{\rho_0>>a} = 4\pi(1-q).
\end{split}
\end{equation}

Let $\Sigma$ be the singular set i.e., $\rho=0$. The first term is independent of regulator $(a)$, the second term depends on $a$ but in the  $a\to 0$ it reduces to the integral curvature on the smooth domain $M_q/\Sigma$ i.e.,
\begin{equation}
\begin{split}
   \underset{a\to 0}{lim} \int_{M_{q(reg)}} \sqrt{g_q} R
    &=4\pi(1-q)+\int_{M_q/\Sigma}R.
\end{split}
\end{equation}
We can write the local representation for $R$,
\begin{equation}
    R^{(q)} = R + 2\frac{(1-q)}{q} \delta(\rho).
\end{equation}
In higher dimensions, for the metric with no extrinsic curvature for boundary $\Sigma$ embedded in $M_1$, the replica metric is the unsquashed cone metric given by
\begin{equation}
    ds_q^2 = e^\sigma(U(\rho,a)d\rho^2+d\psi^2+(\gamma_{ij}(x^i)+h_{ij}(x^i)\rho^2)dx^i dx^j+...)=e^\sigma(d\tilde{s}^2),
\end{equation}
where $\sigma=\sigma_1\rho^2+\sigma_2 \rho^4+...$, we can similarly show, following \cite{Fursaev:1995ef},
 \begin{align}\label{localcurv}
     R^{(q)} &= R+4\pi(1-q) \delta_{\Sigma},\nonumber\\
     R^\mu_{\ \nu}{}^{(q)}&= R^\mu_{\ \nu} +2\pi(1-q)(n^\mu n_\nu) \delta_{\Sigma},\nonumber\\ R^{\mu\nu}_{\ \ \alpha\beta}{}^{(q)}&=R^{\mu\nu}_{\ \ \alpha\beta}+2\pi(1-q)((n^\mu n_\alpha)( n^\nu n_\beta)-(n^\mu n_\beta)( n^\nu n_\alpha)) \delta_{\Sigma},
 \end{align}
where $\int_M \delta_{\Sigma} f=\int_\Sigma f$ and $n^k=n^k_\mu dx^\mu$ are two orthonormal vectors orthogonal to the boundary $\Sigma$ with $n_\mu n_\nu = \sum_{k=1}^2 n^k_\mu n^k_\nu$. Notice that the replica metric has a non zero curvature even if the metric on $M_1$ has zero curvature (for example flat metric on $M_1$). 

The conical singular replica metric expanded around endpoints i.e., \eqref{repmet1}, does not include the $e^\sigma$ conformal factor. Therefore we will only get the $O(q-1)$ terms when computing any geometric quantity. However, since we are computing differences between replica and base manifold quantities, this difference automatically cancels the $O((q-1)^0)$ piece giving us the $O(q-1)$ terms. Therefore, the replica metric without the conformal $e^\sigma$ factor is enough for calculating $r-$scaling of $\Sqft$ or such integral differences between replica and base manifold. In any case the $O((q-1)^0)$ term is simply the quantity evaluated on the smooth domain $M_q/\Sigma$ or on $M_1$. 

If $A^{(q)}$ and $B^{(q)}$ are two quantities on $M_q$, we have

\begin{align}
    &\int_{M_q} A^{(q)}=q\int_{M_1}A+(q-1)\int_{M_q} A^{[1]}+O((q-1)^2)\nonumber\\
     &\quad\quad\quad\quad=q\int_{M_1}A+(q-1)N_1\int_{\Sigma}+O((q-1)^2);\nonumber\\
     &\int_{M_q} B^{(q)}=q\int_{M_1}B+(q-1)\int_{M_q} B^{[1]}+O((q-1)^2)\nonumber\\
     &\quad\quad\quad\quad=q\int_{M_1}B+(q-1)N_2\int_{\Sigma}+O((q-1)^2);
\end{align}
and, 
\begin{equation}
    \int_{M_q} A^{(q)}B^{(q)}= q\int_{M_1}AB+(q-1) \int_{\Sigma}(N_1 B+N_2 A)+O((q-1)^2).
\end{equation}
Using this and \eqref{localcurv} we have
\begin{align}
    &\int_{M_q} R^{(q)}{}^2\Bigg\rvert_{K^i_{\mu\nu}=0}=q\int_{M_1}R^2-8\pi(q-1)\int_{\Sigma} R+O((q-1)^2),\nonumber\\
    &\int_{M_q} R_{\mu\nu}^{(q)}R^{\mu\nu}{}^{(q)}\Bigg\rvert_{K^i_{\mu\nu}=0}=q\int_{M_1}R_{\mu\nu}R^{\mu\nu}-4\pi(q-1)\int_{\Sigma} R_{\mu\nu}n_i^\mu n_i^\nu+O((q-1)^2),\nonumber\\
     &\int_{M_q} R_{\mu\nu\alpha\beta}^{(q)}R^{\mu\nu\alpha\beta}{}^{(q)}\Bigg\rvert_{K^i_{\mu\nu}=0}=\nonumber\\
&q\int_{M_1}R_{\mu\nu\alpha\beta}R^{\mu\nu\alpha\beta}-8\pi(q-1)\int_{\Sigma} R_{\mu\nu\alpha\beta}n_i^\mu n_i^\alpha n_i^\nu  n_i^\beta+O((q-1)^2).
\end{align}
In the case of a general entangling region (general background and foliation) with two non zero extrinsic curvatures of $\Sigma$, $K^{(i)}_{\mu\nu}$ with $i=1,2$, the replica metric takes the form \eqref{repgen}. The integrals of Riemann contractions given above are modified in the case of such squashed cones due to additional contributions from the two extrinsic curvature terms (see \cite{Fursaev:2013fta}). We have,
\begin{align}\label{riemcontrsqcone}
    &\int_{M_q} R^{(q)}{}^2=q\int_{M_1}R^2-8\pi(q-1)\int_{\Sigma} R+O((q-1)^2),\nonumber\\
     &\int_{M_q} R_{\mu\nu}^{(q)}R^{\mu\nu}{}^{(q)}=q\int_{M_1}R_{\mu\nu}R^{\mu\nu}-4\pi(q-1)\int_{\Sigma} \left(R_{\mu\nu}n_i^\mu n_i^\nu-\frac{1}{2} K^2\right)+O((q-1)^2),\nonumber\\
     &\int_{M_q} R_{\mu\nu\alpha\beta}^{(q)}R^{\mu\nu\alpha\beta}{}^{(q)}=\nonumber\\     &q\int_{M_1}R_{\mu\nu\alpha\beta}R^{\mu\nu\alpha\beta}-8\pi(q-1)\int_{\Sigma} \left(R_{\mu\nu\alpha\beta}n_i^\mu n_i^\alpha n_i^\nu  n_i^\beta-\text{Tr }K^2\right)+O((q-1)^2),
\end{align}
where $K^2=\sum_{i=1}^2 K^{(i)}_{\mu\nu}K^{(i)}{}^{\mu\nu}$ and $\text{Tr }K^2=\sum_{i=1}^2 (\text{Tr } K^{(i)})^2$. We have also suppressed the volume element in the integrals above i.e., $\int_{M_q}\equiv\int_{M_q}d^dx \sqrt{g_q}$ and $\int_{\Sigma}\equiv\int_{\Sigma}d^{d-2}x \sqrt{h}$.

Specifically for spherical entangling boundaries in $4$ dimensions where replica metric is given by \eqref{sqcone},
\begin{equation}\label{Ksph}
    \int_\Sigma K^2 =16\pi, \quad \int_\Sigma\text{Tr }K^2 =8\pi,
\end{equation}
and for cylindrical entangling boundaries in 4 diimensions where replica metric is given by \eqref{sqconecyl},
\begin{equation}\label{Kcyl}
    \int_\Sigma K^2 =\int_\Sigma\text{Tr }K^2 = 2\pi\frac{L}{r_0},
\end{equation}
where $L$ is the length of the cylinder in the $z$ direction.

Using \eqref{riemcontrsqcone}, we also note down integrals of certain useful combination of Riemann contractions that show up in trace anomalies in 4 dimensions,
\begin{align}\label{intEI}
&\int_{M_q}d^dx\sqrt{g_q} \left(E_4^{(q)}=R^{(q)}_{\alpha\beta\mu\nu}R^{\alpha\beta\mu\nu}{}^{(q)}-4R^{(q)}_{\mu\nu}R^{\mu\nu}{}^{(q)}+R^{(q)}{}^2 \right)\nonumber\\
&=\int_{M_1}d^dx\sqrt{g} E_4+16\pi^2(1-q) I^\Sigma_{a},\nonumber\\
&\int_{M_q}d^dx\sqrt{g_q} \left(I_4^{(q)}=R^{(q)}_{\alpha\beta\mu\nu}R^{\alpha\beta\mu\nu}{}^{(q)}-2 R^{(q)}_{\mu\nu}R^{\mu\nu}{}^{(q)}+\frac{1}{3}R^{(q)}{}^2\right)\nonumber\\
&=\int_{M_1}d^dx\sqrt{g} I_4+16\pi^2(1-q) I^\Sigma_{c},
\end{align}
where 
\begin{align}\label{intresEI}
    I^\Sigma_{a}&=\frac{1}{2\pi}\int_\Sigma (R_{abab}-2 R_{aa}+R+K^2-\text{Tr }K^2),\nonumber\\
    I^\Sigma_{c}&=\frac{1}{2\pi}\int_\Sigma \left(R_{abab}-R_{aa}+\frac{1}{3} R+\frac{1}{2} K^2-\text{Tr }K^2\right).
\end{align}
$R$ is the Ricci scalar of metric $g_{\mu\nu}$, $n^\mu_a$, $a=1,2$ is a pair of unit vectors orthogonal to the surface $\Sigma$, $R_{aa}= R_{\mu\nu}n^\mu_a n^\nu_a$, $R_{abab}=R_{\alpha\beta\mu\nu}n^\alpha_a n^\beta_b n^\mu_a n^\nu_b$, is projection of the Ricci and Riemann tensor onto the subspace orthogonal to surface $\Sigma$.

Specific values of integrals of Riemmann contraction and $K^2$, Tr $K^2$ in 5 and 6 dimensions for different choices of hypersurface $\Sigma$ have also been calculated in \cite{Fursaev:2013fta}.

\subsubsection{Scaling of entangling region}\label{scaling sec}
 Let $r=r_1$ and $r=r_2$ be the two boundaries of the entangling region with $r_2>r_1$.  On a Minkowski background, we have translation symmetry along the $r-$direction, and hence $\Sqft{}_{\text{,flat}}$ depends only on the difference $(r_2-r_1)$. However, there is no $r-$translation symmetry on non-Minkowski static metric solutions in \eqref{statmetric}. Hence $\Sqft$ on non-Minkowski static spacetimes depends independently on $r_1$ and $r_2$ and not just on the difference $(r_2-r_1)$. Hence, we fix one end of the entangling region and scale the other to capture this dependence. This scaling behavior is also useful in the discussion of Quantum extremal surfaces \cite{Engelhardt:2014gca}. If $r_0$ is the fixed end, we have 
\begin{align}\label{scaling}
  \lvert r - r_0 \rvert \to e^{\alpha}\lvert r - r_0 \rvert \implies r\to r' = r_0 + e^{\alpha} (r-r_0).
\end{align}
This works for both $r>r_0$ and $r<r_0$. 

We will keep the farther end $r_2$ fixed, and scale $r_1$ i.e., $r_2 \to r_2$ and $r_1 \to r_2+e^\alpha(r_1-r_2)$. Under this scaling $\Sqft[r_1,r_2] \to \Sqft[r_2+e^{\alpha}(r_1-r_2),r_2]$. The LHS of (\ref{sqftvar}) is now the difference between both of these upto $\mathcal{O}(\alpha)$ 
\begin{equation}
  \Sqft[r_2+e^{\alpha}(r_1-r_2),r_2]-\Sqft[r_1,r_2]=  -\alpha(\Delta r)\frac{\d \Sqft}{\d r_1} + \mathcal{O}(\alpha^2),
\end{equation}
where $\Delta r =(r_2-r_1)$.

If we however keep $r_1$ fixed and scale $r_2$, i.e., $r_1 \to r_1$ and $r_2 \to r_1+e^\alpha(r_2-r_1)$, we have
\begin{equation}
  \Sqft[r_1,r_1+e^{\alpha}(r_2-r_1)]-\Sqft[r_1,r_2]=  \alpha(\Delta r)\frac{\d \Sqft}{\d r_2} + \mathcal{O}(\alpha^2).
\end{equation}
On the RHS of (\ref{sqftvar}) we implement the same scaling as follows. On the RHS we have $\delta_{\Delta r} g_{\mu \nu}$, $\delta_{\Delta r} g^{(q)}_{\mu \nu}$ and $\delta_{\Delta r}(\d_q g^{(q)}_{\mu \nu})\rvert_{q=1}$. To demonstrate the scaling we will use $f_{\mu \nu}$ as a proxy for $g_{\mu \nu}$, $ g^{(q)}_{\mu \nu}$ and $\d_q g^{(q)}_{\mu \nu}$. Under scaling $f_{\mu \nu}(x) dx^\mu dx^\nu \to f_{\mu \nu}(x') dx'^\mu dx'^\nu=f'_{\mu \nu}(x,\alpha) dx^\mu dx^\nu$ (here $x'^\mu$ denotes scaled coordinates and in our case we are only scaling the $r$ direction keeping $r_0$ fixed). Here we keep one endpoint $r_0$ ($r_1$ or $r_2$) fixed and scale all other points along the $r$ direction as in equation (\ref{scaling}). Then $\delta_{\Delta r} f_{\mu \nu}(x) = f'_{\mu \nu}(x,\alpha)-f_{\mu \nu}(x) $ upto $\mathcal{O}(\alpha)$\\

\begin{align}\label{scalingst} 
   &\delta_{\Delta r} f_{rr}(\tau,r,x^i)= \alpha \left((r-r_0) \d_r + 2 \right)f_{rr}(\tau,r,x^i),\nonumber\\
     &\delta_{\Delta r} f_{r\tau}(\tau,r,x^i)= \alpha \left((r-r_0) \d_r + 1  \right)f_{r\tau}(\tau,r,x^i),\nonumber\\
   &\delta_{\Delta r} f_{\tau\tau}(\tau,r,x^i)= \alpha (r-r_0) \d_r f_{\tau\tau}(\tau,r,x^i),\nonumber\\
   &\delta_{\Delta r} f_{ij}(\tau,r,x^i)= \alpha (r-r_0) \d_r f_{ij}(\tau,r,x^i).
\end{align}

\subsection{Replica metric contribution to \texorpdfstring{$\Sqft$}{}}
The replica contribution in integral $I_1$ in (\ref{sqftvar}) is from the difference between replica and base metric. This contribution is entirely from the regulated conical singularity and is localised around the endpoints of the entangling region. Hence using the replica metric expanded around the boundary, given by (\ref{repmet1}) and \eqref{repmetsq} is enough to capture this. To calculate $I_1$ we first calculate the local representation of $g^{(q)}_{\mu \nu}$ and $\delta_{\Delta r} g^{(q)}_{\mu \nu}$. 

Noting $p(q=1)=1$, in $\{\rho,\psi,y^i\}$ basis, we have,
\begin{align}\label{dqguv}
     \d_q g^{(q)}_{\mu \nu})\rvert_{q=1} =&\d_q U(\rho)\rvert_{q=1}\equiv U^{[1]}(\rho); \quad \mu=\nu=\rho,\nonumber\\
     =&2\rho \cos(\psi) (r_0+\rho\cos(\psi))\log\left(\frac{\rho}{c}\right)  p'(q)\rvert_{q=1} h_{ij}(y);\nonumber\\
     &\mu=i,\nu=j \text{ (flat metric \eqref{sqcone})},\nonumber\\     
    =& 2\rho\left(\cos(\psi)K^1_{ij}+\sin(\psi)K^2_{ij}\right)\log\left(\frac{\rho}{c}\right) p'(q)\rvert_{q-1}; \nonumber\\
    & \mu=i,\nu=j \text{ (general metric \eqref{repgen})},\nonumber\\
    =& 0; \text{ otherwise. }
\end{align}   

In $\{r,\tau,y^i\}$ basis (using \eqref{conversion}), we have,
\begin{align}
     &\d_q g^{(q)}_{\mu \nu})\rvert_{q=1}\nonumber\\
     &=\frac{(r-r_0)^2 U^{[1]}(\rho)}{(r-r_0)^2+(\tau-\tau_0)^2}=\cos(\psi)^2 U^{[1]}(\rho); \quad \mu=\nu=r,\nonumber\\
     &=\frac{(\tau-\tau_0)^2 U^{[1]}(\rho)}{(r-r_0)^2+(\tau-\tau_0)^2}=\sin(\psi)^2 U^{[1]}(\rho); \quad \mu=\nu=\tau,\nonumber\\
     &=\frac{(r-r_0)(\tau-\tau_0) U^{[1]}(\rho)}{(r-r_0)^2+(\tau-\tau_0)^2}=\sin(\psi)\cos(\psi) U^{[1]}(\rho); \quad \mu=r,\nu=\tau \text{ or vice versa}.
\end{align}
All other components are same as that in \eqref{dqguv}. We also have,
\begin{equation}
    \alpha(r-r_0)\d_r f(\rho,\psi)=\alpha\cos(\psi)(\cos(\psi)\rho \d_\rho f-\sin(\psi)\d_\psi f).
\end{equation}
Using this and (\ref{scalingst}) we have,
\begin{align}
&\delta_{\Delta r} (( \d_q g^{(q)}_{\mu \nu})\rvert_{q=1}) \nonumber\\
=& \alpha\left(U^{[1]}(x)\cos(\psi)^2[3-\cos(2\psi)]+x\d_xU^{[1]}(x)\cos(\psi)^4\right); \quad \mu=\nu=r,\nonumber \\
    =&\alpha\frac{\sin(2\psi)^2}{4}\left(-2 U^{[1]}(x)+x\d_xU^{[1]}(x)\right); \quad \mu=\nu=\tau, \nonumber\\
    =&\alpha \cos(\psi)^4\left(2 \tan(\psi)^3U^{[1]}(x)+\tan(\psi)x\d_xU^{[1]}(x)\right); \,\mu=r,\nu=\tau \text{ or vice versa}, \nonumber\\  
    =&2 \alpha \rho  \cos (\psi ) p'(q)\rvert_{q=1}\left(\cos^2(\psi)(r_0+\rho\cos(\psi))+\log\left(\frac{\rho}{c}\right)(r_0+2\rho\cos(\psi))\right) h_{ij};\nonumber\\
    & \mu=i,\nu=j \text{ for }g_{\mu\nu}^{(q)}-\eqref{sqcone},\nonumber\\
    =&2 \alpha \rho  \cos (\psi ) p'(q)\rvert_{q=1}\Bigg\{ \frac{\sin (2\psi )}{2} K_{2 ij} +K_{1ij} \left(\cos ^2(\psi ) + \log \left(\frac{\rho }{c}\right)\right)\nonumber\\
   &+\log \left(\frac{\rho }{c}\right)\Big( \cos(\psi)\left(\sin (\psi ) \rho\d_\rho K_{2 ij}+\cos (\psi )\rho\d_\rho K_{1ij}\right)\nonumber\\
   &+\sin(\psi)(\sin(\psi)\d_\psi K_{2ij}-\cos(\psi)\d_\psi K_{1ij})\Big)\Bigg\};\quad\mu=i,\nu=j \text{ for }g_{\mu\nu}^{(q)}-\eqref{repgen}, \nonumber\\
    =&0; \text{ otherwise},
\end{align}
where $x=\rho/a$. We have kept $U(\rho,a)$ general in the expressions above.

We notice that $\d_q g^{(q)}_{\mu \nu}\rvert_{q=1}$  and $\delta_{\Delta r} (( \d_q g^{(q)}_{\mu \nu})\rvert_{q=1})$ have only an $\mathcal{O}(a^{i>0})$ term. Since the volume element (\ref{volele}) is at least $O(a^2)$, therefore
\begin{align}
    & \int_{\rho=0,\psi=0;\forall y^i}^{\rho=\rho_0,\psi=2\pi q} d^dx \sqrt{g_q}   \d_q g^{(q)}_{\mu \nu}\rvert_{q=1}=0\text{ as } \{a\to0,\rho_0 \to\infty\},\label{metlocalrepint2} \\
    & \int_{\rho=0,\psi=0;\forall y^i}^{\rho=\rho_0,\psi=2\pi q} d^dx \sqrt{g_q} \delta_{\Delta r} ( \d_q g^{(q)}_{\mu \nu})\rvert_{q=1}=0\text{ as } \{a\to0,\rho_0 \to\infty\}.\label{metlocalrepint1}
\end{align}
Although the integral vanishes in the $a\to0$ limit (due to positive powers of $a$ in the integrand), the integral itself diverges before taking the $a\to0$ limit. This divergence is due to the upper limit being $\infty$ in the $\rho$ integral which can be made finite by integrating upto a finite upper limit $\rho_0$, and then taking the limit $\{a\to0, \rho_0\to\infty\}$ such that the terms involving the combination of both tend to zero i.e., $a\to0$ faster than the leading power of $\rho_0$ (or terms involving $\rho_0$) tends to $\infty$. Since one can always choose how fast $a\to 0$, the integral (\ref{metlocalrepint2}) and (\ref{metlocalrepint1}) holds.

We notice that \eqref{metlocalrepint2} is also true for a general replica metric \eqref{repgen} (for entangling surface $x^a=0$; $a=1,2$), as can be seen from \eqref{dqguv} having only $\mathcal{O}(a^{i>0})$ terms. Now, even if we scale $\d_q g_{\mu\nu}^{(q)}\rvert_{q=1}$corresponding to general replica metric \eqref{repgen}, by an general length scaling $\delta_L$, since $\delta_L$ is dimensionless, therefore $\delta_L(\d_q g_{\mu\nu}^{(q)}\rvert_{q=1})$ will also have only $\mathcal{O}(a^{i>0})$ terms, therefore we have
\begin{align}
    \int_{\rho=0,\psi=0;\forall y^i}^{\rho=\infty,\psi=2\pi q} d^dx \sqrt{g_q} \delta_{L} ( \d_q g^{(q)}_{\mu \nu})\rvert_{q=1}=0\text{ as } \{a\to0,\rho_0 \to\infty\}.
\end{align}
Therefore, for the replica integrals, we can use the local representation
\begin{align}
g^{(q)}_{\mu \nu}&= g_{\mu \nu}\text{ up to } \mathcal{O}((q-1)), \label{locrepofmet}\\
    \delta_{\Delta r} (g^{(q)}_{\mu \nu})&=\delta_{\Delta r} (g_{\mu \nu})\text{ up to } \mathcal{O}((q-1))\label{locrepofmetscal},\nonumber\\
     \delta_{L} (g^{(q)}_{\mu \nu})&=\delta_{L} (g_{\mu \nu})\text{ up to } \mathcal{O}((q-1)).
\end{align}
i.e., for the purpose of a replica integral we have $\d_q g^{(q)}_{\mu \nu}\rvert_{q=1}=0$ and $\delta_{\Delta r} ( \d_q g^{(q)}_{\mu \nu})\rvert_{q=1}=0$. Therefore 
\begin{align}
   I_1 \equiv &\frac{1}{2} \int_{M_q} d^dx \sqrt{g} \delta_{\Delta r} ( \d_q g^{(q)}_{\mu \nu})\rvert_{q=1} \left< T^{\mu \nu} \right> \nonumber\\
   +&\frac{1}{4} \int_{M_q} d^dx\sqrt{g} g^{\sigma\kappa} (\d_q g^{(q)}_{\sigma\kappa} ) \rvert_{q=1}  \delta_{\Delta r} g_{\mu \nu}  \left< T^{\mu \nu} \right> =0,\nonumber\\
   I_1' \equiv&\frac{1}{2} \int_{M_q} d^dx \sqrt{g_q} \delta_{\Delta r} ( \d_q g^{(q)}_{\mu \nu})\rvert_{q=1} \left< T^{\mu \nu} \right>  =0.
\end{align}

\subsection{Replica stress tensor contribution to \texorpdfstring{$\Sqft$}{}}
We now have,
\begin{multline}\label{sqftvar2}
       \{-\alpha (r_2-r_1)\} \frac{\partial \Sqft}{\partial r_1} =  \lim_{q \to 1}  \Bigg\{\left(I_2 \equiv\frac{1}{2}\int_{M_q} d^dx \sqrt{g}  \left( \delta_{\Delta r} g_{\mu \nu}\right) \left< T^{\mu \nu} \right>^{[1]}\right) \\
        -\frac{1}{q-1}\Bigg(\int_{M_q} d^dx\sqrt{g_q}\sigma(x)\beta^i(\lambda)\left<\mathcal{O}_i(x)\right>_q-q\int_{M_1} d^dx\sqrt{g}\sigma(x)\beta^i(\lambda)\left<\mathcal{O}_i(x)\right>_{q=1}\Bigg) \Bigg\}. 
\end{multline}
$I_2$ captures the contribution coming from the difference between the replica and base stress tensor expectation value. Using (\ref{scalingst}) we have, 
\begin{fleqn}
\begin{equation}
\begin{split}\label{sqftvar3}
     -&\alpha (r_2-r_1) \frac{\partial \Sqft}{\partial r_1} \\
     =&  \alpha \lim_{q \to 1}\frac{1}{(q-1)}\Bigg\{\int_{M_q}d^dx \sqrt{g}\Bigg[\frac{(r-r_2)}{2}(\d_r g_{rr}) \big< T^{rr} \big>^{[1]}+\Bigg(\frac{(r-r_2)}{2}(\d_rg_{\tau\tau})-g_{\tau\tau}\Bigg)\\
     &\hspace{2.5cm}\big< T^{\tau\tau} \big>^{[1]}-r r_2 h_{ij}\big< T^{ij} \big>^{[1]}+\Bigg(g_{\mu\nu}{}_{, \,q=1}\big< T^{\mu \nu} \big>_q-\big< T^\mu_\mu \big>_{q=1}\Bigg)\Bigg]\\
     &\hspace{1.8cm}-\Bigg(\int_{M_q} d^dx\sqrt{g_q}\beta^i(\lambda)\left<\mathcal{O}_i(x)\right>_q-q\int_{M_1} d^dx\sqrt{g}\beta^i(\lambda)\left<\mathcal{O}_i(x)\right>_{q=1}\Bigg)\Bigg\}.
\end{split}
\end{equation}
\end{fleqn}
where the transformation is such that $\sigma(x)=\alpha$.\\

 We could add $\alpha \underset{q\to1}{\text{lim}}\int_{M_q}d^dx \sqrt{g} \big< T^{\mu \nu} \big>_q (\partial_q g^{(q)}_{\mu \nu})\rvert_{q=1}=0$ (using \eqref{metlocalrepint2} or \eqref{locrepofmet}) in the last line to get the trace of stress tensor on $M_q$.

 Here we have kept the farther end $r_2$ fixed and scaled all other points along the $r$ direction. Alternatively, we could fix $r_1$ and scale all other points. In that case we would get the expression above i.e., \eqref{sqftvar3} with $r_1$ interchanged with $r_2$. 

\paragraph{Scaling $\tau$:}Since $\Sqft$ on a static spacetime is independent of $\tau$, therefore we could also scale the $\tau$ direction suitably with no effect other than writing our expressions more compactly as we will see.

We fix $r_2$ and scale all other points in the $r$ direction as described in section (\ref{scaling sec}) and in addition scale $\tau \to \tau'=e^{\alpha} \tau$. Under this we have $\Sqft[r_1, r_2, \tau] \to \Sqft[r_2+e^\alpha (r_1-r_2), r_2, e^\alpha \tau]$. On the LHS we have

\begin{align}
   \Sqft[r_2+e^\alpha (r_1-r_2), r_2, e^\alpha \tau]- \Sqft[r_1, r_2, \tau]&=-\alpha (\Delta r) \frac{\d \Sqft}{\d r_1} + \tau\frac{\d \Sqft}{\d \tau} \nonumber\\&= -\alpha (\Delta r) \frac{\d \Sqft}{\d r_1}.
\end{align}
since $\frac{\d \Sqft}{\d \tau}=0$ on a static background.

On the RHS of \eqref{sqftvar2} we have to replace $\delta_{\Delta r} g_{\mu \nu}$ with $\delta_{\Delta \tau}\delta_{\Delta r} g_{\mu \nu}$. Where the latter stands for now scaling the $r$ and $\tau$ coordinates together in both the metric and the measure in the line element, and taking the difference between the resulting metric and the unscaled metric upto $\mathcal{O}(\alpha)$. In the static case none of the metric components depend on $\tau$ and there are no cross terms such as $dr d\tau$ in the line element. Therefore scaling $\tau$ along with $r$ in the line element only introduces one non-trivial term for the $\delta_{\Delta \tau}\delta_{\Delta r}g_{\tau\tau}$ component, where there is an additional term $2\alpha g_{\tau\tau}$ coming from scaling $d\tau^2$. We have
\begin{equation}
   \delta_{\Delta \tau}\delta_{\Delta r} g_{\tau \tau} =  \alpha (r-r_2) \d_r g_{\tau\tau}+2\alpha g_{\tau\tau}.
\end{equation}
Introducing this in (\ref{sqftvar2}) we get
\begin{align}\label{sqftvar4}
     &\{-\alpha (r_2-r_1)\} \frac{\partial \Sqft}{\partial r_1} \nonumber\\
     &= \alpha \lim_{q \to 1}\frac{1}{(q-1)}\Bigg\{\int_{M_q}d^dx \sqrt{g_q}\Bigg[\frac{(r-r_2)}{2}(\d_r g_{rr}) \big< T^{rr} \big>^{[1]}
     +\frac{(r-r_2)}{2}(\d_rg_{\tau\tau})\big< T^{\tau\tau} \big>^{[1]}\nonumber\\
     &\hspace{5cm}-r r_2 h_{ij}\big< T^{ij} \big>^{[1]}+\Bigg(\big< T^{\mu}_\mu \big>_q-\big< T^\mu_\mu \big>_{q=1}\Bigg)\Bigg]\nonumber\\
     &\hspace{2.3cm}-\Bigg(\int_{M_q} d^dx\sqrt{g_q}\beta^i(\lambda)\left<\mathcal{O}_i(x)\right>_q-q\int_{M_1} d^dx\sqrt{g}\beta^i(\lambda)\left<\mathcal{O}_i(x)\right>_{q=1}\Bigg)\Bigg\}.
 \end{align}
Instead if we keep only $r_1$ fixed, we get the same expression as above with $r_1 \leftrightarrow r_2$.
 
 Using the dilatation Ward identity, where $\mathcal{A}_q$ is the QFT dilatation anomaly,
 \begin{equation}\label{dilward}
      \left<T^\mu_\mu\right>_q=\mathcal{A}_q-\beta^i(\lambda) \left<\mathcal{O}_i\right>_q.
 \end{equation}

Using \eqref{dilward} in \eqref{sqftvar4}, we have,
\begin{align}\label{sqftvar5}
     &\{-\alpha (r_2-r_1)\} \frac{\partial \Sqft}{\partial r_1} \nonumber\\
     &= \alpha \lim_{q \to 1}\frac{1}{(q-1)}\Bigg\{\int_{M_q}d^dx \sqrt{g_q}\Bigg[\frac{(r-r_2)}{2}(\d_r g_{rr}) \big< T^{rr} \big>^{[1]}
     +\frac{(r-r_2)}{2}(\d_rg_{\tau\tau})\big< T^{\tau\tau} \big>^{[1]}\nonumber\\
     &\hspace{5cm}-r r_2 h_{ij}\big< T^{ij} \big>^{[1]}+\Bigg(\mathcal{A}_q-\mathcal{A}\Bigg)\Bigg]\nonumber\\
     &\hspace{2.3cm}-2\Bigg(\int_{M_q} d^dx\sqrt{g_q}\beta^i(\lambda)\left<\mathcal{O}_i(x)\right>_q-q\int_{M_1} d^dx\sqrt{g}\beta^i(\lambda)\left<\mathcal{O}_i(x)\right>_{q=1}\Bigg)\Bigg\}.
 \end{align}

\subsection{Comments on \texorpdfstring{$\Sqft$ - Local vs global contributions}{} \label{Sqftcomments}}
Let us try to understand the expression for the entanglement entropy \eqref{Sqft1}. Given the specific state and the entangling region, $W_1$ is simply a constant independent of the shape of the entangling region. This is because $e^{W_1}=Z_1$ is the trace of the density matrix obtained by sewing the entangling region. 

The information about the entangling region is in $W^{[1]}$ (or in general also in $W_q-qW_1$) via the replica boundary conditions involved in $Z_q$.

We will refer to terms that depend on every point in spacetime as densities and terms that take a value over the entire manifold (integrated value of densities over the entire manifold) as global terms.

\paragraph{Metric dependent terms localise at the boundary of the entangling region:}The replica boundary conditions lead to conical singularities in the replica metric only at the boundaries of the entangling region. The conical singularity contributes non trivially, locally in a radius $a$ (the concial singularity regulator), and $g_{\mu\nu}^{(q)}-g_{\mu\nu}$ is localised at the boundaries of the entangling region. Terms in $W_q-qW_1$ density that depends only on $g_{\mu\nu}^{(q)}-g_{\mu\nu}$ (metric dependent terms) will also be localised at the boundary (of entangling region) and these give boundary  (of entangling region) dependent terms in  $W_q-qW_1$  (in the limit the regulator $a\to0$). Specifically, these boundary terms contribute to  $W^{[1]}$, leading to dependence on boundary coordinates and hence details of the entangling region in $\Sqft$. More generally replica corrections (i.e., terms with $(q-1)^{i>0}$ coefficient) in density terms on $M_q$ (such as the replica corrections to replica stress tensor  $\langle T^{\mu\nu}\rangle_q -\langle T^{\mu\nu}\rangle$) that depend on only replica corrections to replica metric $g_{\mu\nu}^{(q)}-g_{\mu\nu}$ (metric dependent terms) are localised at the boundary (of entangling region). These contribute to boundary dependence in global terms on $M_q$. Notice that these metric-dependent density terms depend on the conical singularity regulator $(a)$ and hence we will sometimes refer to it as regulator dependent terms.

\paragraph{State dependent terms are non local and receive contributions over the entire entangling region:}There are also non metric dependent, i.e., regulator independent terms in replica corrections to density terms (including $\langle T^{\mu\nu}\rangle_q -\langle T^{\mu\nu}\rangle$). We will refer to them as state-dependent or regulator independent contributions. These provide non-trivial contributions over the entire manifold and not just the entangling boundaries. Therefore global terms also receive contributions from the entire manifold. Although the replica boundary conditions affects the metric only in a small neighborhood around the boundary of the entangling region, quantities on the replica manifold (replica states, replica correlation functions) respecting these boundary conditions can differ from the corresponding quantities on $M_1$ throughout the manifold (boundary conditions can alter solutions throughout the manifold). For example: If we start with flat space on $M_1$ which has translational symmetry along one of the spatial directions in the entangling region, and if the entangling region is finite along that spatial direction then the corresponding replica manifold and metric will now have a conical singularity at the boundary of this entangling region. This breaks the translation symmetry along this spatial direction on the replica manifold and the corresponding quantities (density terms) on the replica manifold will now have terms that reflect this broken symmetry which would be trivial on $M_1$. Outside the entangling region, since the fields are not identified between different copies of $M_1$, we do not expect densities to have any replica corrections in this region. Therefore, in general the replica corrections (i.e., terms with $(q-1)^{i>0}$ coefficient) in densities (such as $W_q-qW_1$ density,  $\langle T^{\mu\nu}\rangle_q -\langle T^{\mu\nu}\rangle$), sourced by the replica state is expected to have non-trivial contributions throughout the entangling region only, and not just the conical singular endpoints. These contribute terms that depend on the full entangling region in replica corrections to global terms.  Since regulator independent density terms receive contributions throughout the entangling region only, non-local dependencies such as dependence on the dimensions (example: length, volume) of the entangling region in $\Sqft$, for example, come from such terms. 

The metric/ regulator dependent contributions localised at the boundary completely determine the leading UV/ divergent behavior in $\Sqft$, while the regulator independent (state dependent) terms contribute all the finite non local terms in $\Sqft$.

\section{Replica stress tensor} \label{sec:six}
In expression \eqref{sqftvar2}, that captures how $\Sqft$ varies with the scaling of the entangling region, given the metric and the entangling region, the only quantity in $I_2$ that we have not explicitly evaluated in terms of quantities on $M_1$ is the replica corrections to the replica stress tensor, specifically $\langle T^{\mu\nu}\rangle^{[1]}$. In this section, we discuss the stress tensor expectation value on the replica manifold and express them in terms of stress tensor expectation on $M_1$.
Results in section \eqref{section6} apply to QFT on a static background\footnote{In non-static cases the expression remains same as \eqref{sqftvar} or \eqref{sqftvar0}, except one has to include variations of length scale along the other finite $y^i$ directions in the entangling region.}. For the discussion on Replica stress tensor, we will restrict our attention to conformal field theories on a $d-$dimensional general background on manifold $M_1$ unless specified otherwise later. The theory on the resulting replica manifold $M_q$ will also be a CFT as can be seen from the replica partition function in \eqref{replicadensitytrace} with boundary conditions \eqref{BC}.

In general, for a state in a CFT on $M_q$ (including $M_1$) one can express the (replica)stress tensor as the sum of a traceless part ($t^{\mu \nu}{}^{(q)}$) and a non zero trace part ($\chi^{\mu \nu}{}^{(q)}[g_{\mu\nu}^{(q)}]$) 
\begin{equation}\label{stresssplit}
\left\langle T^{\mu \nu}\right\rangle_q=t^{\mu \nu}{}^{(q)}+\chi^{\mu \nu}{}^{(q)}[g_{\mu\nu}^{(q)}].
\end{equation}
$t^{\mu \nu}{}^{(q)}$ characterises the specific state, referred to as state part of the replica stress tensor expectation (not to be confused with state dependent part or regulator independent part since we can have regulator dependent terms in $t^{\mu \nu}{}^{(q)}$ which together with other terms are traceless), while $\chi^{\mu \nu}{}^{(q)}[g_{\mu\nu}^{(q)}]$ captures the anomaly and the (replica) Casimir or zero-point energy. It depends only on the (replica) metric $g_{\mu\nu}^{(q)}$, and we will refer to it as the geometric part of the replica stress tensor expectation (not necessarily same as the metric dependent or regulator dependent part). 

\subsection{Replica stress tensor expectation - Trace and Conservation equations}

For this discussion, we will be inherently working with the renormalised stress tensor. Since the theory on $M_q$ is a CFT (with replica boundary conditions) given we have a CFT on $M_1$, we have the following constraints on the replica stress tensor expectation,

\paragraph{Trace equations}
\begin{align}
    t^\mu_{\ \mu}{}^{(q)}&=0 \text{ (Tracelessness of state part of replica stress tensor)},\label{reptrlesseqn0}\\
    \chi^\mu_\mu{}^{(q)}[g_{\mu\nu}^{(q)}]&=\mathcal{A}[g_{\mu\nu}^{(q)}]\quad \text{(Trace anomaly for a CFT on $M_q$)}.\label{reptranomeqn0}
\end{align}
These two equations imply,
\begin{equation}\label{fullreptrace0}
    \langle T^\mu_{\ \mu}\rangle_q=\mathcal{A}[g_{\mu\nu}^{(q)}].
\end{equation}
The replica trace equations above hold only for states in CFTs.

\paragraph{Conservation equations}
\begin{align}\label{repconseqn0}
    \nabla_\mu{}^{(q)} \langle T^{\mu\nu} \rangle_q&=\nabla_\mu{}^{(q)}(t^{\mu \nu}{}^{(q)}+\chi^{\mu \nu}{}^{(q)}[g_{\mu\nu}^{(q)}])=0, 
    \end{align}
where $\nabla_\mu{}^{(q)}$ is the covariant derivative wrt $g_{\mu\nu}^{(q)}$ and we have conservation of the full replica stress tensor expectation. 

The replica conservation equation above holds for any QFT. The state and geometric parts are not separately conserved. Also, notice that the decomposition into state and geometric parts is not unique but the split is convenient in some of the analysis below. One can always add traceless terms $t'_{\mu\nu}{}^{(q)}$ to $t_{\mu\nu}^{(q)}$ and $\chi'_{\mu\nu}{}^{(q)}$ to $\chi_{\mu\nu}^{(q)}$ satisfying 
\begin{equation}\label{nonuniquedecomp}
\nabla_\mu^{(q)}(t'_{\mu\nu}{}^{(q)}+\chi'_{\mu\nu}{}^{(q)})=0.
\end{equation}
The state part can also have regulator dependent terms which are traceless. 

The replica stress tensor expectation $\langle T_{\mu\nu} {}\rangle_q$ differs from base stress tensor expectation $\langle T_{\mu\nu} {}\rangle$ only in a localised region around the endpoints of the entangling region for purely metric dependent terms and over the entire entangling region for state-dependent (regulator independent) terms. The replica metric, regulating function $U(\rho,a)$, Christoffels, and the state and geometric parts of the replica stress tensor can be expressed as a series in $(q-1)$
\begin{equation}\label{series}
    \begin{split}
    g{}^{(q)}_{\mu \nu}&=g_{\mu \nu} {}^{[0]}+(q-1)g_{\mu \nu} {}^{[1]}+\mathcal{O}((q-1)^2),\\
    U(\rho,a) &= 1+(q-1)U^{[1]}(\rho,a) +\mathcal{O}((q-1)^2), \\
        \Gamma^\mu_{\ \nu\sigma}{}^{(q)}&=\Gamma^\mu_{\ \nu\sigma} {}^{[0]}+(q-1)\Gamma^\mu_{\ \nu\sigma} {}^{[1]}+\mathcal{O}((q-1)^2),\\
        \chi_{\mu\nu} {}^{(q)}&=\chi_{\mu\nu} {}^{[0]}+(q-1)\chi_{\mu\nu} {}^{[1]}+\mathcal{O}((q-1)^2),\\
        t_{\mu\nu} {}^{(q)}&=t_{\mu\nu} {}^{[0]}+(q-1)t_{\mu\nu} {}^{[1]}+\mathcal{O}((q-1)^2).
    \end{split}
\end{equation}
\textbf{Conservation} of replica stress tensor expectation gives
\begin{equation}
    \nabla_\mu{}^{(q)}\langle T^\mu_{\ \sigma}\rangle_q=\partial_\mu\langle T^\mu_{\ \sigma}\rangle_q+\Gamma^\mu_{\ \mu\lambda}{}^{(q)}\langle T^\lambda_{\ \sigma}\rangle_q-\Gamma^\lambda_{\ \mu\sigma}{}^{(q)}\langle T^\mu_{\ \lambda}\rangle_q=0,
\end{equation}
\begin{align}
    \nabla_\mu{}^{(q)}\langle T^\mu_{\ \sigma}\rangle_q&=\partial_\mu\left(g^{\mu a}{}^{(q)}\langle T_{a\sigma}\rangle_q\right)+\left(\Gamma^\mu_{\ \mu\lambda}{}^{[0]}+(q-1)\Gamma^\mu_{\ \mu\lambda}{}^{[1]}\right)\left(g^{\lambda a}{}^{(q)}\langle T_{a \sigma}\rangle_q\right)\nonumber\\
    &-\left(\Gamma^\lambda_{\ \mu\sigma}{}^{[0]}+(q-1)\Gamma^\lambda_{\ \mu\sigma}{}^{[1]}\right)\left(g^{\mu a}{}^{(q)}\langle  T_{a \lambda}\rangle_q\right)=0.
\end{align}
At $O(q-1)$
\begin{align}\label{gencons}
    &\partial_\mu\left(g^{\mu a}{}^{[0]}\langle T_{a\sigma}\rangle^{[1]}\right)+\Gamma^\mu_{\ \mu\lambda}{}^{[0]}g^{\lambda a}{}^{[0]}\langle T_{a \sigma}\rangle^{[1]}
   -\Gamma^\lambda_{\ \mu\sigma}{}^{[0]}g^{\mu a}{}^{[0]}\langle  T_{a \lambda}\rangle^{[1]}\nonumber\\
   =&-\partial_\mu\left(g^{\mu a}{}^{[1]}\langle T_{a\sigma}\rangle^{[0]}\right)-\Gamma^\mu_{\ \mu\lambda}{}^{[0]}g^{\lambda a}{}^{[1]}\langle T_{a \sigma}\rangle^{[0]}-\Gamma^\mu_{\ \mu\lambda}{}^{[1]}g^{\lambda a}{}^{[0]}\langle T_{a \sigma}\rangle^{[0]}\nonumber\\
   &+\Gamma^\lambda_{\ \mu\sigma}{}^{[0]}g^{\mu a}{}^{[1]}\langle  T_{a \lambda}\rangle^{[0]}+\Gamma^\lambda_{\ \mu\sigma}{}^{[1]}g^{\mu a}{}^{[0]}\langle  T_{a \lambda}\rangle^{[0]}.
\end{align}
Metric compatibility $\nabla_{\mu}^{[0]}g^{\mu a}{}^{[0]}=0$ on $M_1$ implies
\begin{equation}
    \partial_\mu g^{\mu a}{}^{[0]}+\Gamma^\mu_{\ \mu\lambda}{}^{[0]}g^{\lambda a}{}^{[0]}+\Gamma^a_{\ \mu\lambda}{}^{[0]}g^{\mu\lambda}{}^{[0]}=0.
\end{equation}
Using this, LHS in \eqref{gencons} reduces to
\begin{align}\label{LHS}
    LHS=&g^{\mu a}{}^{[0]}\partial_\mu\langle T_{a\sigma}\rangle^{[1]}+\langle T_{a\sigma}\rangle^{[1]}\left(\partial_\mu g^{\mu a}{}^{[0]}+\Gamma^\mu_{\ \mu\lambda}{}^{[0]}g^{\lambda a}{}^{[0]}\right)
   -\Gamma^\lambda_{\ \mu\sigma}{}^{[0]}g^{\mu a}{}^{[0]}\langle  T_{a \lambda}\rangle^{[1]}\nonumber\\
   =&g^{\mu a}{}^{[0]}\partial_\mu\langle T_{a\sigma}\rangle^{[1]}-\Gamma^a_{\ \mu\lambda}{}^{[0]}g^{\mu\lambda}{}^{[0]}\langle T_{a\sigma}\rangle^{[1]}
   -\Gamma^\lambda_{\ \mu\sigma}{}^{[0]}g^{\mu a}{}^{[0]}\langle  T_{a \lambda}\rangle^{[1]}.
\end{align}
Metric compatibility $\nabla_{\mu}^{(q)}g^{\mu a}{}^{(q)}=0$ on $M_q$ implies
\begin{equation}
    \partial_\mu g^{\mu a}{}^{(q)}+\Gamma^\mu_{\ \mu\lambda}{}^{(q)}g^{\lambda a}{}^{(q)}+\Gamma^a_{\ \mu\lambda}{}^{(q)}g^{\mu\lambda}{}^{(q)}=0.
\end{equation}
At $O(q-1)$
\begin{align}
    \partial_\mu g^{\mu a}{}^{[1]}+\Gamma^\mu_{\ \mu\lambda}{}^{[0]}g^{\lambda a}{}^{[1]}+\Gamma^a_{\ \mu\lambda}{}^{[0]}g^{\mu\lambda}{}^{[1]}
    +\Gamma^\mu_{\ \mu\lambda}{}^{[1]}g^{\lambda a}{}^{[0]}+\Gamma^a_{\ \mu\lambda}{}^{[1]}g^{\mu\lambda}{}^{[0]}=0.
\end{align}
Using this, RHS in \eqref{gencons} reduces to
\begin{align}\label{RHS}
    RHS=&-g^{\mu a}{}^{[1]}\partial_\mu \langle T_{a\sigma}\rangle^{[0]}-\left(\partial_\mu g^{\mu a}{}^{[1]}+\Gamma^\mu_{\ \mu\lambda}{}^{[0]}g^{\lambda a}{}^{[1]}\right)\langle T_{a \sigma}\rangle^{[0]}-\Gamma^\mu_{\ \mu\lambda}{}^{[1]}g^{\lambda a}{}^{[0]}\langle T_{a \sigma}\rangle^{[0]}\nonumber\\
   &+\Gamma^\lambda_{\ \mu\sigma}{}^{[0]}g^{\mu a}{}^{[1]}\langle  T_{a \lambda}\rangle^{[0]}+\Gamma^\lambda_{\ \mu\sigma}{}^{[1]}g^{\mu a}{}^{[0]}\langle  T_{a \lambda}\rangle^{[0]}\nonumber\\
   =&-g^{\mu a}{}^{[1]}\partial_\mu \langle T_{a\sigma}\rangle^{[0]}+\left(\Gamma^a_{\ \mu\lambda}{}^{[0]}g^{\mu\lambda}{}^{[1]}
    +\Gamma^\mu_{\ \mu\lambda}{}^{[1]}g^{\lambda a}{}^{[0]}+\Gamma^a_{\ \mu\lambda}{}^{[1]}g^{\mu\lambda}{}^{[0]}\right)\langle T_{a \sigma}\rangle^{[0]}\nonumber\\
   &-\Gamma^\mu_{\ \mu\lambda}{}^{[1]}g^{\lambda a}{}^{[0]}\langle T_{a \sigma}\rangle^{[0]}+\Gamma^\lambda_{\ \mu\sigma}{}^{[0]}g^{\mu a}{}^{[1]}\langle  T_{a \lambda}\rangle^{[0]}+\Gamma^\lambda_{\ \mu\sigma}{}^{[1]}g^{\mu a}{}^{[0]}\langle  T_{a \lambda}\rangle^{[0]}\nonumber\\
   =&\left(-g^{\mu a}{}^{[1]}\partial_\mu+\Gamma^a_{\ \mu\lambda}{}^{[0]}g^{\mu\lambda}{}^{[1]}
    +\Gamma^a_{\ \mu\lambda}{}^{[1]}g^{\mu\lambda}{}^{[0]}\right)\langle T_{a \sigma}\rangle^{[0]}\nonumber\\
   &+\left(\Gamma^\lambda_{\ \mu\sigma}{}^{[0]}g^{\mu a}{}^{[1]}+\Gamma^\lambda_{\ \mu\sigma}{}^{[1]}g^{\mu a}{}^{[0]}\right)\langle  T_{a \lambda}\rangle^{[0]}.
\end{align}
Combining both sides of equality in \eqref{gencons}, given by \eqref{LHS} and \eqref{RHS},
\begin{equation}\label{repdeqn}
 \addtolength{\fboxsep}{5pt}
   \boxed{
   \begin{gathered}
   \left(g^{\mu a}{}^{[0]}\partial_\mu-\Gamma^a_{\ \mu\lambda}{}^{[0]}g^{\mu\lambda}{}^{[0]}\right)\langle T_{a\sigma}\rangle^{[1]}
   -\Gamma^\lambda_{\ \mu\sigma}{}^{[0]}g^{\mu a}{}^{[0]}\langle  T_{a \lambda}\rangle^{[1]} \\
   =-\left(g^{\mu a}{}^{[1]}\partial_\mu-\Gamma^a_{\ \mu\lambda}{}^{[0]}g^{\mu\lambda}{}^{[1]}
    -\Gamma^a_{\ \mu\lambda}{}^{[1]}g^{\mu\lambda}{}^{[0]}\right)\langle T_{a \sigma}\rangle^{[0]}+\left(\Gamma^\lambda_{\ \mu\sigma}{}^{[0]}g^{\mu a}{}^{[1]}+\Gamma^\lambda_{\ \mu\sigma}{}^{[1]}g^{\mu a}{}^{[0]}\right)\langle  T_{a \lambda}\rangle^{[0]}
    \end{gathered}
    }
\end{equation}   

This is of the form
\begin{equation}\label{genconsrep}
    L^a{}^{[0]}\langle T_{a\sigma}\rangle^{[1]}-M^{\lambda a}{}_\sigma{}^{[0]} \langle  T_{a \lambda}\rangle^{[1]}=- L^a{}^{[1]}\langle T_{a\sigma}\rangle^{[0]}+M^{\lambda a}{}_\sigma{}^{[1]} \langle  T_{a \lambda}\rangle^{[0]}.
\end{equation}
where 
\begin{align}
     L^a{}^{(q)}&=g^{\mu a}{}^{(q)}\partial_\mu-\Gamma^a_{\ \mu\lambda}{}^{(q)}g^{\mu\lambda}{}^{(q)},\\
     M^{\lambda a}{}_\sigma{}^{(q)}&=\Gamma^\lambda_{\ \mu\sigma}{}^{(q)}g^{\mu a}{}^{(q)}.
\end{align}
Equation \eqref{repdeqn} are $d$ coupled first-order linear differential equation involving $O(q-1)$ replica stress tensor corrections $\langle T_{\mu\nu}\rangle^{[1]}$ on the LHS, and on RHS we have the sources: stress tensor on $M_1$ i.e., $\langle T_{\mu\nu}\rangle^{[0]}$, $O(q-1)$ replica metric corrections $g_{\mu\nu}^{[1]}$, metric on $M_1$. We can also write \eqref{repdeqn} in an integral form using Stokes theorem. However to solve for $\langle T_{\mu\nu}\rangle^{[1]}$, we need the explicit form of the sources along with more linear equations involving $\langle T_{\mu\nu}\rangle^{[1]}$ (since we have $d(d+1)/2)$ unknowns and $(d+1)$ linear equations). In the presence of certain symmetries we can reduce the number of unknowns involved. We will see in the following sections that in the static case the replica conservation and trace equations are enough. 

\paragraph{Note on homogenous solution to $\langle  T_{\mu \nu}\rangle^{[1]}$:}\label{homogenote}We also notice that the conservation equations on $M_1$ i.e.,
\begin{align}
    \nabla_{\mu}^{[0]}\langle T^\mu_{\ \sigma} \rangle^{[0]} &=0,
\end{align}
implies
\begin{equation}\label{basedeqn}
    L^a{}^{[0]}\langle T_{a\sigma}\rangle^{[0]}-M^{\lambda a}{}_\sigma{}^{[0]} \langle  T_{a \lambda}\rangle^{[0]}=0,
\end{equation}
i.e., it is the same as the replica conservation equation \eqref{genconsrep} with no sources and with $\langle  T_{\mu \nu}\rangle^{[1]}$ replaced by $ \langle T_{\mu \nu}\rangle^{[0]}$. However, notice that the homogenous solution to $\langle T^\mu_{\ \sigma} \rangle^{[1]}$ is not only $\langle T^\mu_{\ \sigma} \rangle^{[0]}$ since the presence of a conical singularity in $M_q$ at the boundary breaks any translation symmetry along $\rho$ in $M_1$. Therefore terms that we usually set to zero in $\langle T^\mu_{\ \sigma} \rangle^{[0]}$ due to translation symmetry on $M_1$ can be non trivial in $\langle T^\mu_{\ \sigma} \rangle^{[1]}$ on $M_q$. 

In the case of squashed cone the translation symmetry along $\psi$ is also broken and we only have the discrete symmetry i.e., $\psi \to \psi+2\pi k$ where $k\in \mathbb{Z}$ which might introduce $\psi$ dependence and more terms on $M_q$. Even if there were no translation symmetry on $M_1$, to begin with, the homogenous solution is still modified due to the presence of conical and curvature singularities (in squashed cone) on $M_q$. In other words we need to be careful when fixing the integration constants in the general homogenous solution to $\langle T^\mu_{\ \sigma} \rangle^{[1]}$(they are not the same as those in $\langle T^\mu_{\ \sigma} \rangle^{[0]}$ on $M_1$). The homogenous solution to $\langle T^\mu_{\ \sigma} \rangle^{[1]}$ has some more terms in addition to $\langle T^\mu_{\ \sigma} \rangle^{[0]}$.

\subsection{Local and non local contributions}\label{metstate}

As discussed before in section \eqref{Sqftcomments} the replica corrections to the replica stress tensor i.e., $\langle T_{\mu\nu}\rangle_q-\langle T_{\mu\nu}\rangle$  are localised around the boundary of the entangling region for purely metric dependent terms; and around the full entangling region for regulator independent (state dependent) terms.    

Notice that the regulated replica metric $g_{\mu\nu}^{(q)}$ i.e., \eqref{repgen} is in general valid (is a local expansion) in only a small neighborhood around the boundary of the entangling region. Hence, once we make use of replica metric \eqref{repgen} in the replica conservation equation \eqref{repconseqn0}, the $O(q-1)$ corrections to the replica stress tensor $\langle T_{\mu\nu}\rangle^{[1]}$ calculated from the corresponding replica conservation equations will also be valid only in a small neighborhood of the boundary of the entangling region. 

This is enough to extract the $O(q-1)$ corrections to the purely metric dependent parts of the replica stress tensor since this by definition depends only on $(g_{\mu\nu}^{(q)}-g_{\mu\nu})$ which is non zero only in a small neighborhood (determined by the distance from the conical singularity set by the regulator $(a)$) around the conical singular endpoints (since the replica metric differs from the metric on $M_1$ only in a small neighborhood around the conical singular endpoints). The localised contributions around $\rho=0$, of the $O(q-1)$ corrections to the purely metric dependent parts of the replica stress tensor, can be explicitly calculated using \eqref{localrep}. For the purely metric dependent parts of the solution $\langle  T_{\mu \nu}\rangle^{[1]}$ we also have
\begin{equation}\label{bcond1}
    \langle  T_{\mu \nu}\rangle^{[1]}\rvert_{\rho>>a}\to 0,
\end{equation} 
where $a$ is the regulator. For integrals involving such terms it is enough to perform the $\rho$ integral between $[0,\rho_0>>a]$ since the integrand is zero for $\rho>>a$.

The regulator independent (state-dependent) replica stress tensor (at all orders in $(q-1)$) are non trivial due to broken symmetries on $M_q$ that were originally present in $M_1$ background metric (because of conical singularity or states that break the symmetry on background metric on $M_1$). They are non trivial over the entire entangling region and do not have a regulating radius $(a)$ dependence. Since these terms must go to zero as $q\to 1$ (since symmetries are restored as $q\to1$), they cannot have a non trivial $O((q-1)^0)$ term i.e., they do not have a non trivial counterpart on $M_1$. This is unlike the regulator dependent case where we also had a non trivial $O(1)$ term.  

In the case of a disconnected boundary (at $r=r_1$ and $r=r_2$) we have two conical singularities, one at each boundary. The net contribution to the $O(q-1)$ replica stress tensor is the sum of the contributions from each conical singularity. We perform the integral restricted around the entangling region. As mentioned before, the replica boundary conditions do not affect the different copies of $M_1$ outside the entangling region. For the regulator independent parts of the replica stress tensor, we consider the contribution around $r_1$ in the semi disc of radius $(r_2-r_1)/2$, with $r\geq r_1$ centered around $r_1$, and contribution around $r_2$ in semi disc of radius $(r_2-r_1)/2$ with $r\leq r_2$ centered around $r_2$. We perform the integrals for the regulator-independent parts of the replica stress tensor restricted to these semi discs. On a flat background on $M_1$, due to translational invariance, the $O(q-1)$ replica stress tensor has identical contributions from each conical singularity and the net contribution is the sum of individual conical contribution, therefore in this case the integral is the same as integrating over a disc of radius $(r_2-r_1)/2$ around one of the conical singularities.

 \subsection{Replica stress tensor - Static metric}
 So far in the replica conservation equations we have not made a choice of the metric on $M_1$. Although the $(d+1)$ - replica conservation and trace equations hold true on any general background, on static backgrounds we can exploit certain symmetries to reduce the number of unknown replica stress tensor components to match the number of equations. 
 
 The replica conservation and trace equations can be explicitly expanded for the most general metric \eqref{repgen}. However, since all components of the replica metric are non-trivial in the general case, it is rather easy to expand it for the particular case of interest as and when necessary. On static backgrounds we can work in a gauge in which - $g^{(q)}_{\rho i}=g^{(q)}_{\psi i}=g^{(q)}_{\rho\psi}=0$, see \eqref{repmetstat}, we will now explicitly write down the replica conservation and trace equations in this case.
 
 Note that $\nabla_\mu{}^{(q)}$ is non-trivial, and on replica manifold one needs Christoffels $\Gamma^\mu_{\ \nu\sigma}{}^{(q)}$ to explicitly write down the conservation equations on $M_q$, for instance: metric with zero $\Gamma^\mu_{\ \nu\sigma}$ on $M_1$ have non zero $\Gamma^\mu_{\ \nu\sigma}{}^{(q)}$. For the replica metric $g{}^{(q)}_{\mu\nu}$ corresponding to the static background on $M_1$, the non-trivial Christoffel symbols are:
\begin{equation}\label{chris}
    \begin{split}
        &\Gamma^\rho_{\ \rho\rho}{}^{(q)}=\frac{1}{2}U^{-1}\partial_\rho U,\nonumber\\
        &\Gamma^\rho_{\ \psi\psi}{}^{(q)}=-U^{-1}\rho,\\
        &\Gamma^\psi_{\ \rho\psi}{}^{(q)}=\frac{1}{\rho},\nonumber\\
        &\Gamma^i_{\ \rho k}{}^{(q)}=\frac{1}{2}g^{ij}{}^{(q)}\partial_\rho g_{jk}^{(q)},\\
        \end{split}
        \hspace{1cm}
        \begin{split}
        &\Gamma^i_{\ \psi k}{}^{(q)}=\frac{1}{2}g^{ij}{}^{(q)}\partial_\psi g_{jk}^{(q)},\\
        &\Gamma^\rho{}_{ij}{}^{(q)}=-\frac{1}{2}U^{-1}\partial_{\rho} g_{ij}^{(q)},\\
        &\Gamma^\psi{}_{ij}{}^{(q)}=-\frac{1}{2\rho^2}\partial_{\psi} g_{ij}^{(q)},\\
        &\Gamma^i_{\ kl}{}^{(q)}=\Gamma^i_{\ kl}{}^{[0]},      
    \end{split}
\end{equation}
with
\begin{align}
    g^{(q)}_{ij}=&\delta_{ij}, \text{ or}\nonumber \\
    =&(r_0+\rho^p c^{1-p}\cos(\psi))^2 h_{ij}(y), \text{ or}\nonumber \\
    =&(r_0+\rho^p c^{1-p}\cos(\psi))^2; \quad i=j=\phi,
     \text{ and } \delta_{ij}; \text{ otherwise}, \text{ or}\nonumber\\
    =& (\gamma_{ij}+2\rho^p c^{1-p}\cos(\psi) K^1_{ij}+\dots).
\end{align}
The first three equalities correspond to planar, spherical, and cylindrical entangling regions in flat space respectively, and the last equality corresponds to a static background. We have,
\begin{align}
\begin{split}
    &L^a{}^{[0]}\langle T_{a\sigma}\rangle^{[1]}\\
    &=\left(\d_\rho-\frac{\Gamma^\rho{}_{\psi\psi}}{\rho^2}-\Gamma^\rho{}_{ij}g^{ij}\right)\langle T_{\rho\sigma}\rangle^{[1]}+\left(\frac{1}{\rho^2}\d_\psi-\Gamma^\psi{}_{ij}g^{ij}\right)\langle T_{\psi\sigma}\rangle^{[1]}+\left(g^{ik}\d_k-\Gamma^i{}_{jk}g^{jk}\right)\langle T_{i\sigma}\rangle^{[1]}\nonumber\\
    &=\left(\d_\rho+\frac{1}{\rho}+\Gamma^i{}_{i\rho}\right)\langle T_{\rho\sigma}\rangle^{[1]}+\frac{1}{\rho^2}\left(\d_\psi+\Gamma^i{}_{i\psi}\right)\langle T_{\psi\sigma}\rangle^{[1]}+\left(\d_k+\Gamma^i{}_{ik}\right)\langle T^k_{\sigma}\rangle^{[1]},
\end{split}
\end{align}
here we have used $-\Gamma^\rho{}_{ij}g^{ij}=\Gamma^i{}_{i\rho}$ and $-\rho^2\Gamma^\psi{}_{ij}g^{ij}=\Gamma^i{}_{i\psi}$. 

In general, explicit coordinate  ($\{\rho,\psi\}$) dependence of $L^a{}^{(q)}\langle T_{a \lambda}\rangle^{(q)}$ can be written but $M^{\lambda a}{}_{\sigma}^{(q)}\langle T_{a \lambda}\rangle^{(q)}$ depends on the choice of $\sigma$. We will write down the full explicit replica stress tensor conservation equations in subsequent sections. 

\subsubsection{Replica stress tensor: Geometric part} \label{repgeomstress}
The geometric part of stress tensor on $M_1$ i.e., $\chi_{\mu\nu}[g_{\mu\nu}]$ and its trace i.e., the trace anomaly $\chi_{\ \mu}^{\mu}[g_{\mu\nu}]=\mathcal{A}[g_{\mu\nu}]$ are well known. They are non-trivial for an even-dimensional CFT on a curved background and are given by
\begin{align}\label{evendanom}
    \left\langle \chi_{\ \mu}^{ \mu}\right\rangle=\frac{1}{(4 \pi)^{d / 2}}\left(\sum_j c_{d j} I_j^{(d)}-(-)^{\frac{d}{2}} a_d E^{(d)}+\sum_j d_{d j} D^i J_i^{(d)}\right),
\end{align}
where $E^{(d)}$ is the Euler density in $d$ dimensions (type A anomaly), $I_j^{(d)}$ are independent Weyl invariants (type B anomalies), and the last term denotes the type D anomalies which are total derivatives that could be canceled by the Weyl variation of local covariant counterterms (scheme dependent). We can express $E^{(d)}$ in the form, 
\begin{equation}
E^{(d)}=\frac{1}{2^{d / 2}} \delta_{\mu_1 \cdots \mu_d}^{\nu_1 \cdots \nu_d} R_{\nu_1 \nu_2}^{\mu_1 \mu_2} \cdots R_{\nu_{d-1} \nu_d}^{\mu_{d-1} \mu_d}.
\end{equation}
For example in $d=2$
\begin{equation}
    \chi_{\mu\nu}[g_{\mu\nu}]^{d=2}=-\frac{c}{48\pi}Rg_{\mu\nu}.
\end{equation}
In $d=4$ there is one Weyl invariant and in $d=6$ there are three Weyl invariants,
\begin{equation}\label{4danom}
    \chi_{\ \mu}^\mu[g_{\mu\nu}]^{d=4}=-\frac{a}{(4 \pi)^2} E^{(4)}+\frac{c}{(4 \pi)^2} C_{\mu \nu \rho \sigma} C^{\mu \nu \rho \sigma}.
\end{equation}
When $d$ is odd
\begin{equation}
    \chi_{\mu\nu}[g_{\mu\nu}]^{\text{$d$ odd}}=0.
\end{equation}
In general, for a conformally flat background $\overline{g}_{\mu\nu}(x)=e^{2\sigma(x)} \eta_{\mu\nu}=\Omega^2 \eta_{\mu\nu}$ (with stress tensor vacuum expectation value normalized to 0 in flat space) the following relation between the vacuum stress tensor expectation (the only contribution is the geometric part) and the trace anomaly holds \cite{PhysRevD.25.1499, Herzog:2013ed},
\begin{equation}
    \frac{\delta \sqrt{-\bar{g}}\left\langle\bar{T}^{\mu \nu}(x)\right\rangle}{\delta \sigma\left(x^{\prime}\right)}=2 \frac{\delta \sqrt{-\bar{g}}\left\langle\bar{T}_\lambda^\lambda\left(x^{\prime}\right)\right\rangle}{\delta \bar{g}_{\mu \nu}(x)}.
\end{equation}
The solution of which is given by
\begin{align}
& \left\langle\bar{T}_\nu^\mu\right\rangle=\Omega^{-4}\left\langle T_\nu^\mu\right\rangle-\frac{a_4}{(4 \pi)^2}\left[\left(4 \bar{R}_\rho^\lambda \bar{W}^{\rho \mu}{ }_{\lambda \nu}-2 \bar{H}_\nu^\mu\right)-\Omega^{-4}\left(4 R_\rho^\lambda W^{\rho \mu}{ }_{\lambda \nu}-2 H_\nu^\mu\right)\right] \nonumber \\
& -\frac{\gamma}{6(4 \pi)^2}\left[I_\nu^\mu-\Omega^{-4} I_\nu^\mu\right]-8 \frac{c_4}{(4 \pi)^2}\left[\bar{D}^\rho \bar{D}_\lambda\left(\bar{W}^{\rho \mu}{ }_{\lambda \nu} \ln \Omega\right)+\frac{1}{2} \bar{R}_\rho^\lambda \bar{W}^{\rho \mu}{ }_{\lambda \nu} \ln \Omega\right],
\end{align}
where,
\begin{align}
H_{\mu \nu} & \equiv-\frac{1}{2}\left[g_{\mu \nu}\left(\frac{R^2}{2}-R_{\lambda \rho}^2\right)+2 R_\mu^\lambda R_{\nu \lambda}-\frac{4}{3} R R_{\mu \nu}\right], \nonumber\\
I_{\mu \nu} & \equiv 2 D_\mu D_\nu R-2 g_{\mu \nu} D^2 R-2 R R_{\mu \nu}+\frac{1}{2} g_{\mu \nu} R^2.
\end{align}
\paragraph{Replica geometric part (solution):}We can get the geometric part of the replica stress tensor by taking
\begin{equation}\label{geomrepstress}
    \chi_{\mu\nu}^{(q)}[g_{\mu\nu}^{(q)}]=\chi_{\mu\nu}[g_{\mu\nu}\to g_{\mu\nu}^{(q)}],
\end{equation}
since it immediately solves equation \eqref{reptranomeqn0}.
\\ \\
Therefore, we have
\begin{align}
     \chi_{\mu\nu}^{(q)}[g_{\mu\nu}^{(q)}]^{d=2}&=-\frac{c}{48\pi}R^{(q)}g_{\mu\nu}^{(q)}.
\end{align}
We have,
\begin{equation}
    \chi_{\mu\nu}^{[1]}=\frac{\d \chi_{\mu\nu}}{\d g_{\sigma\eta}}g_{\sigma\eta}^{[1]}.
\end{equation}

\subsubsection{Replica stress tensor: State part} 
We now determine the traceless replica state part and the full replica stress tensor from that on the base. We assume that the states respect certain symmetries of the background. For example, on static backgrounds, we assume that the states satisfy $\langle T_{ij}\rangle=0$ $\forall$ $i\neq j$ where $y^i \in \Sigma$, since we have spherical or planar or hyperbolic symmetry along $y^i{}'s$ and hence $g_{ij}=0$ $\forall$ $i\neq j$. On a static metric we also have $g_{\rho i}=0=g_{\psi i}$\footnote{This is not true in non static spacetimes.}. We consider states that respect these symmetries as well i.e., states that satisfy $\langle T_{\rho i}\rangle=0=\langle T_{\psi i}\rangle$. These states are of interest in the context of the black hole problem since we expect the average behavior of the hawking radiation to also respect this symmetry. While calculating $\Sqft$ in our examples, we consider states that are common to all theories, such as the vacuum and thermal states.

We will work with the base manifold stress tensor expectation for a general state on $M_1$ (i.e., we consider a general, state part of the base stress tensor $t_{\mu\nu}$ denoted as $t_{\mu\nu}^{[0]}$) unless specified. \\ \\
\textbf{Tracelessness} of $t^{(q)}_{\mu\nu}$ i.e., \eqref{reptrlesseqn0} gives
\begin{equation}
    t^\mu_{\ \mu}{}^{(q)}=g^{\mu\nu} {}^{(q)} t_{\mu\nu} {}^{(q)}=0.
\end{equation}
Using \eqref{series}, and considering only the $\mathcal{O}(q-1)$ term, we have
\begin{equation}
    g^{\mu\nu}{}^{[0]}  t_{\mu\nu} {}^{[1]}+g^{\mu\nu} {}^{[1]} t_{\mu\nu} {}^{[0]}=0.
\end{equation}
This gives
\begin{equation}\label{trless}
    t_{\rho\rho}^{[1]}+\frac{1}{\rho^2}t_{\psi\psi}^{[1]}+g^{ij}t_{ij}^{[1]}=U^{[1]}(\rho,a)t_{\rho\rho}^{[0]}\textcolor{black}{-g^{ij}{}^{[1]}t_{ij}^{[0]}}.
\end{equation}
The trace equation for the full replica stress tensor i.e., \eqref{fullreptrace0} gives,
\begin{equation}\label{fullreptrace1}
     \langle T^\rho_{\ \rho} \rangle_q+\frac{1}{\rho^2} \langle T_{\psi\psi}\rangle_q+\langle T^i_{\ i}\rangle_q=\mathcal{A}[g_{\mu\nu}^{(q)}].
\end{equation}
\textbf{Conservation} of replica stress tensor expectation i.e., \eqref{repconseqn0} gives
\begin{equation}
    \nabla_\mu{}^{(q)}\langle T^\mu_{\ \sigma}\rangle_q=\partial_\mu\langle T^\mu_{\ \sigma}\rangle_q+\Gamma^\mu_{\ \mu\lambda}{}^{(q)}\langle T^\lambda_{\ \sigma}\rangle_q-\Gamma^\lambda_{\ \mu\sigma}{}^{(q)}\langle T^\mu_{\ \lambda}\rangle_q=0.
\end{equation}
\begin{enumerate}
    \item For $\sigma=\rho$
    \begin{multline}\label{rhocons}
        \nabla_\mu{}^{(q)}\langle T^\mu_{\ \rho}\rangle_q=0=(\partial_\rho+\Gamma^\psi_{\ \psi\rho}{}^{(q)}+\Gamma^i_{\ i\rho}{}^{(q)})\langle T^\rho_{\ \rho}\rangle_q+(\partial_\psi+\Gamma^i_{\ i\psi}{}^{(q)})\langle T^\psi_{\ \rho}\rangle_q\\
        +(\partial_k+\Gamma^i_{\ ik}{}^{(q)})\langle T^k_{\ \rho}\rangle_q-\Gamma^\psi_{\ \psi\rho}{}^{(q)}\langle T^\psi_{\ \psi}\rangle_q-\Gamma^i_{\ k\rho}{}^{(q)}\langle T^k_{\ i}\rangle_q.
    \end{multline}
Using the replica Christoffels \eqref{chris} and series expanding in $(q-1)$ using \eqref{series} and considering only the $\mathcal{O}(q-1)$ term, we have  
 \begin{align}
      &\left(\partial_\rho+\frac{1}{\rho}+\Gamma^i_{\ i\rho}\right)\langle T_{\rho \rho}\rangle{}^{[1]}+\frac{1}{\rho^2}(\partial_\psi+\Gamma^i_{\ i\psi})\langle T_{\psi \rho}\rangle{}^{[1]}-\frac{1}{\rho^3}\langle T_{\psi \psi}\rangle{}^{[1]}+(\partial_k+\nonumber\\
        &\Gamma^i_{\ ik})\langle T^k_{\ \rho}\rangle{}^{[1]}-\Gamma^i_{\ k\rho}\langle T^k_{\ i}\rangle{}^{[1]}= \left(\d_\rho+\frac{1}{\rho}+\Gamma^i_{\ i\rho}\right)(U^{[1]}\langle T_{\rho \rho}\rangle{}^{[0]})-\textcolor{black}{\Gamma^i_{\ i\rho}{}^{[1]}\langle T_{\rho \rho}\rangle{}^{[0]}}\nonumber\\
        &\textcolor{black}{-\frac{1}{\rho^2}\Gamma^i_{\ i\psi}{}^{[1]}\langle T_{\psi \rho}\rangle{}^{[0]}+ \Gamma^i_{\ k\rho}{}^{[1]}\langle T^k_{\ i}\rangle{}^{[0]} }.
    \end{align}
    
    If we use the trace equation for the full replica stress tensor i.e., \eqref{fullreptrace1} to eliminate $\langle T^\rho_\rho\rangle_q$, we get
    
\begin{align}\label{repconsexprho}
&\frac{1}{\rho^2}(\d_\psi+\Gamma^i_{\ i \psi})\langle T _{\rho\psi}\rangle^{[1]}-\frac{1}{\rho^2}(\d_\rho+\Gamma^i_{\ i \rho})\langle T_{\psi\psi} \rangle^{[1]}-\left(\d_\rho+\frac{1}{\rho}+\Gamma^i_{\ i \rho}\right) \Big(\langle T^k_{\ k} \rangle^{[1]}\nonumber\\
&-\mathcal{A}^{[1]}[g_{\mu\nu}^{(q)}]\Big)
+(\d_k+\Gamma^i_{\ ik})\langle T^k_{\ \rho} \rangle^{[1]}-\Gamma^i_{\ k\rho}\langle T^k_{\ i} \rangle^{[1]}=\textcolor{black}{\frac{1}{\rho^2}(\Gamma^i_{\ i \rho}{}^{[1]}\langle T_{\psi\psi} \rangle^{[0]}}\nonumber\\
&\textcolor{black}{-\Gamma^i_{\ i \psi}{}^{[1]}\langle T_{\rho\psi} \rangle^{[0]})+\Gamma^i_{\ i \rho}{}^{[1]}\left(\langle T^k_{\ k} \rangle^{[0]}-\mathcal{A}[g_{\mu\nu}]\right)}\textcolor{black}{+\Gamma^i_{\ k \rho}{}^{[1]}\langle T^k_{\ i} \rangle^{[0]}}.
\end{align}  

Using \eqref{stresssplit} implies,
\begin{align}\label{stateconsexprho}
&\frac{1}{\rho^2}(\d_\psi+\Gamma^i_{\ i \psi})t^{[1]}_{\rho\psi}-\frac{1}{\rho^2}(\d_\rho+\Gamma^i_{\ i \rho})t^{[1]}_{\psi\psi}-\left(\d_\rho+\frac{1}{\rho}+\Gamma^i_{\ i \rho}\right) t^k_{\ k}{}^{[1]}+(\d_k
\nonumber\\&+\Gamma^i_{\ ik})t^k_{\ \rho}{}^{[1]}-\Gamma^i_{\ k\rho}t^k_{\ i}{}^{[1]}=
-\frac{1}{\rho^2}(\d_\psi+\Gamma^i_{\ i \psi})\chi^{[1]}_{\rho\psi}+\frac{1}{\rho^2}(\d_\rho+\Gamma^i_{\ i \rho})\chi^{[1]}_{\psi\psi}\\
&+\left(\d_\rho+\frac{1}{\rho}+\Gamma^i_{\ i \rho}\right) \left(\chi^k_{\ k}{}^{[1]}-\mathcal{A}^{[1]}[g_{\mu\nu}^{(q)}]\right)-(\d_k+\Gamma^i_{\ ik})\chi^k_{\ \rho}{}^{[1]}+\Gamma^i_{\ k\rho}\chi^k_{\ i}{}^{[1]}\nonumber\\
&\textcolor{black}{+\frac{1}{\rho^2}(\Gamma^i_{\ i \rho}{}^{[1]}\langle T_{\psi\psi} \rangle^{[0]}-\Gamma^i_{\ i \psi}{}^{[1]}\langle T_{\rho\psi} \rangle^{[0]})+\Gamma^i_{\ i \rho}{}^{[1]}\left(\langle T^k_{\ k} \rangle^{[0]}-\mathcal{A}[g_{\mu\nu}]\right)}\textcolor{black}{+\Gamma^i_{\ k \rho}{}^{[1]}\langle T^k_{\ i} \rangle^{[0]}},\nonumber
\end{align}  
where we have collected the source terms (i.e., those expressed in terms of the stress tensor on the base manifold) on the RHS.
    \item For $\sigma=\psi$
    \begin{multline}\label{psicons}
        \nabla_\mu{}^{(q)}\langle T^\mu_{\ \psi}\rangle_q=0=(\partial_\rho+\Gamma^\rho_{\ \rho\rho}{}^{(q)}+\Gamma^i_{\ i\rho}{}^{(q)}-\Gamma^\rho_{\ \psi\psi}{}^{(q)}g_{\rho\rho}{}^{(q)}g^{\psi\psi}{}^{(q)})\langle T^\rho_{\ \psi}\rangle_q\\+(\partial_\psi+\Gamma^i_{\ i\psi}{}^{(q)})\langle T^\psi_{\ \psi}\rangle_q
        +(\partial_k+\Gamma^i_{\ ik}{}^{(q)})\langle T^k_{\ \psi}\rangle_q-\Gamma^i_{\ k\psi}{}^{(q)}\langle T^k_{\ i}\rangle_q.
    \end{multline}
    
Similarly using \eqref{chris} and \eqref{series} and retaining only the $O(q-1)$ terms, we get
\begin{align}\label{repconsexppsi}
    &\left(\d_\rho+\frac{1}{\rho}+\Gamma^i_{\ i \rho}\right)\langle T_{\rho\psi}\rangle^{[1]}+\frac{1}{\rho^2}(\d_\psi+\Gamma^i_{\ i\psi})\langle T_{\psi\psi}\rangle^{[1]}+(\d_k+\Gamma^i_{\ ik})\langle T^k_{\ \psi}\rangle^{[1]}\nonumber\\
    &-\Gamma^i_{\ k\psi}\langle T^k_{\ i}\rangle^{[1]}=\left(\frac{1}{2}\d_\rho U^{[1]}+U^{[1]}\d_\rho+\frac{U^{[1]}}{\rho}+\Gamma^i_{\ i \rho}U^{[1]}\right)\langle T_{\rho\psi}\rangle^{[0]}\nonumber\\
    & \textcolor{black}{-\Gamma^i_{\ i\rho}{}^{[1]}\langle T_{\rho\psi}\rangle^{[0]}-\frac{1}{\rho^2}\Gamma^i_{\ i\psi}{}^{[1]}\langle T_{\psi\psi}\rangle^{[0]}+\Gamma^i_{\ k\psi}{}^{[1]}\langle T^k_{\ i}\rangle^{[0]}}.
\end{align}
Using \eqref{stresssplit} implies,
\begin{align}\label{stateconsexppsi}
   &\left(\d_\rho+\frac{1}{\rho}+\Gamma^i_{\ i \rho}\right)t^{[1]}_{\rho\psi}+\frac{1}{\rho^2}(\d_\psi+\Gamma^i_{\ i\psi})t^{[1]}_{\psi\psi}+(\d_k+\Gamma^i_{\ ik})t^k_{\ \psi}{}^{[1]}-\Gamma^i_{\ k\psi}t^k_{\ i}{}^{[1]}=\nonumber\\
  &-\left(\d_\rho+\frac{1}{\rho}+\Gamma^i_{\ i \rho}\right)\chi^{[1]}_{\rho\psi}-\frac{1}{\rho^2}(\d_\psi+\Gamma^i_{\ i\psi})\chi^{[1]}_{\psi\psi}-(\d_k+\Gamma^i_{\ ik})\chi^k_{\ \psi}{}^{[1]}+\Gamma^i_{\ k\psi}\chi^k_{\ i}{}^{[1]}\nonumber\\
  &+  \left(\frac{1}{2}\d_\rho U^{[1]}+U^{[1]}\d_\rho+\frac{U^{[1]}}{\rho}+\Gamma^i_{\ i \rho}U^{[1]}\right)\Big(\langle T_{\rho\psi}\rangle^{[0]}=t^{[0]}_{\rho\psi}+\chi^{[0]}_{\rho\psi}\Big)\nonumber\\& \textcolor{black}{-\Gamma^i_{\ i\rho}{}^{[1]}\langle T_{\rho\psi}\rangle^{[0]}-\frac{1}{\rho^2}\Gamma^i_{\ i\psi}{}^{[1]}\langle T_{\psi\psi}\rangle^{[0]}+\Gamma^i_{\ k\psi}{}^{[1]}\langle T^k_{\ i}\rangle^{[0]}}.
\end{align}
    \item For $\sigma=i$
    \begin{align}\label{icons}
        &\nabla_\mu{}^{(q)} \langle T^\mu_{\ i}\rangle_q=0=(\partial_\rho+\Gamma^\rho_{\ \rho\rho}{}^{(q)}+\Gamma^\psi_{\ \psi\rho}{}^{(q)}+\Gamma^k_{\ k\rho}{}^{(q)})\langle T^\rho_{\ i}\rangle_q-\Gamma^k_{\ \rho i}{}^{(q)}\langle T^\rho_{\ k}\rangle_q\nonumber\\
        &+(\partial_\psi+ \Gamma^k_{\ k\psi}{}^{(q)})\langle T^\psi_{\ i}\rangle_q
        -\Gamma^k_{\ \psi i}{}^{(q)}\langle T^\psi_{\ k}\rangle_q+(\partial_k+\Gamma^l_{\ lk}{}^{(q)})\langle T^k_{\ i}\rangle_q-\Gamma^k_{\ li}{}^{(q)}\langle T^l_{\ k}\rangle_q\nonumber\\
        &-\Gamma^\rho_{\ ki}{}^{(q)}\langle T^k_{\ \rho}\rangle_q-\Gamma^\psi_{\ ki}{}^{(q)}\langle T^k_{\ \psi}\rangle_q.
    \end{align}
    
    Similarly using \eqref{chris} and \eqref{series} and retaining only the $O(q-1)$ terms, we get
 \begin{align}\label{repconsexpi}
        &\left(\d_\rho+\frac{1}{\rho}+\Gamma^k_{\ k\rho}\right)\langle T_{\rho i}\rangle^{[1]}+\frac{1}{\rho^2}(\d_{\psi}+\Gamma^k_{\ k\psi})\langle T_{\psi i}\rangle^{[1]}+(\d_k+\Gamma^l_{\ lk})\langle T^k_{\ i}\rangle^{[1]}\nonumber\\
        &-\Gamma^k_{\ \rho i}\langle T_{\rho k}\rangle^{[1]}-\Gamma^k_{\ \psi i}\frac{1}{\rho^2}\langle T_{\psi k}\rangle^{[1]}-\Gamma^k_{\ li}\langle T^l_{\ k}\rangle^{[1]}
        +f^{[0]}_{ki}(\rho,\psi,x^i) \langle T^k_{\ \rho}\rangle^{[1]}\nonumber\\
        &-\Gamma^\psi_{\ ki} \langle T^k_{\ \psi}\rangle^{[1]}=\left(\frac{\d_\rho U^{[1]}}{2}+\frac{U^{[1]}}{\rho}+\Gamma^k_{\ k \rho}U^{[1]}+U^{[1]}\d_\rho\right)\langle T_{\rho i}\rangle^{[0]}\\
        &-\Gamma^k_{\ \rho i}U^{[1]}\langle T_{\rho k}\rangle^{[0]}+U^{[1]}f^{[0]}_{ki}(\rho,\psi,x^i)\langle T^k_{\rho}\rangle^{[0]}\textcolor{black}{-\Gamma^k_{\ k \rho}{}^{[1]}\langle T_{\rho i}\rangle^{[0]}}\textcolor{black}{+\Gamma^k_{\ \rho i}{}^{[1]}\langle T_{\rho k}\rangle^{[0]}}\nonumber\\
        &\textcolor{black}{-\frac{1}{\rho^2}\Gamma^k_{\ k\psi}{}^{[1]}\langle T_{\psi i}\rangle^{[0]}+\frac{1}{\rho^2}\Gamma^k_{\ \psi i}{}^{[1]}\langle T_{\psi k}\rangle^{[0]} 
        -f^{[1]}_{ki}(\rho,\psi,x^i)\langle T^k_{\ \rho}\rangle{}^{[0]}+\Gamma^\psi_{\ ki}{}^{[1]}\langle T^k_{\ \psi}\rangle{}^{[0]}},\nonumber
    \end{align}
     where $f^{(q)}_{ki}(\rho,\psi,x^i)=(1/2)\d_\rho g^{(q)}_{ki}=f^{[0]}_{ki}(\rho,\psi,x^i)+(q-1)f^{[1]}_{ki}(\rho,\psi,x^i)+O((q-1)^2)$. Notice that there are no $\langle T_{ij}\rangle{}^{[0]}$ sources in the $\sigma=i$ equation. This is because $\Gamma^i_{\ jk}{}^{(q)}=\Gamma^i_{\ jk}{}^{[0]}$. We also notice that in $d=2,3,4$, $\chi_{\mu i}^{[0]}=0$ for $\mu\in\{\rho,\psi\}$ and hence $\chi_{\mu i}^{[1]}=0$ for $\mu\in\{\rho,\psi\}$ (see section \eqref{repgeomstress}).
\end{enumerate}

 Tracelessness and conservation gives us $(d+1)$ equations to solve for $\langle T_{\mu\nu} \rangle^{[1]}$ (and hence $t_{\mu\nu} {}^{[1]}$) in terms of $\langle T_{\mu\nu} \rangle$ (and $t_{\mu\nu} {}^{[0]}$) for any state satisfying the symmetry requirements. Although, there are $d(d+1)/2$ independent components of $\langle T_{\mu\nu} \rangle^{[1]}$ (or $t_{\mu\nu} {}^{[1]}$) and only $(d+1)$ equations to solve for them, under the presence of certain symmetries we can reduce the number of independent components to match the number of constraint equations. This is true for a static metric on $M_1$. Static metric on $M_1$ that are solutions to Einstein field equations with $\Lambda>0$(dS), $\Lambda<0$(AdS), $\Lambda=0$(flat), have spherical or planar or hyperbolic symmetry along $y^i{}'s \in \Sigma$ hence $g_{ij}=0$ for $i\neq j$. On a static metric we also have $g_{\rho i}=0=g_{\psi i}$\footnote{This is not true in non static spacetimes.}. We consider states that respect these symmetries as well i.e., states that satisfy 
 \begin{align}
    &\langle T_{\rho i}\rangle=0=\langle T_{\psi i}\rangle.
 \end{align}
These states are of interest in the context of a black hole problem since we expect the average behavior of the Hawking radiation to also respect this symmetry. 
 
Therefore, all the source terms in the $i-$replica conservation equation i.e., \eqref{repconsexpi} vanish, giving us only the homogenous solution. The replica metric also has the symmetries discussed above i.e., it has spherical or planar or hyperbolic symmetry along $y^i{}'s \in \Sigma$ (there are no non trivial replica corrections along $y^i{}'s$ \footnote{In non static case if we take a finite entangling region along $y^i{}'s$, this won't be true.}) hence $g^{(q)}_{ij}=0$ for $i\neq j$ and $g^{(q)}_{\rho i}=0=g^{(q)}_{\psi i}$. Therefore, for the states considered on $M_1$, we have 
\begin{align}\label{staticconstraint}
    &\langle T_{\rho i}\rangle^{[1]}=0=\langle T_{\psi i}\rangle^{[1]},\\
    &\langle T_{ij}\rangle^{[1]}=\langle T_{ij}\rangle^{[0]} \quad \forall\, i\neq j.
 \end{align}
With this the $\sigma=i$ conservation equation on $M_q$ \eqref{icons} gives,
\begin{equation}
        (\d_k+\Gamma^l_{\ lk})\langle T^k_{\ i}\rangle^{[1]}-\Gamma^k_{\ li}\langle T^l_{\ k}\rangle^{[1]}=0.
\end{equation}
\eqref{staticconstraint} further implies $t_{\mu i}^{[1]}=-\chi_{\mu i}^{[1]}$ for $\mu \in\{\rho,\psi\}$. Since in $d=2,3,4$, $\chi_{\mu i}^{[0]}=0=\chi_{\mu i}^{[1]}$ for $\mu\in\{\rho,\psi\}$, therefore in $d=2,3,4$
 \begin{align}
   &t_{\rho i}^{[1]}=t_{\psi i}^{[1]}=0, \\
   &t_{ij}^{[1]}=\langle T_{ ij} \rangle^{[0]}-\chi_{ij}^{[1]}.
\end{align}

On static spacetime since $g_{ij}=0$ for $i\neq j$ (due to symmetries in $\Sigma$) we have $\langle T_{ij}\rangle^{[0]}=0$ for $i\neq j$ (since the only tensors in $\Sigma$ are $g_{ij}$ and tensors formed using it, therefore $\langle T_{ij}\rangle^{[0]}$ must be proportional to $g_{ij}$ or tensors formed using it). We do not consider states that violate this. For such states \eqref{icons} further reduces to,
 \begin{equation}\label{sigmaieqn}
 \addtolength{\fboxsep}{5pt}
   \boxed{
   \begin{gathered}
 \text{For a fixed $i$: }       (\d_i+\sum_l \Gamma^l_{\ li})\langle T^i_{\ i}\rangle^{[1]}-\sum_{l} \Gamma^l_{\ li}\langle T^l_{\ l}\rangle^{[1]}=0.
 \end{gathered}
}
\end{equation}
We also note $\Gamma^l_{\ li}=\frac{1}{2}g^{ll}\d_i g_{ll}$ (no sum over $l$).\\ \\
We consider certain specific cases below. For $g_{ij}=\delta_{ij}$ or $d\leq 3$ we have $\Gamma^i_{\ jk}=0$. In this case, we simply have,
\begin{equation}
   \text{For a fixed $i$: }  \d_i\langle T^i_{\ i}\rangle^{[1]}=0.
\end{equation}
In $d=4$ static black holes in AdS, dS and flat we can choose coordinates such that $g_{ij}dy^idy^j=r^2(d\theta^2+h_{\phi\phi}(\theta)d\phi^2) $, where
\begin{align}
   h_{\phi\phi}(\theta)&=\sin^2(\theta); \quad k=1 \text{ (spherical)},\\
   &=\sinh^2(\theta) \text{ or } \cosh^2(\theta) \text{ or } e^{-2\theta}; \quad k=-1 \text{ (hyperbolic)},\\
   g_{ij}dy^idy^j&=\frac{r^2}{l^2}(d\theta^2+d\phi^2); \quad k=0 \text{ (planar)},
\end{align}
here $k$ is the parameter introduced in \eqref{hortopo}. We have
\begin{align}
    &(\d_\theta+\frac{1}{2h_{\phi\phi}}\d_{\theta}h_{\phi\phi})\langle T^\theta_{\ \theta}\rangle^{[1]}-\frac{1}{2h_{\phi\phi}}\d_{\theta}h_{\phi\phi} \langle T^\phi_{\ \phi}\rangle^{[1]}=0,\\
    &\d_\phi \langle T^\phi_{\ \phi}\rangle^{[1]}=0.
\end{align}

Therefore in the static case the only non-trivial replica stress tensor expectation corrections are $\{\langle T_{\rho \rho}\rangle{}^{[1]},\langle T_{\rho \psi}\rangle{}^{[1]},\langle T_{\psi \psi}\rangle{}^{[1]}, \langle T_{\ ii}\rangle{}^{[1]} \}$ - a total of $(d+1)$ components determined by the $(d+1)$ linear equations - the trace anomaly condition i.e., \eqref{fullreptrace1} and the $d$ conservation equations \eqref{repconsexprho}, \eqref{repconsexppsi}, \eqref{sigmaieqn}. We have, \\
\\
\textbf{$O(q-1)$ Trace condition:}
\begin{equation}\label{trlessfull}
  \boxed{  \langle T_{\rho\rho}\rangle^{[1]}+\frac{1}{\rho^2}\langle T_{\psi\psi}\rangle^{[1]}+\langle T^i_{\ i}\rangle^{[1]}=U^{[1]}(\rho,a)\langle T_{\rho\rho}\rangle^{[0]}+\mathcal{A}^{[1]}[g_{\mu\nu}^{(q)}].}
\end{equation}
\textbf{$O(q-1)$ $\rho$-component of Conservation:} 

 \begin{equation}
   \boxed{
   \begin{gathered}
     \left(\partial_\rho+\frac{1}{\rho}+\Gamma^i_{\ i\rho}\right)\langle T_{\rho \rho}\rangle{}^{[1]}+\frac{1}{\rho^2}(\partial_\psi+\Gamma^i_{\ i\psi})\langle T_{\psi \rho}\rangle{}^{[1]}-\frac{1}{\rho^3}\langle T_{\psi \psi}\rangle{}^{[1]}
        \\-\sum_i \Gamma^i_{\ i\rho}\langle T^i_{\ i}\rangle{}^{[1]}= \left(\d_\rho+\frac{1}{\rho}+\Gamma^i_{\ i\rho}\right)(U^{[1]}\langle T_{\rho \rho}\rangle{}^{[0]})\textcolor{black}{-\Gamma^i_{\ i\rho}{}^{[1]}\langle T_{\rho \rho}\rangle{}^{[0]}}
        \\\textcolor{black}{-\frac{1}{\rho^2}\Gamma^i_{\ i\psi}{}^{[1]}\langle T_{\psi \rho}\rangle{}^{[0]}+\sum_i \Gamma^i_{\ i\rho}{}^{[1]}\langle T^i_{\ i}\rangle{}^{[0]} },
        \end{gathered}
        }
    \end{equation}
or using trace equation to eliminate $\langle T_{\rho\rho}\rangle_q$,
\begin{equation}\label{rhoconsfull}
\addtolength{\fboxsep}{5pt}
   \boxed{
   \begin{gathered}
  \frac{1}{\rho^2}(\d_\psi+\Gamma^i_{\ i \psi})\langle T _{\rho\psi}\rangle^{[1]}-\frac{1}{\rho^2}(\d_\rho+\Gamma^i_{\ i \rho})\langle T_{\psi\psi} \rangle^{[1]}
  \\-\left(\d_\rho+\frac{1}{\rho}+\Gamma^i_{\ i \rho}\right) \left(\langle T^k_{\ k} \rangle^{[1]}-\mathcal{A}^{[1]}[g_{\mu\nu}^{(q)}]\right)-\sum_i \Gamma^i_{\ i\rho}\langle T^i_{\ i} \rangle^{[1]}
  =\textcolor{black}{\frac{1}{\rho^2}(\Gamma^i_{\ i \rho}{}^{[1]}\langle T_{\psi\psi} \rangle^{[0]}
  }
  \\\textcolor{black}{-\Gamma^i_{\ i \psi}{}^{[1]}\langle T_{\rho\psi} \rangle^{[0]})+\Gamma^i_{\ i \rho}{}^{[1]}\left(\langle T^k_{\ k} \rangle^{[0]}-\mathcal{A}[g_{\mu\nu}]\right)}\textcolor{black}{+\sum_i\Gamma^i_{\ i \rho}{}^{[1]}\langle T^i_{\ i} \rangle^{[0]}}.
  \end{gathered}
}
\end{equation} 
\textbf{$O(q-1)$ $\psi$-component of Conservation:}
\begin{equation}\label{psiconsfull}
  \addtolength{\fboxsep}{5pt}
   \boxed{
   \begin{gathered}
     \left(\d_\rho+\frac{1}{\rho}+\Gamma^i_{\ i \rho}\right)\langle T_{\rho\psi}\rangle^{[1]}+\frac{1}{\rho^2}(\d_\psi+\Gamma^i_{\ i\psi})\langle T_{\psi\psi}\rangle^{[1]}-\sum_i\Gamma^i_{\ i\psi}\langle T^i_{\ i}\rangle^{[1]}\\=
   \left(\frac{1}{2}\d_\rho U^{[1]}+U^{[1]}\d_\rho+\frac{U^{[1]}}{\rho}+\Gamma^i_{\ i \rho}U^{[1]}\right)\langle T_{\rho\psi}\rangle^{[0]}\\ \textcolor{black}{-\Gamma^i_{\ i\rho}{}^{[1]}\langle T_{\rho\psi}\rangle^{[0]}-\frac{1}{\rho^2}\Gamma^i_{\ i\psi}{}^{[1]}\langle T_{\psi\psi}\rangle^{[0]}+\sum_i\Gamma^i_{\ i\psi}{}^{[1]}\langle T^i_{\ i}\rangle^{[0]}}.
    \end{gathered}
    }
    \end{equation}
We can now use \eqref{sigmaieqn} to solve for $\langle T_{ii}\rangle {}^{[1]}$, then \eqref{rhoconsfull} and \eqref{psiconsfull} to solve for $\langle T_{\rho\psi}\rangle {}^{[1]}$ and $\langle T_{\psi\psi}\rangle {}^{[1]}$ in terms of $\langle T_{\rho\psi}\rangle {}^{[0]}$, $\langle T_{\psi\psi}\rangle {}^{[0]}$, $\langle T_{ii}\rangle {}^{[0]}$ and $U^{[1]}(\rho,a)$, $p(q)$. \eqref{trlessfull} can then be used to solve for $\langle T_{\rho\rho}\rangle {}^{[1]}$ in terms of $\langle T_{\mu\nu}\rangle {}^{[0]}$, $U^{[1]}(\rho,a)$, $p(q)$. The scaling of the R\'{e}nyi entropies which depends on $\langle T_{\mu\nu}\rangle_q$, and EE \eqref{sqftvar2} which depends only on $\langle T_{\mu\nu}\rangle {}^{[1]}$, can be expressed completely in terms of $\langle T_{\mu\nu}\rangle {}^{[0]}$ for a general state (i.e., stress tensor on $M_1$ only).

Since $\chi_{\mu\nu}^{[1]}$ is already known \eqref{repgeomstress}, we could also directly calculate $t_{\mu\nu}^{[1]}$ using the conservation, trace equations above and \eqref{stresssplit}. Terms with $\chi_{\mu\nu}^{[1]}$ now act as additional source terms.

\subsection{Sources - Stress Tensor on \texorpdfstring{$M_1$}{}}
The sources in the trace and conservation equations \eqref{trlessfull}-\eqref{psiconsfull} are $\langle T_{\rho\psi}\rangle {}^{[0]}$, $\langle T_{\psi\psi}\rangle {}^{[0]}$, $\langle T_{\rho\rho}\rangle {}^{[0]}$, $\langle T_{ii}\rangle^{[0]}$ (i.e., stress tensor on base manifold $M_1$) and $\mathcal{A}^{[1]}[g_{\mu\nu}^{(q)}]$.

\paragraph{Flat space, vacuum:} 
\begin{equation}
    \langle T_{\mu\nu} \rangle^{[0]}=0=t_{\mu\nu} ^{[0]}.
\end{equation}
We can choose renormalisation scheme such that $\chi_{\mu\nu}[\eta_{\mu\nu}]=0$.

\paragraph{Curved space, vacuum:}
\begin{equation}
    \langle T_{\mu\nu} \rangle^{[0]}=\chi_{\mu\nu}^{[0]}.
\end{equation}

\paragraph{Thermal/equilibrium fluid stress tensor:}

The stress tensor at temperature $T_{QFT}$ in a CFT is of particular interest. Some of the static backgrounds are also a black hole with inherent temperature $T_{BH}$, which will be equivalent to $T_{QFT}$ in (quasi) thermal equilibrium.

The finite temperature stress tensor is expressed more elegantly as an equilibrium relativistic fluid tensor, with a conformal equation of state.

In general, the relativistic fluid tensor can be expressed as \cite{Romatschke:2009kr}

\begin{equation}
    t_{\mu\nu}^{[0]}=\epsilon u_\mu u_\nu +p(g_{\mu\nu}+u_\mu u_\nu),
\end{equation}
where $\epsilon$ is the energy density, $p$ the pressure, $u_\mu u^\mu=-1$. 
In a static gauge ($x^\mu=\{\tau, r,x^i\}$), one can choose $\hat{u}=\d/\d\tau$ or $u_\mu=\delta_{\mu}^ \tau$. We have,
\begin{equation}
     t_{\mu\nu}^{[0]}=(\epsilon+p) \delta^\tau_\mu \delta^\tau_\nu+pg_{\mu\nu},
\end{equation}
i.e.,
\begin{align}
    &t_{rr}^{[0]}=p g_{rr},\nonumber\\
    &t_{\tau\tau}^{[0]}= \epsilon+p(1+g_{\tau\tau}),\nonumber\\
    &t_{r\tau}^{[0]}= p g_{r \tau},\nonumber\\
    &t_{ij}^{[0]}=p g_{ij},
\end{align}
or using \eqref{conversion}
\begin{align}\label{thermalrhopsi}
    &t_{\rho\rho}^{[0]}=p g_{rr}\cos(\psi)^2+(\epsilon+p(1+g_{\tau\tau}))\sin(\psi)^2+pg_{r\tau}\sin(2\psi),\nonumber\\
    &t_{\psi\psi}^{[0]}= p g_{rr}\rho^2\sin(\psi)^2+(\epsilon+p(1+g_{\tau\tau}))\rho^2\cos(\psi)^2-pg_{r\tau}\rho^2\sin(2\psi),\nonumber\\
    &t_{\rho\psi}^{[0]}= -p g_{rr}\rho\sin(\psi)\cos(\psi)+(\epsilon+p(1+g_{\tau\tau}))\rho\sin(\psi)\cos(\psi)+pg_{r\tau}\rho\cos(2\psi),\nonumber\\
    &t_{ij}^{[0]}=p g_{ij}.
\end{align}
In $(-1,+1,+1,...,+1)$ signature where $u_\mu u^\mu=-1$, tracelessness $t^\mu_{\ \mu}{}^{[0]}=0$ implies,
\begin{equation}
    \epsilon=(d-1)p.
\end{equation}
This is the well-known equation of state for a conformal theory. Therefore in a CFT we can rewrite the state part of the thermal stress tensor as,  
\begin{equation}
    t_{\mu\nu}^{[0]}= p(g_{\mu\nu}+d  u_\mu u_\nu)= p(g_{\mu\nu}+d  \delta^\tau_\mu \delta^\tau_\nu).
\end{equation}
In the Euclidean $(+1,+1,+1,...,+1)$ signature we have $u_\mu u^\mu=1$, and $\epsilon=-(d+1)p$, which implies,
\begin{equation}\label{fluidstress}
    t_{\mu\nu}^{[0]}{}^E= p(g_{\mu\nu}-d  u_\mu u_\nu)=p(g_{\mu\nu}-d  \delta^\tau_\mu \delta^\tau_\nu),
\end{equation}
where in the second equality we have chosen a static gauge ($x^\mu=\{\tau, r,x^i\}$), with $\hat{u}=\d/\d\tau$ or $u_\mu=\delta^\tau_{\mu}$. 

In flat space we can choose a gauge in which $g_{\mu\nu}=\delta_{\mu\nu}$, with $p=CT^d$.

The conservation equation gives \cite{Romatschke:2009kr},
\begin{equation}
    \nabla^\mu  t^{[0]}_{\mu\nu}=0=\d_\nu p+dD_u p\, u_\nu+dp\,a_\nu,
\end{equation}
where
\begin{align}
    &D_u p=u_\mu\nabla^\mu p,\\
    &a_\nu=u_\mu\nabla^\mu u_\nu.
\end{align}
This is often projected along ($u^\nu\nabla^\mu  t^{[0]}_{\mu\nu}$) and transverse ($\Delta^{\alpha \nu}\nabla^\mu  t^{[0]}_{\mu\nu}$ $(\Delta^{\alpha \nu}=g^{\alpha \nu}+u^\alpha u^\nu)$) to the fluid. For a general relativistic fluid, these projections give
\begin{align}
    &-D_u \epsilon - (\epsilon+p)(\nabla^\nu u_\nu)=0,\\
    &(\epsilon+p)a_\nu+\nabla^\perp_\nu p=0; \quad \nabla^\perp_\nu=\Delta^{\mu\nu}\nabla_\nu.
\end{align}
In our context $\epsilon+p=dp$ and thus,
\begin{align}
   & \nabla^\nu u_\nu = -\frac{d-1}{d} D_u(\ln p),\\
   & a_\nu = -\frac{1}{d} \nabla^\perp_\nu(\ln p).
\end{align}
The theory of relativistic hydrodynamics i.e., derivative expansion around equilibrium is well established; these conservation relations are corrected order by order.

We will now explicitly solve the conservation and trace equations on $M_q$ in certain examples.

\section{Replica stress tensor - vacuum states} \label{sec:seven}
In this section, we consider a CFT vacuum state on a $d-$dimensional metric. Let us first consider unsquashed cones in $M_q$. These correspond to static metrics on $M_1$ that have zero extrinsic curvature for boundary $\Sigma$ embedded in $M_1$. Among solutions to Einstein field equations, this is trivially true for all metrics in 2 dimensions, 3 dimensional metrics of the form 
\begin{equation}\label{3dimfinitelen}
   g_{\tau\tau}(r) d\tau^2+g_{rr}(r) dr^2+\phi_0^2 dy^2,
\end{equation}
with the entangling region $\tau$ constant, $r_1\leq r \leq r_2$ spanning the range of $y^i$'s, and flat backgrounds in cartesian coordinates in $d\geq 2$ dimensions i.e., metrics of form \eqref{flatmet2} with planar entangling region - $\tau$ constant, $x_1\leq x \leq x_2$ spanning the range of $z^i$'s i.e., $M^{d-2}$. On these backgrounds all christoffels with $y^i$ or $z^i$ component(s) vanish. The replica metric is given by the unsquashed conical metric \eqref{repmet1}. Along the $\Sigma$ components the replica metric is the same as the metric on $M_1$. The Killing vectors on $g_{\mu\nu}^{(q)}$ are $\frac{1}{\rho} \d_{\psi}$, $\d_i$. 

\subsection{Connected entangling boundary with no extrinsic curvature}
In this case the full replica trace and conservation equations given by \eqref{fullreptrace1}, \eqref{rhocons}, \eqref{psicons}, \eqref{icons} reduce to:  

\textbf{Trace condition:}
\begin{equation}\label{fptreqn}
   \langle T^\rho_{\ \rho}\rangle{}_q+\langle T^\psi_{\ \psi}\rangle_q+\langle T^i_{\ i}\rangle{}_q=\mathcal{A}[g_{\mu\nu}^{(q)}].
\end{equation}

\textbf{$\rho$-component of Conservation:}
    \begin{equation}\label{fprhocomp}
      \left(\partial_\rho+\frac{1}{\rho}\right)\langle T^\rho_{ \ \rho}\rangle{}_q+\frac{1}{\rho^2}\partial_\psi\langle T_{ \rho \psi}\rangle{}_q-\frac{1}{\rho}\langle T^\psi_{\ \psi}\rangle{}_q
        =0,
    \end{equation}
    
    or using trace equation to eliminate $\langle T^\rho_{\rho}\rangle_q$,
\begin{equation}\label{fprhotr}
  \frac{1}{\rho^2}\partial_\psi\langle T_{ \rho \psi}\rangle{}_q-\left(\d_\rho+\frac{2}{\rho}\right)\langle T^\psi_{\ \psi} \rangle{}_q-\left(\d_\rho+\frac{1}{\rho}\right) \left(\langle T^k_{\ k} \rangle{}_q-\mathcal{A}[g_{\mu\nu}^{(q)}]\right)=0.
\end{equation}

    \textbf{$\psi$-component of Conservation:}
\begin{equation}\label{fppsicons}
     \left(U^{-1}\d_\rho-\frac{1}{2}U^{-2}\d_\rho U+\frac{U^{-1}}{\rho}\right)\langle T_{\rho \psi}\rangle{}_q+\d_\psi\langle T^\psi_{\ \psi}\rangle{}_q=0.
    \end{equation}
    
    \textbf{$i$-component of Conservation:}
\begin{equation}\label{fpicomp}
     \text{For a fixed $i$: }  \d_i\langle T^i_{\ i}\rangle{}_q=0.
\end{equation}
Let us consider the vacuum state on $M_1$, for which $\langle T_{\mu\nu}\rangle^{[0]}=0$. The solutions must be independent of $\psi$ and $y^i$ since $\frac{1}{\rho} \d_{\psi}$, $\d_i$ are Killing vectors on $M_q$ in the unsquashed case and the vacuum state respects this symmetry \footnote{In general we consider the theory on $M_q$ and the states in the spectrum that respect global symmetries.}. Since $\mathcal{A}[g_{\mu\nu}^{(q)}]$ depends on only $g_{\mu\nu}^{(q)}$ it is independent of $\psi$ in the unsquashed case. Considering that the vacuum state itself does not introduce any preferred tensor or dimensionful parameter, and $g_{\rho\psi}^{(q)}=0$ and that the only invariant vectors on $M_q$ are $\frac{1}{\rho} \d_{\psi}$, $\d_i$, we can conclude $\langle T^\rho_{\ \psi}\rangle{}_q=0$. We also note that $\langle T^\mu_{\ \nu}\rangle{}_q$ has mass dimension $d$ in $d-$dimensions.  

Since all replica solutions are independent of $\psi$, \eqref{fptreqn}-\eqref{fpicomp} gives,

\begin{align}\label{fppsiindep}
 \text{Trace: }  &\langle T^\rho_{\ \rho}\rangle{}_q+\langle T^\psi_{\ \psi}\rangle_q+\langle T^i_{\ i}\rangle{}_q=\mathcal{A}[g_{\mu\nu}^{(q)}],\nonumber\\
 \text{$\rho$ conservation: }  &\left(\partial_\rho+\frac{1}{\rho}\right)\langle T^\rho_{ \ \rho}\rangle{}_q-\frac{1}{\rho}\langle T^\psi_{\ \psi}\rangle{}_q=0,\nonumber \\   
 \text{$\psi$ conservation: } &\left(\d_\rho+\frac{1}{2}U^{-1}\d_\rho U+\frac{1}{\rho}\right)\langle T^\rho_{\ \psi}\rangle{}_q=0,\nonumber\\
  \text{$i$ conservation: } &\d_i\langle T^i_{\ i}\rangle{}_q=0  \text{; for a fixed $i$}, 
\end{align}
using trace equation to eliminate $\langle T^\rho_{\ \rho}\rangle_q$ in $\rho$ conservation equation,
\begin{equation}\label{fprhotrpsiindep}
  \text{$\rho$ conservation and trace: }  \left(\d_\rho+\frac{2}{\rho}\right)\langle T^\psi_{\ \psi} \rangle{}_q+\left(\d_\rho+\frac{1}{\rho}\right) \left(\langle T^k_{\ k} \rangle{}_q-\mathcal{A}[g_{\mu\nu}^{(q)}]\right)=0.
\end{equation}
Solving the $\psi$ conservation equation above, we get,
\begin{equation}
    \langle T^\rho_{\ \psi}\rangle{}_q=\frac{a_1}{\rho\sqrt{U}}.
\end{equation}
For this to be dimensionally correct, the constant $a_1$ must have mass dimension $(d-1)$. Since there are no dimensionful constants when we consider the vacuum on $M_1$ therefore the only solution to the above case is $a_1=0$. 

 We see that in the absence of anomaly ($d$ odd), the PDEs i.e., \eqref{fppsiindep} and hence its solutions are independent of the regulator $(a)$. The only dimensionful parameter is $\rho$. Therefore, in the absence of anomaly the solution $\langle T^\mu_{\ \nu}\rangle{}_q\propto \frac{1}{\rho^d}$ in $d-$dimensions, with the dimensionless proportionality constant depending on $d$ and $q$. This constant has to be of $O(q-1)$ since $\langle T^\mu_{\ \nu}\rangle{}_{q=1}=0$. 

 Alternatively, conformal invariance under $g_{\mu\nu}^{(q)}\to \omega^2 g_{\mu\nu}^{(q)}$ for constant $\omega$ implies that under this transformation $T^\mu_{\ \nu}{}_{(q)}\to \omega^{-d} T^\mu_{\ \nu}{}_{(q)}$. Under the absence of anomaly the vev $\langle T^\mu_{\ \nu}\rangle_{q, ren}$ obeys the same transformation law. We also have  
\begin{equation}
    ds_q^2 = \omega^2\left(d\left(\frac{\rho}{\omega}\right)^2+\left(\frac{\rho}{\omega}\right)d\phi^2+\delta_{ij}d\left(\frac{y^i}{\omega}\right)d\left(\frac{y^j}{\omega}\right)\right).
\end{equation}
Therefore,  $\langle  T^\mu_{\ \nu}\rangle_{q, ren}=\omega^{-d} \langle  T^\mu_{\ \nu}\left(\frac{\rho}{\omega}\right)\rangle_{q, ren}$ which implies $\langle T^\mu_{\ \nu}\rangle^{[1]}\propto \frac{1}{\rho^d}$.

In the presence of anomaly there is now an additional dimensionful parameter $a$ (conical singularity regulator) and hence $\langle T^\mu_{\ \nu}\rangle{}_q$ now has an additional term which depends on dimensionful parameters $\rho$ and $a$ and dimensionless constants $d$ and $q$. This is not necessarily only the geometric part of the replica stress tensor ($\chi^\mu_{\nu}{}^{(q)}$) since the state part can also have $a$ dependence due to reasons specified around \eqref{nonuniquedecomp} (i.e., non-uniqueness in state and geometric decomposition).

Equation \eqref{fpicomp} implies
\begin{equation}\label{masslessCFTi}
   \langle T^i_{\ j}\rangle{}_q=g_1(\rho,a) \delta^i_{\ j}=\left(g(\rho,a)+\frac{C_d}{\rho^d}\right)\delta^i_{\ j}.
\end{equation}
We have assumed that the functional form is same in every direction in $\Sigma$. This is true since the replica $q$ and $a$ dependence is same along all directions in $\Sigma$ in $g_{\mu\nu}^{(q)}$ (it is trivial for the unsquashed case) and the vacuum state on $M_1$ has no preference in directions in $\Sigma$. We also have,
\begin{align}
   \langle T^\psi_{\ \psi}\rangle{}_q=g_2(\rho,a).
\end{align}
We define $f(\rho,a)=\mathcal{A}[g_{\mu\nu}^{(q)}]-(d-2)g(\rho,a)$. The $\rho$ conservation and trace equation 
\eqref{fprhotrpsiindep} implies,
\begin{equation}
\begin{split}
   \frac{1}{\rho^2} \d_\rho(\rho^2\langle T^\psi_{\ \psi}\rangle{}_q)&= \left(\d_\rho+\frac{1}{\rho}\right) \left(\mathcal{A}[g_{\mu\nu}^{(q)}]-\langle T^k_{\ k}\rangle{}_q\right)\\
   &=\left(\d_\rho+\frac{1}{\rho}\right) \left(f(\rho,a)-(d-2)\frac{C_d}{\rho^d}\right),
\end{split}
\end{equation}
which implies
\begin{equation}\label{masslessCFTpsi}
    \langle T^\psi_{\ \psi}\rangle{}_q=(1-d)\frac{C_d(1-\delta_{d,2})}{\rho^d}+f-\frac{1}{\rho^2}\int d\rho \rho f+\frac{c_1}{\rho^2}.
\end{equation}
The trace equation along with \eqref{masslessCFTi}, \eqref{masslessCFTpsi} implies,
\begin{equation}\label{masslessCFTrho}
  \langle T^\rho_{\ \rho}\rangle{}_q=\frac{C_d(1-\delta_{d,2})}{\rho^d}+\frac{1}{\rho^2}\int d\rho \rho f-\frac{c_1}{\rho^2}.
\end{equation}
In $d\neq 2$ we have the constant $c_1=0$ (using dimensional analysis for renormalised stress tensor). Therefore in $d\geq2$ we have,
\begin{align}\label{fprhopsisol}
     &\langle T^\psi_{\ \psi}\rangle{}_q=(1-d)\frac{C_d}{\rho^d}+f-\frac{1}{\rho^2}\int d\rho \rho f,\nonumber\\
     &\langle T^\rho_{\ \rho}\rangle{}_q=\frac{C_d}{\rho^d}+\frac{1}{\rho^2}\int d\rho \rho f,
\end{align}
where we have identified $C_2 $ with $ -c_1$ in $d=2$.

In the solutions above there is an arbitrary function $g(\rho,a)$ which encodes the regulator $a$ dependence in $\langle T_{ii}\rangle{}_q$. The non trivial terms involving the coefficient $C_d(q)$ in the replica corrections to stress tensor vev show up because the presence of a conical singularity in $M_q$ at the boundary breaks any translation symmetry along $(r,\tau)$ that is present on $M_1$.  

In the above case we have considered a CFT (i.e., it satisfies trace condition \eqref{fptreqn}). If we however consider a QFT, the stress tensor on the replica and base manifold satisfy only the conservation equations i.e., in this case we have \eqref{fprhocomp}, \eqref{fppsicons}, \eqref{fpicomp} which reduce to \eqref{fppsiindep}. If the QFT has no other dimensionful parameters, in this case the non trivial solutions are given by: 
\begin{align}\label{masslessQFT}
   &\langle T^\psi_{\ \psi}\rangle{}{}_q=\frac{Q_d(q)}{\rho^d}+F(\rho,a,q),\nonumber\\
   &\langle T^\rho_{\ \rho}\rangle{}_q=\frac{Q_d(q)}{(1-d)\rho^d}+\frac{1}{\rho}\int d\rho F(\rho,a,q),\nonumber\\
  &\langle T^i_{\ j}\rangle{}_q=\left(\frac{R_d (q)}{\rho^d}+ G(\rho,a,q)\right)\delta^i_{\ j}.
\end{align}
In the presence of dimensionful parameters such as mass the dependence on $\rho$ in the $a$ independent term changes, and hence the coefficients depending on $q$ and $d$ change respecting the conservation equations \eqref{fprhocomp}, \eqref{fppsicons}, \eqref{fpicomp}. 

For example, consider the case of a free QFT with mass $m$ (with an $m^2$ dependent term in Lagrangian density). In such cases the $m$ dependence that shows up in the vev of the stress tensor is an overall $m^2$ factor. Using the conservation equations i.e., in this case \eqref{fprhocomp}, \eqref{fppsicons}, \eqref{fpicomp} which reduce to \eqref{fppsiindep}, we have,
\begin{align}\label{massive}
   &\langle T^\psi_{\ \psi}\rangle{}_q=\frac{m^2 M_d(q)}{\rho^{d-2}}+F_1(\rho,a,q),\nonumber\\
   &\langle T^\rho_{\ \rho}\rangle{}_q=\frac{m^2 M_d(q)}{(3-d)\rho^{d-2}}+\frac{1}{\rho}\int d\rho F_1(\rho,a,q),\nonumber\\
   &\langle T^i_{\ j}\rangle{}_q=\left(\frac{m^2 S_d(q)}{\rho^{d-2}}+ G_1(\rho,a,q)\right)\delta^i_{\ j}.
\end{align}
For the traceless part of $\langle T^\mu_{\ \nu}\rangle{}_q$ we have,
\begin{equation}\label{massive trless}
    S_d(q)=\frac{(d-4)}{(3-d)(d-2)}M_d(q).
\end{equation}
The terms with the regulator $a$ dependence (purely replica metric-dependent terms) are localised around the endpoints of the entangling region. The terms without $a$ dependence (state-dependent terms) describe the state and are non-local i.e., it has contribution over the entire entangling region. We treat both terms appropriately as described in section \eqref{metstate}.

In general the regulator $a$ independent/ state-dependent part of $\left\langle T^\mu{ }_\nu\right\rangle{}_q$ in $d-$dimensions is of form,
\begin{align}\label{aindeprepstress}
&\left\langle T^\mu{ }_\nu\right\rangle{}_{q,\text{ $a$=0}}
=\frac{1}{\rho^d} \Bigg[C_d(q)\operatorname{diag}((1-d),1,1,1,...,1)\nonumber\\
&+ \operatorname{diag}\left(Q_d(q),\frac{Q_d(q)}{1-d},R_d(q),R_d(q),...,R_d(q)\right) \nonumber \\
& +(m\rho)^2 M_d(q) \operatorname{diag}\left(1,\frac{1}{3-d},\frac{d-4}{(3-d)(d-2)},...,\frac{d-4}{(3-d)(d-2)}\right)\nonumber\\
&+(m\rho)^2   \operatorname{diag}\left(N_d(q),\frac{N_d(q)}{3-d},S_d(q),S_d(q),...,S_d(q)\right)\Bigg],
\end{align}
where the basis is in order $(\psi,\rho,y^i)$. We will denote the coefficients in the $O(q-1)$ contribution $\left\langle T^\mu{ }_\nu\right\rangle{}^{[1]}_{\text{ $a$=0}}$ using $A^{[1]}$ where $A$ is the corresponding $q$-dependent coefficient in $\left\langle T^\mu{ }_\nu\right\rangle{}_{q,\text{$ a$=0}}$.

We can also directly solve the replica trace and conservation equations \eqref{fptreqn}-\eqref{fpicomp} at $O(q-1)$ to get $\left\langle T^\mu{ }_\nu\right\rangle{}^{[1]}_{\text{ $a$=0}}$ directly in terms of $\left\langle T^\mu{ }_\nu\right\rangle{}^{[0]}_{\text{ $a$=0}}$. It is worth noting these as well,
\\

\textbf{$O(q-1)$ Trace condition:}
\begin{equation}
   \langle T^\rho_{\ \rho}\rangle^{[1]}+\langle T^\psi_{\ \psi}\rangle^{[1]}+\langle T^i_{\ i}\rangle^{[1]}=\mathcal{A}^{[1]}[g_{\mu\nu}^{(q)}].
\end{equation}

\textbf{$O(q-1)$ $\rho$-component of Conservation:}
    \begin{equation}
      \left(\partial_\rho+\frac{1}{\rho}\right)\langle T^\rho_{\  \rho}\rangle{}^{[1]}+\frac{1}{\rho^2}\partial_\psi\langle T_{\psi \rho}\rangle{}^{[1]}-\frac{1}{\rho}\langle T^\psi_{\ \psi}\rangle{}^{[1]}
        =0,
    \end{equation}
   
    or using trace equation to eliminate $\langle T_{\rho\rho}\rangle_q$,
\begin{equation}
  \frac{1}{\rho^2}\d_\psi\langle T _{\rho\psi}\rangle^{[1]}-\frac{1}{\rho^2}\d_\rho\left(\rho^2\langle T^\psi_{\ \psi} \rangle^{[1]}\right)-\left(\d_\rho+\frac{1}{\rho}\right) \left(\langle T^k_{\ k} \rangle^{[1]}-\mathcal{A}^{[1]}[g_{\mu\nu}^{(q)}]\right)=0.
\end{equation} 

    \textbf{$O(q-1)$ $\psi$-component of Conservation:}
\begin{equation}
     \left(\d_\rho+\frac{1}{\rho}\right)\langle T_{\rho\psi}\rangle^{[1]}+\d_\psi\langle T^\psi_{\ \psi}\rangle^{[1]}=
   \left(\frac{1}{2}\d_\rho U^{[1]}+U^{[1]}\d_\rho+\frac{U^{[1]}}{\rho}\right)\langle T_{\rho\psi}\rangle^{[0]}.
    \end{equation}
  
    \textbf{$O(q-1)$ $i$-component of Conservation:}
\begin{equation}
     \text{For a fixed $i$: }  \d_i\langle T^i_{\ i}\rangle^{[1]}=0.
\end{equation}
Written in this form we notice that the only non zero stress tensor sources on $M_1$ is $\langle T_{\rho\psi}\rangle^{[0]}$. This is because in the full replica trace and conservation equations for unsquashed conical case i.e., \eqref{fptreqn}-\eqref{fpicomp}, the replica $q-$dependence via $U(\rho,a)$ is only associated with coefficient of $\langle T_{\rho\psi}\rangle_q$ and $\mathcal{A}[g_{\mu\nu}^{[q]}]$. We also notice that 
\begin{align}
   & \langle T^\rho_{\ \mu}\rangle{}^{[1]}=-U^{[1]}\langle T_{\rho\mu}\rangle{}^{[0]}+\langle T_{\rho\mu}\rangle{}^{[1]}; \quad\mu\in\{\rho,\psi\},\nonumber\\
   & \langle T^\psi_{\ \mu}\rangle{}^{[1]}=\frac{1}{\rho^2}\langle T_{\psi \mu}\rangle{}^{[1]},\nonumber \\
   & \langle T^i_{\ j}\rangle{}^{[1]}=g^{ik}\langle T_{kj}\rangle{}^{[1]},
\end{align}
therefore when both indices are lowered we also have an additional stress tensor source on $M_1$ $:\langle T_{\rho\rho}\rangle^{[0]}$.

\paragraph{Cosmic string literature:}\label{cosmicstringlit}The study of the vacuum polarisation is also important in condensed matter systems with conical defects. For example, in graphitic cones one has $q = 5/6, 2/3, 1/2, 1/3, 1/6$. All these types of cones have been experimentally observed \cite{Graphitic}.

The exact form of the proportionality constant (depending on $q$) depends on the specific theory. A similar conical singular structure as on an unsquashed $M_q$ arises in the presence of a cosmic string. The non zero vev on $M_q$ (including that of the energy momentum tensor) induced by this conical singularity has been well studied in the cosmic string literature.  

In $d=4$ the vev of the renormalised energy momentum tensor on $M_q$ for the massless scalar case has been considered in \cite{PhysRevD.35.3779, Moretti:1997dx}, the massive scalar case in \cite{Iellici:1997ud}, the massless and massive fermionic case in \cite{BezerradeMello:2012ajq} 
\begin{align}\label{scalar}
&\left\langle T^\mu_{\ \nu}\right\rangle{}^{scalar}_q\nonumber\\
&= \frac{1}{1440 \pi^2 \rho^4}\Bigg[\left(\frac{1}{q^4}-1\right) \operatorname{diag}(-3,1,1,1)+20(6 \xi-1)\left(\frac{1}{q^2}-1\right) \operatorname{diag}\left(\frac{3}{2},\frac{-1}{2},1,1\right) \nonumber \\
&\hspace{2.3cm}+15(m \rho)^2\left(\frac{1}{q}-1\right)\left(12 \xi-1-\frac{1}{q}\right) \operatorname{diag}(-1,1,0,0)\nonumber\\
&\hspace{2.3cm}-15(m \rho)^2\left(\frac{1}{q^2}-1\right) \operatorname{diag}(0,0,1,1)\Bigg],
\end{align}
where the basis is in order $(\psi,\rho,y^1,y^2)$ and $\xi$ is a parameter which fixes the coupling of the scalar field to gravity by means of the curvature $R$. We can map this result to solutions we have obtained via conservation and trace equations. In the above expression $\operatorname{diag}(-3,1,1,1)$ term corresponds to the massless CFT solutions \eqref{masslessCFTi}, \eqref{masslessCFTpsi}, \eqref{masslessCFTrho}. Note that for $\xi=1/6$ with $m=0$ we recover the CFT result; $\operatorname{diag}\left(\frac{3}{2},\frac{-1}{2},1,1\right)$ term corresponds to the massless scalar QFT solution \eqref{masslessQFT}; $\operatorname{diag}(0,0,1,1)$ term to the massive scalar solution \eqref{massive}; and $\operatorname{diag}(-1,1,0,0)$ term to the massive traceless scalar solution \eqref{massive trless}. Comparing \eqref{aindeprepstress} with the expression above we can read off the exact factors and $q$ dependence in $C_d(q), Q_d(q), M_{d}(q), N_{d}(q), R_{d}(q), S_{d}(q)$.

A more general expression for a CFT in $d=4$ is given by
\begin{equation}\label{4dimgenplrepstr}
    \left\langle T_\nu^\mu\right\rangle_q=\frac{f(q)}{1440 \pi^2 \rho^4} \operatorname{diag}(-3,1,1,1),
\end{equation}
where
\begin{equation}
    f(q)=\frac{1}{16 q^4}\left(1-q^2\right)\left[2\left(1+\frac{q^2}{3}\right) a+\left(1-q^2\right)(2 b-c)\right],
\end{equation}
with
\begin{align}
& a=12\left(n_s+6 n_f+12 n_v\right), \nonumber\\
& b=-4\left(n_s+\frac{11}{2} n_f+62 n_v\right), \nonumber\\
& c=-240 n_v .
\end{align}
 for a free field multiple containing $n_s$ conformal scalars, $n_f$ Dirac fermions and $n_v$ gauge vector fields.

 Comparing \eqref{aindeprepstress} with this we can read off
 \begin{equation}\label{Cd}
     C_d(q)=\frac{f(q)}{1440 \pi^2 }.
 \end{equation}
 
Expressions in $d-$dimensions for casimir effect on $M_q$ for free massless scalar field is considered in \cite{PhysRevD.36.3095}. In $d=4$ the vev of the renormalised energy momentum tensor on $M_q$ for the electromagnetic field (photon) is considered in \cite{Moretti:1996wd,PhysRevD.45.4486} and linearised gravitational field (graviton) in \cite{PhysRevD.45.4486}
\begin{align}\label{photon}
    \left\langle T^\mu_{\ \nu} \right\rangle{}^{photon}_q=\frac{\left(\frac{1}{q^2}-1\right)\left(\frac{1}{q^2}+11\right)}{720 \pi^2 \rho^4} \operatorname{diag}(-3,1,1,1),
\end{align}
\begin{align}\label{graviton}
\left\langle T^\mu_{\ \nu}  \right\rangle{}^{graviton}_q=\frac{\left(\frac{1}{q}-1\right)}{720 \pi^2 \rho^4} \operatorname{diag}[-3 g(q),g(q),f(q), f(q)], 
\end{align}
where
\begin{align}
    &f(q)=\frac{121}{q^3}+\frac{121}{q^2}+\frac{1421}{q}-1459,\nonumber\\
    &g(q)=\frac{121}{q^3}+\frac{121}{q^2}-\frac{829}{q}+251.
\end{align}
The photon case corresponds to the massless CFT solutions \eqref{masslessCFTi}, \eqref{masslessCFTpsi}, \eqref{masslessCFTrho}, and the graviton case corresponds to the massless QFT solution \eqref{masslessQFT}.

\paragraph{Conical $N$ point correlations in terms of flat space $(N+1)$ point correlations: } When $g_{\mu\nu}$ is flat and the entangling surface is planar, \cite{Rosenhaus:2014woa,Smolkin:2014hba} gives the $N-$point vacuum correlation function on $M_q$ in terms of specific $(N+1)$ point vacuum correlation functions on $M_1$ involving an additional Rindler (modular) Hamiltonian insertion (see \eqref{smolkin}). For a CFT the $O(q-1)$ result in \cite{Rosenhaus:2014woa}, \cite{Smolkin:2014hba} matches with the results we got in \eqref{aindeprepstress} expanded upto $O(q-1)$ with ($Q_d^{[1]}=M_{d}^{[1]}=N_{d}^{[1]}=R_{d}^{[1]}=S_{d}^{[1]}=0$). \cite{Rosenhaus:2014woa, Smolkin:2014hba} make use of the CFT stress tensor 2-point function $\langle T_{\mu\nu}T_{22}\rangle$ fixed by conformal symmetry \cite{Osborn:1993cr,Erdmenger:1996yc}. The non zero coefficient $C_d^{[1]}$ for $d=4$ can be read off from comparing with stress tensor one point function directly calculated on conical backgrounds given by \eqref{4dimgenplrepstr},
\begin{align}\label{Cd1}
    C_{d=4}^{[1]}=-\pi^2 \frac{C_T}{120},
\end{align}
where $C_T$ is the charge related to the $B-$type conformal anomaly. For a free field multiple containing $n_s$ conformal scalars, $n_f$ Dirac fermions and $n_v$ gauge vector fields we have
\begin{equation}\label{CT}
    C_T=\frac{1}{3\pi^4}(n_s + 6n_f + 12n_v).
\end{equation}

The modular Hamiltonian is known only for certain symmetric geometries, such as spherical \cite{Casini:2010kt} and planar regions in flat space, and is not known for generic entangling boundaries and backgrounds which limits its use in determining correlation functions on $M_q$. However, the replica conservation and trace equations on static backgrounds that we have established, determines $\langle T_{\mu\nu}\rangle_q$ for any state and entangling surface on static backgrounds. We have $(d+1)$ equations for $(d+1)$ unknowns -  $\langle T_{\rho\rho}\rangle_q, \langle T_{\psi\psi}\rangle_q, \langle T_{\rho\psi}\rangle_q, \langle T_{ii}\rangle_q $.

\subsection{Disconnected entangling boundaries with no extrinsic curvature}
\paragraph{Multiple Conical Singularities literature:} The results so far in this section hold for spaces with one conical singularity. But when we have an entangling region with a disconnected boundary such as the one we have (i.e., on static metric \eqref{statmetric}, $r=r_1$ and $r=r_2$ spanning $y^i$ on a $\tau$ constant slice are two disconnected boundaries), $M_q$ has two conical singularities, one at each boundary. If we have an entangling region with multiple disconnected boundaries then we have the manifold $M_q$ with multiple conical singularities, one at each boundary. We can always have a local replica metric and replica stress tensor around each conical singularity but we would like to have a global replica metric and replica stress tensor, since the terms that do not depend on the conical regulator $(a)$ can contribute in the entire entangling region and hence non locally. Finding a single chart that captures the global replica manifold and replica metric with multiple conical singularities might be difficult or not possible for a general curved background in $M_1$, but it is possible when the metric on $M_1$ is flat, which is the case we will be focussing on.   

Flat space with multiple conical singularities and correlators on them have been discussed in literature on spaces with multiple cosmic strings where there is a conical singularity at the location of each cosmic string \cite{Letelier:1987iu,PhysRevD.55.3903}.

Consider a planar entangling region $x=x_1$ spanning $z^i$ on a $\tau=\tau_0$ slice on flat metric \eqref{flatpl} on $M_1$. The replica manifold $M_q$ has one conical singularity at $x=x_1$, $\tau=\tau_0$ and the corresponding replica metric around the conical singularity is given by,
\begin{equation}
    ds_q^2{}_\text{around bdy}=d\rho^2+\rho^2 d\psi^2+\delta_{ij}dz^i dz^j; \quad \rho \in [0,\infty), \psi \in [0,2\pi q),
\end{equation}
where $x-x_1=\rho \cos(\psi)$ and $\tau-\tau_0=\rho\sin(\psi)$. With a change of coordinates $\rho=q r^q$ and $\psi=q \theta$ we can write the metric in the form
\begin{equation}
    ds_q^2{}_\text{around bdy}=e^{-4 V(r)}(dr^2+r^2 d\theta^2)+\delta_{ij}dz^i dz^j; \quad r \in [0,\infty), \psi \in [0,2\pi),
\end{equation}
where $V(r)=2\lambda \log(r)$ and $\lambda=(q-1)/4$. The advantage of doing this is that the $q$ dependence is now only in $\lambda$ and not in the range of any of the coordinates.

The generalization to replica metric with multiple (say $N$) conical singularities (one at each $\{x_i,\tau_i\}$, $i\in \{1,2,...,N\}$), corresponding to $M_1$ with flat metric \eqref{flatpl} is given by,
\begin{equation}
    ds_q^2{}_\text{around bdy}=e^{-4 \sum_{i=1}^N V_i(r_i)}(dr^2+r^2 d\theta^2)+\delta_{ij}dz^i dz^j; \quad r \in [0,\infty), \psi \in [0,2\pi),
\end{equation}
where $V_i=2\lambda_i \log(r_i)$, $r_i=\sqrt{(x-x_i)^2+(\tau-\tau_i)^2}$ is the distance from each conical sigularity, $\lambda_i=(q_i-1)/4$. In our case $q_i=q$ and $\tau_i=\tau_0$ $\forall \,i$.

For a CFT on $M_1$ we can now use the replica conservation and trace equations $\nabla_{\mu}^{(q)}\langle T^\mu_{\ \nu}\rangle_q=0$, $T^\mu_{\ \mu}=\mathcal{A}^{(q)}[g_{\mu\nu}^{(q)}]$. The homogenous solution in $\{r,\tau,y^i\}$ basis is given by
\begin{align}
    \left\langle T_{\ \nu}^{\mu}\right\rangle_q &= C\sum_{i=1}^N \frac{\lambda_i(q)}{r_i^4}\left(\begin{array}{cccc}
1-\frac{d(\tau-\tau_i)^2}{r_i^2} & \frac{d(x-x_i)(\tau-\tau_i)}{r_i^2} \\
\frac{d(x-x_i)(\tau-\tau_i)}{r_i^2} & (1-d)+\frac{d(\tau-\tau_i)^2}{r_i^2} 
\end{array}\right); \quad \{\mu,\nu\} \in \{r,\tau\},\nonumber\\
&= C\sum_{i=1}^N \frac{\lambda_i(q)}{r_i^4}\delta^\mu_{\ \nu}; \quad \{\mu,\nu\} \in \{i,j\},
\end{align}
where $C$ is a constant. In the case of $N$ conical singularities the net contribution is the sum of contribution from each conical singularity. 

\paragraph{Replica stress tensor vev:}
We will now consider the replica stress tensor vev in the case of two disconnected planar entangling boundaries on flat space. In the case of an entangling region with disconnected boundary $x=x_1$ and $x=x_2$ spanning $z^i{}'s$ on a constant time slice, we have two conical singularities at the two disconnected boundaries. We also have an additional dimensionful parameter $L=x_2-x_1$ on which the $\langle T^\mu_{\ \nu} \rangle_q$ can depend on. Since the boundaries $\Sigma$ has no extrinsic curvature, we still have $\d_\psi$ and $\d_i$ as killing vectors on $M_1$ and $M_q$. We will now solve \eqref{fppsiindep} taking these into account. The $i-$conservation equation implies 
\begin{equation}
    \langle T^i_{\ i}\rangle_q=\tilde g_1(\rho,a,L)= (d-2)\left(\frac{C_d(q)}{\rho^d} K(\rho, L)+\tilde g(\rho,a,L)\right),
\end{equation}
where $C_d(q)$ is a $O(q-1)$ function of $q$, and $K(\rho,L)$ is a dimensionless function capturing the $L$ dependence in the regulator independent part. Using \eqref{fprhotrpsiindep},we get
\begin{align}
      \langle T^\psi_{\ \psi}\rangle_q &=\frac{1}{\rho^2}\int d\rho \rho \d_\rho(\rho(\mathcal{A}^{(q)}-  \langle T^i_{\ i}\rangle_q))\nonumber\\
      &=\frac{-1}{\rho^2}\left(\frac{(d-2)C_d(q)}{\rho^{d-2}}K(\rho,L)-\int d\rho (d-2)\frac{C_d(q)}{\rho^{d-1}}K(\rho,L)\right)+\frac{1}{\rho^2}\int d\rho \rho \d_\rho (\rho \tilde f(\rho,a))\nonumber\\
      &=\frac{(1-d)C_d(q)}{\rho^{d}}K(\rho,L)+\frac{1}{\rho^2}\int d\rho \d_\rho K(\rho,L) \frac{C_d(q)}{\rho^{d-2}}+\tilde f-\frac{1}{\rho^2}\int d\rho \rho \tilde f,
\end{align}
where $\tilde f(\rho,a)=\mathcal{A}^{(q)}-(d-2)\tilde g(\rho,a)$.
The trace equation implies,
\begin{align}
  \langle T^\rho_{\ \rho}\rangle{}_q &=\mathcal{A}^{(q)}-\langle T^\psi_{\ \psi}\rangle_q-\langle T^i_{\ i}\rangle_q\nonumber\\
  &=\frac{C_d(q)}{\rho^{d}}K(\rho,L)-\frac{1}{\rho^2}\int d\rho \d_\rho K(\rho,L) \frac{C_d(q)}{\rho^{d-2}}+\frac{1}{\rho^2}\int d\rho \rho \tilde f.
\end{align}
The function $K(\rho,L)$ must satisfy the following properties,
\begin{enumerate}
    \item It is dimensionless since $\langle T^\mu_{\ \nu}\rangle_q$ has mass dimension $d$ which implies $K(\rho, L)$ can only depend on the ratio $\rho/L$.
    \item For $\rho>>L$, $K(\rho/L) \to 1$, since far away from the entangling region it must be indistinguishable from a single conical singularity. 
\end{enumerate}
Both these conditions imply $K(\rho,L)$ must be of the form,
\begin{equation}
    K(\rho,L)=\sum_{i\geq 0}A_i\left(\frac{L}{\rho}\right)^i;\quad i\in \mathbb{Z}.
\end{equation}
We have,
\begin{align}\label{planardisconnvev}
    \langle T^r_{\ r}\rangle_q\rvert_{a=0}&=\langle T^\rho_{\ \rho}\rangle_q \cos^2(\psi)+\langle T^\psi_{\ \psi}\rangle_q \sin^2(\psi)\nonumber\\
    &=\frac{C_d(q)}{\rho^{d}}K(\rho,L)[1-d\sin^2(\psi)]-\frac{\cos(2\psi)}{\rho^2}\int d\rho \d_\rho K(\rho,L) \frac{C_d(q)}{\rho^{d-2}}. 
\end{align}

\section{Entanglement entropy - vacuum states} \label{sec:eight}
We will now calculate the scaling of the R\'{e}nyi and EE for a vacuum state on a $d$ dimensional background, and an entangling boundary with no extrinsic curvature. We make use of the results for the replica stress tensor vev from the previous section. Using \eqref{sqftvar2}, the scaling of $\Sqft$ in this case is given by,
\begin{equation}
    -\alpha\Delta r \frac{\d S_{\text{$d$ dim}}}{\d r_1}=\underset{q\to1}{lim}\frac{1}{2}\int_{M_q}d^dx\sqrt{g}\left( \delta_{\Delta r} g_{rr} \langle T^{rr}\rangle^{[1]}+\delta_{\Delta r} g_{\tau\tau} \langle T^{\tau\tau}\rangle^{[1]}\right),
\end{equation}
where we have kept $r_2$ fixed and scaled $r_1$. We can instead scale $r_2$ and keep $r_1$ fixed simply by interchanging $r_1$ and $r_2$ in the expression above. We recover the scaling of the EE in the limit $q\to1$. Notice also that $\delta_{\Delta r} g_{r\tau}=\delta_{\Delta r} g_{ij}=0$ in static backgrounds with zero extrinsic curvature for $\Sigma$. 

On a flat background metric, we have translation symmetry along $r$ or $x$, therefore $\Sqft$ only depends on $\Delta r$ or $\Delta x$. Therefore it is enough to consider the scaling of $\Sqft$ wrt scaling the entire interval $\Delta r \to e^\alpha \Delta r$ or $\Delta x \to e^\alpha \Delta x$, as opposed to keeping one of the boundaries fixed and scaling the other. On scaling the metric, we have
\begin{align}
    \delta_{\Delta r}g_{\mu\nu}&=2\alpha g_{rr}; \quad \mu=\nu=r,\\
    &=0;\quad \text{otherwise}.
\end{align}
In flat space vacuum CFT state we have $\langle T_{\mu\nu}\rangle^{[0]}=0$, therefore
\begin{equation}  
\langle T^\mu_{\hspace{0.2cm}\nu}\rangle^{[1]}=g^{\mu\alpha}{}^{[1]}\langle T_{\alpha\nu}\rangle^{[0]}+g^{\mu\alpha}{}^{[0]}\langle T_{\alpha \nu}\rangle^{[1]}=g^{\mu\alpha}{}^{[0]}\langle T_{\alpha \nu}\rangle^{[1]},
\end{equation}
i.e., for the vacuum state we can raise and lower indices in $\langle T^\mu_{\hspace{0.2cm}\nu}\rangle^{[1]}$ simply by using the metric on the base manifold $M_1$,
therefore,
\begin{equation}\label{Sflatddimscaling}
    \Delta r \frac{\d S_{\text{$d$ dim}}}{\d \Delta r}=\underset{q\to1}{lim}\int_{M_q}d^dx\sqrt{g} \langle T^{r}_r\rangle^{[1]}.
\end{equation}
Notice that the scaling is now with respect to $\Delta r$. This is because in flat space $\delta_{\Delta r} g^{\mu\nu}$ is independent of whether we scale wrt $r_1$ or $r_2$ or $r_2-r_1=\Delta r$. This reflects the translational symmetry along $r$ in flat space.

As seen in the previous section, the replica stress tensor has regulator ($a$) dependent terms and regulator-independent terms, which needs to be treated appropriately when performing replica integrals \eqref{metstate}. We will separately calculate the contributions from both kind of terms. The complete result for scaling of R\'{e}nyi and EE is the sum of both contributions.

\subsection{Regulator independent contribution}
Let us now consider only the regulator ($a$) independent contribution to the replica stress tensor vev. These are non local and contribute over the entire entangling region. 

\paragraph{Approach 1 - As the sum of individual conical stress tensor:}
We first consider the regulator ($a$) independent contribution to the replica stress tensor vev given by \eqref{aindeprepstress}. In $\{r,\tau,y^i\}$ basis we have
\begin{align}
   \langle T^r_{r}\rangle^{[1]}_{\text{$a$=0}}=&\langle T^\rho_{\ \rho}\rangle{}^{[1]}_{\text{$a$=0}}\cos^2(\psi)+\langle T^\psi_{\ \psi}\rangle{}^{[1]}_{\text{$a$=0}}\sin^2(\psi)\nonumber\\
   =&\frac{1}{\rho^d}\left[C_d^{[1]}+\frac{Q_d^{[1]}}{1-d}\right][1-d\sin^2(\psi)]+\frac{m^2}{\rho^{(d-2)}}(M_d^{[1]}+N_d^{[1]})\left[\frac{1+(2-d)\sin^2(\psi)}{(3-d)}\right],\nonumber\\
    \langle T^\tau_{\tau}\rangle^{[1]}_{\text{$a$=0}}=&\langle T^\rho_{\ \rho}\rangle{}^{[1]}_{\text{$a$=0}}\sin^2(\psi)+\langle T^\psi_{\ \psi}\rangle{}^{[1]}_{\text{$a$=0}}\cos^2(\psi)\nonumber \\
     =&\frac{1}{\rho^d}\left[C_d^{[1]}+\frac{Q_d^{[1]}}{1-d}\right][(1-d)+d \sin^2(\psi)]+\frac{m^2}{\rho^{(d-2)}}(M_d^{[1]}+N_d^{[1]})\left[1+\frac{(d-2)}{(3-d)}\sin^2(\psi)\right],\nonumber\\
     \langle T^r_{\tau}\rangle^{[1]}_{\text{$a$=0}}=&\langle T_r^{\tau}\rangle^{[1]}_{\text{$a$=0}}=\left(\langle T^\rho_{\ \rho}\rangle{}^{[1]}_{\text{$a$=0}}-\langle T^\psi_{\ \psi}\rangle{}^{[1]}_{\text{$a$=0}}\right)\cos(\psi)\sin(\psi)\nonumber \\
     =&\left(\frac{d}{\rho^d}\left[C_d^{[1]}+\frac{Q_d^{[1]}}{1-d}\right]+\frac{(d-2) m^2}{(3-d)\rho^{(d-2)}}\left(M_d^{[1]}+N_d^{[1]}\right)\right)\cos(\psi)\sin(\psi),\nonumber\\
     \langle T^i_{j}\rangle^{[1]}_{a=0}=&\frac{\delta^i_{j}}{\rho^d}\left(C_{d}^{[1]}+R_{d}^{[1]}+(m\rho)^2\left[\frac{(d-4)M_d^{[1]}}{(3-d)(d-2)}+S_d^{[1]}\right]\right).
\end{align}
We have $C_d(q)$ given by \eqref{Cd}. From this we see that $C_d^{[1]}=\frac{-(n_s+6n_f+12n_v)}{360\pi^2}$. We can read off the other undetermined coefficients depending on $q$ by comparing our result \eqref{aindeprepstress} with the result in literature obtained by directly computing on conical manifolds for scalar case, given by \eqref{scalar}.

In the case of a disconnected boundary (at $r=r_1$ and $r=r_2$) we have two conical singularity, one at each boundary. As we have just seen in the case of multiple conical singularities, the net contribution to the $O(q-1)$ replica stress tensor is the sum of contribution from each conical singularity. We perform the integral restricted around the entangling region. In $\{r,\tau\}$ plane these are semi discs of radius $(r_2-r_1)/2$, with $r\geq r_1$ around $r_1$ and $r\leq r_2$ around $r_2$. Since the $O(q-1)$ replica stress tensor has identical contributions from each conical singularity and the net contribution is the sum of individual conical contribution, this integral is the same as integrating over a disc of radius $(r_2-r_1)/2$ around one of the conical singularities.

\begin{figure}[H]
\centering
\includegraphics[scale=0.7]{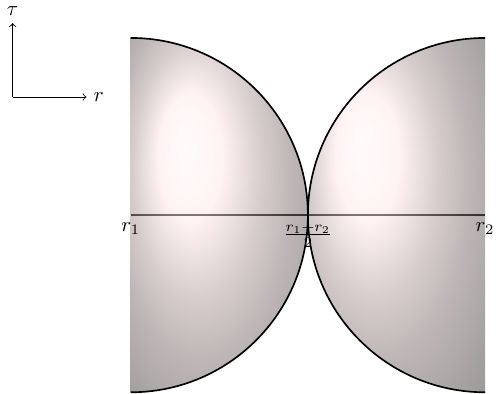}
\caption{Region of integration}
\label{Rofint1}
\end{figure}
  
Using \eqref{Sflatddimscaling}, the non local contribution to the entanglement entropy $S_{\text{$d$ dim, bulk}}$  (i.e, contribution depending on the entire entangling region and not just the entangling boundary) in $d-$dimensional flat space with an entangling region of size $\Delta r$ with a zero extrinsic curvature boundary is 
\begin{align}\label{Sflatddimscaling1}
    &\Delta r \frac{\d S_{\text{$d$ dim, bulk}}}{\d \Delta r}\Bigg\rvert_{\text{reg indep}}\nonumber\\
    =&\underset{q\to1}{lim}\int_{\Sigma}d^{d-2}x \sqrt{h}\int_{0}^{2\pi q}d\psi\int_{\epsilon}^{\frac{r_2-r_1}{2}}d\rho \rho \langle T^{r}_r\rangle_{a=0}^{[1]}\nonumber\\
    =&2\pi Vol(M_{d-2})\int_{\epsilon}^{\frac{r_2-r_1}{2}}d\rho \left(\frac{A_d}{\rho^{d-1}}+\frac{m^2 B_d }{\rho^{d-3}}\right)\nonumber\\
    =&2\pi A_2\log\left(\frac{\Delta r}{2\epsilon}\right)+\frac{\pi}{4} m^2 B_2(\Delta r)^2; \quad d=2,\nonumber\\
    =& 2\pi Vol(M_{d-2}) \left\{A_4\left(-\frac{2}{(\Delta r)^2}+\frac{1}{2\epsilon^2}\right)+m^2 B_4\log\left(\frac{\Delta r}{2\epsilon}\right)\right\}; \quad d=4,\nonumber\\
    =&2\pi Vol(M_{d-2}) \left\{\frac{A_d}{2-d}\left(\frac{2^{d-2}}{(\Delta r)^{d-2}}-\frac{1}{\epsilon^{d-2}}\right)+\frac{m^2 B_d}{4-d}\left(\frac{2^{d-4}}{(\Delta r)^{d-4}}-\frac{1}{\epsilon^{d-4}}\right)\right\}; \nonumber\\
     &\text{otherwise},
\end{align}
where $\Delta r=r_2-r_1$ and $Vol(M_{d-2})=\int_{\Sigma}d^{d-2}x \sqrt{h}$. In 3 dimensional flat space with $\Sigma$ spanning a finite length $2\pi\phi_0$ as in \eqref{3dimfinitelen}, $Vol(M_{d-2})=2\pi\phi_0$.

We have,
\begin{equation}
    A_d=\frac{(2-d)}{2}\left(C_d^{[1]}+\frac{Q_d^{[1]}}{(1-d)}\right), \quad B_d=\frac{(d-4)(M_d^{[1]}+N_d^{[1]})}{2(d-3)}.
\end{equation}

In $d=4$, we can read off the specific values of the coefficients from the cosmic string literature \eqref{cosmicstringlit}. For the scalar case from \eqref{scalar} we have we have 
\begin{equation}
    C_4^{[1]}=\frac{-1}{360 \pi^2}, \quad Q_4^{[1]}=\frac{1}{24 \pi^2}(1-6\xi), \quad M_4^{[1]}=\frac{-1}{48 \pi^2}(1-6\xi),\quad N_4^{[1]}=0.
\end{equation}
For photon from \eqref{photon}
\begin{equation}
    C_4^{[1]}=\frac{-1}{30 \pi^2}, \quad Q_4^{[1]}= M_4^{[1]}=N_4^{[1]}=0.
\end{equation}
For graviton from \eqref{graviton}
\begin{equation}
    Q_4^{[1]}=\frac{-7}{5 \pi^2}, \quad C_4^{[1]}= M_4^{[1]}=N_4^{[1]}=0.
\end{equation}

\paragraph{Approach 2 - Using replica stress tensor with explicit $L$ dependence:}Using the result for the replica stress tensor vev in the case of disconnected planar boundaries, given by \eqref{planardisconnvev}, the regulator dependent contribution to the scaling of the Entanglement entropy is given by,
\begin{align}
 &\Delta r \frac{\d S_{\text{$d$ dim, bulk}}}{\d \Delta r}\Bigg\rvert_{\text{reg indep}}=\underset{q\to1}{lim}\int_{M_q}d^dx\sqrt{g} \langle T^{r}_r\rangle^{[1]}\nonumber\\
    &=\underset{q\to1}{lim}\int_{\Sigma}d^{d-2}x \sqrt{h}\int_{0}^{2\pi q}d\psi\int_{\epsilon}^{\alpha L}d\rho \rho \Bigg(\frac{C_d^{[1]}}{\rho^{d}}K(\rho,L)[1-d\sin^2(\psi)]\nonumber\\
     &\hspace{6cm}-\frac{\cos(2\psi)}{\rho^2}\int d\rho \d_\rho K(\rho,L) \frac{C_d^{[1]}}{\rho^{d-2}}\Bigg)\nonumber\\
     &=(2-d)\pi  Vol(M_{d-2})\int_{\epsilon}^{\alpha L}d\rho \rho \left(\frac{C_d^{[1]}}{\rho^{d}}\left[K(\rho,L)=\sum_{i}A_i\left(\frac{L}{\rho}\right)^i\right]\right)\nonumber\\
    &=(2-d)\pi  Vol(M_{d-2})C_d^{[1]}\sum_{i\geq 0}\left(\frac{1}{L^{d-2}} \frac{A_i}{\alpha^{d-2+i}(2-d-i)}-\frac{A_i L^i}{\epsilon^{d-2+i}(2-d-i)}\right);\,i\in \mathbb{Z}\nonumber\\
    &=(2-d)\pi  Vol(M_{d-2})C_d^{[1]}\sum_{i\geq 0}\left(\frac{1}{L^{d-2}} \frac{B_i}{\alpha^{d-2+i}}-\frac{B_i}{\epsilon^{d-2}}\frac{L^i}{\epsilon^i}\right);\quad i\in \mathbb{Z},
\end{align}
where $B_i=A_i/(2-d-i)$. Notice that the scaling is now with respect to $\Delta r$. This is because in flat space $\delta_{\Delta r} g^{\mu\nu}$ is independent of whether we scale wrt $r_1$ or $r_2$ or $r_2-r_1=\Delta r$. This reflects the translational symmetry along $r$ in flat space. We have the finite part of EE given by,
\begin{align}
    S_{\text{$d$ dim, bulk}}\Big\rvert_{\text{reg $a$ indep, finite}}=\frac{\pi Vol(M_{d-2})C_d^{[1]}}{L^{d-2}}\sum_{i\geq 0}\left(\frac{B_i}{\alpha^{d-2+i}}\right).
\end{align}
By comparing the coefficient of the finite part with the expression we have in \cite{Landgren:2024ccz}, we find that
\begin{equation}
   \pi C_d^{[1]}\sum_{i\geq 0}\frac{B_i}{\alpha^{d-2+i}}=2^{d-3} \pi ^{\frac{d}{2}-1}\frac{l^{d-1}}{(2-d) G} \, _2F_1\left(\frac{1}{2},-\frac{d-2}{2 (d-1)};\frac{d}{2 (d-1)};1\right)  \left(\frac{\Gamma \left(\frac{1}{2 (d-1)}\right)}{\Gamma
   \left(\frac{d}{2 (d-1)}\right)}\right)^{2-d},
\end{equation}
where the ratio $l^{d-1}/G$ is related to the central charge $c$ in the case of even $d$ and to the number of degrees of freedom in the case of odd $d$.

\paragraph{Approach 3:} When there are two disconnected boundaries, the net contribution to the replica stress tensor is the sum of contribution from each boundary with a conical singularity. In general the coordinates are well defined in only a small local patch around each conical singularity and we need multiple charts to cover the entire manifold in the case of two conical singularities. If $\{\rho_1,\psi_1, y^i\}$ and $\{\rho_2,\psi_2,y^i\}$ are the two local coordinates around each boundary conical singularity, we have 
\begin{align}
    \langle T^r_{\,r}\rangle{}_{a=0}=C_{d}(q)\left(\frac{1}{\rho_1^d}(1-d\sin^2\psi_1)+\frac{1}{\rho_2^d}(1-d\sin^2\psi_2)\right).
\end{align}
We will now consider a new set of coordinates $\{\rho,\psi\}$ with $\rho=0$ at $r=(r_1+r_2)/2$ and $\psi=0$ along the axis running from $r_1$ to $r_2$, we have
\begin{align}
    &\rho_1=\sqrt{\frac{L^2}{4}+L \rho  \cos (\psi )+\rho ^2},\nonumber\\ 
    &\rho_2=\sqrt{\frac{L^2}{4}-L \rho  \cos (\psi )+\rho ^2},\nonumber\\
    &\psi_1=\pm\cos ^{-1}\left(\frac{L+2 \rho  \cos (\psi )}{2 \sqrt{\frac{L^2}{4}+L \rho  \cos (\psi )+\rho ^2}}\right)+2 \pi  q,\nonumber\\
    &\psi_2= \pm\cos ^{-1}\left(\frac{2 \rho  \cos (\psi )-L}{2 \sqrt{\frac{L^2}{4}-L \rho  \cos (\psi )+\rho ^2}}\right)+2 \pi q; \quad q\in \mathbb{Z}.
\end{align}
We can now calculate the EE with the integral restricted to the entangling region $\rho \in [0,\frac{L=(r_2-r_1)}{2}]$ and $\psi \in [0,2\pi q)$. This again gives
\begin{align}
    S_{\text{$d$ dim, bulk}}\Big\rvert_{\text{reg $a$ indep, finite}}\propto \frac{1}{L^{d-2}}.
\end{align}
\begin{figure}[H]
\centering
\includegraphics[scale=0.7]{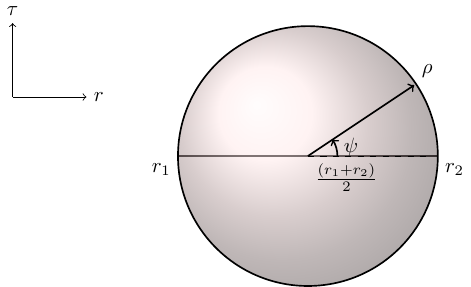}
\caption{Region of integration}
\label{Rofint}
\end{figure}

For the regulator-independent replica stress tensor, when there are two disconnected boundaries, it is not very clear what the right region of integration is, although all the approaches we have considered give the same $\Delta r$ or $L$ dependence. However, for the regulator-dependent replica stress tensor parts, the replica corrections vanish beyond the radius set by the regulator and their integral only picks up the local contribution around each boundary.

\subsection{Regulator dependent contribution}
Let us now consider only the regulator ($a$) dependent contribution to the replica stress tensor vev. The $O(q-1)$ terms in these, contribute locally around each conical singularity to the replica integrals. These are given by the regulator $a$ dependent parts of the solutions \eqref{masslessCFTi}, \eqref{masslessCFTpsi}, \eqref{masslessCFTrho},
\begin{align}
   &\langle T^i_{\ j}\rangle-\langle T^i_{\ j}\rangle\rvert_{a=0}=g(\rho,a)\delta^i_{\ j},\nonumber\\
   &\langle T^\psi_{\ \psi}\rangle{}-\langle T^\psi_{\ \psi}\rangle{}\rvert_{a=0}=f-\frac{1}{\rho^2}\int d\rho \rho f,\nonumber\\
   &\langle T^\rho_{\ \rho}\rangle{}-\langle T^\rho_{\ \rho}\rangle{}\rvert_{a=0}=\frac{1}{\rho^2}\int d\rho \rho f,\nonumber\\
    &\langle T_{\rho\psi}\rangle=0=t_{\rho\psi},
\end{align}
where $f(\rho,a)=\mathcal{A}^{(q)}-g(\rho,a)$. In 2 dimensions we have $g(\rho,a)=0$. In $d>2$ for a planar entangling boundary, we can always choose $g(\rho,a)=0$ (since the manifold is $C_2 \times \Sigma$ and $g_{ij}^{(q)}$ has no $a$ dependence). Using the coordinate transformation \eqref{conversion} we have, 
\begin{align}\label{regdepstress}
\langle T^r_r\rangle{}^{[1]}-\langle T^r_{r}\rangle{}^{[1]}\rvert_{a=0}&=\frac{\cos(2\psi)}{\rho^2}\int d\rho \rho \mathcal{A}^{[1]}+\mathcal{A}^{[1]}\sin^2(\psi).
\end{align}

\subsubsection{2 dimensional CFT}
Let us consider a CFT on a 2-dimensional static metric. 
 \begin{equation}\label{2dimmet}
     ds^2_{M_1}=g_{\tau\tau}(r)d\tau^2 +g_{rr}(r)dr^2,
 \end{equation}
where $r \in [0,\infty)$, $\tau \in (-\infty, \infty)$. We consider the entangling region $r \in [r_1,r_2]$ at $\tau=\tau_0$. 

The replica metric in a local region around the endpoint $r=r_0 \,(r_1\text{ or }r_2)$ is given by
\begin{equation}\label{repmet2dim}
    ds^2_{M_q}\text{ around endpoint} = U(\rho,a)d\rho^2+\rho^2 d\psi^2,
\end{equation}
where $\rho \in [0,\infty)$ and $\psi \in [0,2\pi q)$. We have $(r-r_0)=\rho \cos\psi$, $(\tau-\tau_0)=\rho\sin\psi$, and the conical regulating function $U(\rho,a)\rvert_{\rho \to 0} = q^2$, $U(\rho,a)\rvert_{\rho>>a} = 1$. The non trivial contribution in the difference of replica and base quantities is from locally around the conical singularity.

In 2-dimensions the geometric part of the stress tensor is given by 
\begin{equation}
    \chi_{\mu\nu}[g_{\mu\nu}]=\left(n=-\frac{c}{48\pi}\right)Rg_{\mu\nu},
\end{equation}
where $c$ is the central charge. 

Therefore we have (see section \eqref{repgeomstress}),
\begin{align}
    \chi_{\mu\nu}^{(q)}[g_{\mu\nu}^{(q)}]&=-\frac{c}{48\pi}R^{(q)}g_{\mu\nu}^{(q)}.
\end{align}
From the replica metric around the entangling boundary given by \eqref{repmet2dim}, we have
\begin{align}\label{rep2dimgeom}
    R^{(q)}-R&=\frac{\d_{\rho}U}{\rho U^2},\nonumber\\
    \mathcal{A}^{(q)}- \mathcal{A}&=2nR^{(q)}=2n\frac{\d_{\rho}U}{\rho U^2}=\left(\mathcal{A}^{[1]}=2n \frac{\d_{\rho}U^{[1]}}{\rho }\right)(q-1)+O((q-1)^2),\nonumber\\
    \chi_{\rho\rho}^{(q)}- \chi_{\rho\rho}&= n \frac{\d_{\rho}U}{\rho U} = \left( \chi_{\rho\rho}^{[1]}=n \frac{\d_{\rho}U^{[1]}}{\rho }\right)(q-1)+O((q-1)^2),\nonumber\\
    \chi_{\psi\psi}^{(q)}-\chi_{\psi\psi}&=n \frac{\rho \d_{\rho}U}{U^2 } = \left( \chi_{\psi\psi}^{[1]}=n \rho \d_{\rho}U^{[1]}\right)(q-1)+O((q-1)^2),\nonumber\\
    \chi_{\rho\psi}^{(q)}-\chi_{\rho\psi}&=0.
\end{align}
We have skipped the conformal factor $e^\sigma$ in the replica metric which captures $O(q-1)^0$ terms from $M_1$ as well. This is fine since we are only interested in the $O(q-1)$ terms. Notice $\chi^\rho_{\rho}{}^{[1]}=\chi^\psi_{\psi}{}^{[1]}$.

\paragraph{2 dimensional CFT vacuum state on a flat background:}We will focus on the vacuum state on a flat background. On a flat background we have $\chi_{\mu\nu}^{[0]}=0$ but $\chi_{\mu\nu}^{(q)}\neq0$. 
In a  $d=2$ CFT, we now consider the regulator dependent part of the $O(q-1)$ replica stress tensor sourced by the trace anomaly $\mathcal{A}^{(q)}[g_{\mu\nu}^{(q)}]$ on $M_q$. We have
\begin{align}\label{2dimTrs1}
   &\langle T^\psi_{\ \psi}\rangle{}^{[1]}-\langle T^\psi_{\ \psi}\rangle{}^{[1]}\rvert_{a=0}=\mathcal{A}^{[1]}-\frac{1}{\rho^2}\int d\rho \rho \mathcal{A}^{[1]}=\frac{2n}{\rho^2}\left(\rho\d_\rho U^{[1]}-U^{[1]}\right),\nonumber\\
   &\langle T^\rho_{\ \rho}\rangle{}^{[1]}-\langle T^\rho_{\ \rho}\rangle{}^{[1]}\rvert_{a=0}=\frac{1}{\rho^2}\int d\rho \rho \mathcal{A}^{[1]}=\frac{2n  U^{[1]}}{\rho^2},\nonumber\\
    &\langle T_{\rho\psi}\rangle=0=t_{\rho\psi},
\end{align}
and
\begin{align}\label{2dimregdepstress}
\langle T^r_r\rangle{}^{[1]}-\langle T^r_{r}\rangle{}^{[1]}\rvert_{a=0}&=\frac{\cos(2\psi)}{\rho^2}\int d\rho \rho \mathcal{A}^{[1]}+\mathcal{A}^{[1]}\sin^2(\psi)\nonumber\\
&=\frac{2n }{\rho^2}( U^{[1]}\cos(2\psi)+\rho\d_\rho U^{[1]}\sin^2(\psi)).
\end{align}
Around each boundary/endpoint of the entangling region, the local representation of $F_{\mu\nu}^{[1]}$ (the superscript is used to indicate that it is an $O(q-1)$ coefficient) is given by (see \eqref{localrep})
\begin{equation}
    \underset{q\to1}{lim}\,\,\frac{\delta(\rho)}{2\pi q}\int_{M_q} 
 (\sqrt{U} \text{ or } 1)\rho\, d\rho\, d\psi\, F_{\mu\nu}^{[1]}.
\end{equation}
The choice of $\sqrt{g_q}$ or $\sqrt{g}$ in the volume element gives same result in the limit $q \to 1$.

The local representation of $\langle T^r_r\rangle{}^{[1]}-\langle T^r_{r}\rangle{}^{[1]}\rvert_{a=0}$ around each boundary of the entangling region is,
\begin{equation}
    \langle T^r_r\rangle{}^{[1]}-\langle T^r_{r}\rangle{}^{[1]}\rvert_{a=0}=-2n\delta(\rho).
\end{equation}
We have,
\begin{align}\label{2dimregdep}
  \Delta r \frac{\d S_{\text{CFT}_2}}{\d \Delta r}\Bigg\rvert_{\text{reg dep}}=& \underset{q\to1}{lim}\left(\int_{M_q}d^2x\sqrt{g}  \langle T^r_r\rangle{}^{[1]}-\langle T^r_{r}\rangle{}^{[1]}\rvert_{a=0}\right)\nonumber\\
    =&\underset{q\to1}{lim}\Bigg[\int_{\rho=0,\psi=0}^{\rho=\infty,\psi=2\pi q} \rho d\rho d\psi \Big(-2n(\delta(\rho)\rvert_{\text{around }r_1}+\delta(\rho)\rvert_{\text{around }r_2})\Big)\Bigg]\nonumber\\
    =&-8\pi n=\frac{c}{6}.
\end{align}
Putting together this and the regulator independent replica  stress tensor contribution to EE in 2 dim CFT given by \eqref{Sflatddimscaling1}, we have
\begin{equation}
      S_{\text{CFT}_2}=\frac{c}{6}\log\left(\frac{\Delta r}{\epsilon_{\text{UV}}}\right).
\end{equation}
This is the well known result for EE of vacuum on a 2 dimensional flat background.

Alternatively we can also look at the decomposition of the $O(q-1)$ replica corrections into state and geometric parts. We also have the state part of the replica stress tensor corrections $t_{\mu\nu}^{[1]}$ given by,
\begin{align}
    t^\rho_{\ \rho}{}^{[1]}&= \langle T^\rho_{\ \rho}\rangle{}^{[1]}-\chi^\rho_{\ \rho}{}^{[1]}=- t^\psi_{\ \psi}{}^{[1]}=\frac{C_{2}^{[1]}}{\rho^2}+n\left(\frac{2U^{[1]}}{\rho^2}-\frac{\d_\rho U^{[1]}}{\rho}\right).
\end{align}

Notice that the state part of the replica stress tensor correction $t_{\mu\nu}^{[1]}$ also has metric dependent terms. 

Using the coordinate transformation \eqref{conversion} we have, 
\begin{align}
   \chi_{rr}^{[1]}&=\chi_{\tau\tau}^{[1]}=n \frac{\d_\rho U^{[1]}}{\rho},\\
  \chi_{r\tau}^{[1]}&=0,\\
\mathcal{A}^{[1]}&=2n \frac{\d_\rho U^{[1]}}{\rho}.
\end{align}

Similarly,
\begin{align}
    t^r_{r}{}^{[1]}&=-t^\tau_{\tau}{}^{[1]}= t^\rho_{\ \rho}{}^{[1]}\cos(2\psi),\nonumber\\
   t^r_{\tau}{}^{[1]}&=t^\tau_{r}{}^{[1]}=t^\rho_{\ \rho}{}^{[1]}\sin(2\psi).
\end{align}
We have $\langle T^\mu_{\nu}\rangle{}^{[1]}=t^\mu_{\nu}{}^{[1]}+\chi^\mu_{\nu}{}^{[1]}$.

The local representation of $\chi_{\mu\nu}^{[1]}$ around each boundary of the entangling region is,
\begin{align}
   \chi_{rr}^{[1]}&=\chi_{\tau\tau}^{[1]}=-2n \delta(\rho),\\
    \chi_{r\tau}^{[1]}&=0,\\
    \mathcal{A}^{[1]}&=-4n \delta(\rho).
\end{align}
Similarly the local representation of $t_{\mu\nu}^{[1]}$ around each boundary of the entangling region is, 
\begin{align}\label{localrep2dimstate}    t^r_{r}{}^{[1]}&=t^\tau_{\tau}{}^{[1]}=t^r_{\tau}{}^{[1]}=0.
\end{align}
In flat space vacuum CFT state we have $\chi_{\mu\nu}^{[0]}=t_{\mu\nu}^{[0]}=0$, therefore
\begin{equation}  \chi^\mu_\nu{}^{[1]}=g^{\mu\alpha}{}^{[1]}\chi_{\alpha\nu}^{[0]}+g^{\mu\alpha}{}^{[0]}\chi_{\alpha\nu}^{[1]}=g^{\mu\alpha}{}^{[0]}\chi_{\alpha\nu}^{[1]},\nonumber\\
t^\mu_\nu{}^{[1]}=g^{\mu\alpha}{}^{[1]}t_{\alpha\nu}^{[0]}+g^{\mu\alpha}{}^{[0]}t_{\alpha\nu}^{[1]}=g^{\mu\alpha}{}^{[0]}t_{\alpha\nu}^{[1]}.
\end{equation}
We have,
\begin{align}
  \Delta r \frac{\d S_{\text{CFT}_2}}{\d \Delta r}=& \underset{q\to1}{lim}\left(\int_{M_q}d^2x\sqrt{g} \chi^r_{r}{}^{[1]}\right)\nonumber\\
    =&\underset{q\to1}{lim}\Bigg[\int_{\rho=0,\psi=0}^{\rho=\infty,\psi=2\pi q} \rho d\rho d\psi \Big(-2n(\delta(\rho)\rvert_{\text{around }r_1}+\delta(\rho)\rvert_{\text{around }r_2})\Big)\Bigg]\\
    =&-8\pi n=\frac{c}{6}.
\end{align}
which implies
\begin{equation}
      S_{\text{CFT}_2}=\frac{c}{6}\log\left(\frac{\Delta r}{\epsilon_{\text{UV}}}\right).
\end{equation}

\paragraph{Massive scalar field on a flat background:}In 2 dimensional QFTs on flat space, in the presence of mass $m$ the regulator-independent part of the replica stress tensor \eqref{aindeprepstress} has an additional $m$ dependent term which additionally contributes to the EE as can be seen in \eqref{Sflatddimscaling1},
\begin{equation}
     \Delta r \frac{\d S_{\text{CFT}_2}}{\d \Delta r}=\frac{c}{6}+\frac{\pi}{4} m^2 (M_2^{[1]}+N_2^{[1]})(\Delta r)^2,
\end{equation}
which implies
\begin{equation}
      S_{\text{CFT}_2}=\frac{c}{6}\log\left(\frac{\Delta r}{\epsilon_{\text{UV}}}\right)+\frac{\pi}{8} m^2 (M_2^{[1]}+N_2^{[1]})(\Delta r)^2.
\end{equation}
Similarly for the fermionic field one has to use the additional contributions in the regulator-independent part of the replica stress tensor given in \cite{BezerradeMello:2012ajq}.

\paragraph{2 dimensional CFT on a general background:} In case of a general metric, we have  
\begin{align}
    &-\alpha \Delta r \frac{\d S_{\text{CFT}_2}}{\d r_1}=\underset{q\to1}{lim}\frac{1}{2}\int_{M_q}d^dx\sqrt{g}\left( \delta_{\Delta r} g^{rr} \langle T_{rr}\rangle^{[1]}+\delta_{\Delta r} g^{\tau\tau} \langle T_{\tau\tau}\rangle^{[1]}\right)\nonumber\\
    =&\underset{q\to1}{lim}\frac{1}{2}\int_{M_q}d^dx\sqrt{g}\left(\left( \delta_{\Delta r} g^{rr} \chi_{rr}^{[1]}+\delta_{\Delta r} g^{\tau\tau} \chi_{\tau\tau}^{[1]}\right)+\left( \delta_{\Delta r} g^{rr} t_{rr}^{[1]}+\delta_{\Delta r} g^{\tau\tau} t_{\tau\tau}^{[1]}\right)\right)\nonumber\\
    =&\underset{q\to1}{lim}\frac{1}{2}\int_{M_q}\rho d\rho d\psi\alpha\Big(2 g^{rr}+(r-r_2)\d_rg^{rr} +(r-r_2)\d_r g^{\tau\tau} \Big)\Big(-2n(\delta(\rho)\rvert_{r_1}+\delta(\rho)\rvert_{r_2})\Big)\nonumber\\
    &+\underset{q\to1}{lim}\frac{1}{2}\int_{M_q}d^dx\sqrt{g}\left(\delta_{\Delta r} g^{rr} t_{rr}^{[1]}+\delta_{\Delta r} g^{\tau\tau} t_{\tau\tau}^{[1]}\right)\nonumber\\
    =&-2\pi n\left\{ 2 (g^{rr}(r_1)+g^{rr}(r_2))+(r_1-r_2)\d_rg^{rr}\rvert_{r_1} +(r_1-r_2)\d_r g^{\tau\tau}\rvert_{r_1} \right\}\nonumber\\
    &+\underset{q\to1}{lim}\frac{1}{2}\int_{M_q}d^dx\sqrt{g}\left( \delta_{\Delta r} g^{rr} t_{rr}^{[1]}+\delta_{\Delta r} g^{\tau\tau} t_{\tau\tau}^{[1]}\right).
\end{align}
For the state dependent part one has to solve \eqref{fptreqn}-\eqref{fpicomp}, where we have a non zero source $\chi_{\mu\nu}^{[0]}$ and in the case of general state $t_{\mu\nu}^{[0]}$. Notice that the scaling of EE above is not wrt $\Delta r$, but wrt the boundary coordinate $r_1$. This reflects the lack of translation symmetry along $r$, on a general metric. 

\subsubsection{Even dimensional CFT\label{evendimplanar}} In odd dimensional CFTs since the trace anomaly $\mathcal{A}[g_{\mu\nu}]=0=\mathcal{A}^{(q)}[g_{\mu\nu}^{(q)}]$ therefore there are no regulator dependent contributions to EE. The scaling of the EE is completely given by \eqref{Sflatddimscaling1}. 

In even dimensional CFTs the trace anomaly on $M_1$ is non zero and is given by \eqref{evendanom}. The anomaly on the replica manifold $\mathcal{A}^{(q)}[g_{\mu\nu}^{(q)}]$ is given by the same expression but with $g_{\mu\nu}$ replaced by $g_{\mu\nu}^{(q)}$,
\begin{align}
  \mathcal{A}^{(d,q)}[g_{\mu\nu}^{(q)}]=\left\langle \chi_{\ \mu}^{ \mu}\right\rangle_{d,q}=\frac{1}{(4 \pi)^{d / 2}}\left(\sum_j c_{d j} I_j^{(d,q)}-(-)^{\frac{d}{2}} a_d E^{(d,q)}+\sum_j d_{d j} D^i J_i^{(d,q)}\right).
\end{align}
In general, the replica metric in this case locally around the entangling boundary is given by \eqref{repmet1},
\begin{equation}
    ds_q^2\text{ expanded around boundary}\textcolor{blue}{, reg}=g^{(q)}_{\mu \nu}dx^\mu dx^\nu|_{bdy} =\textcolor{blue}{U(\rho,a)} d\rho^2 + \rho^2 d\psi^2 +\delta_{ij}dz^i dz^j.
\end{equation}
We notice that the replica metric has no $\psi$ dependence and hence the replica anomaly in any even dimensional CFT, for this replica metric, is also independent of $\psi$. From \eqref{regdepstress}, we have
\begin{align}
\langle T^r_r\rangle{}^{[1]}-\langle T^r_{r}\rangle{}^{[1]}\rvert_{a=0}&=\frac{\cos(2\psi)}{\rho^2}\int d\rho \rho \mathcal{A}^{[1]}+\mathcal{A}^{[1]}\sin^2(\psi).
\end{align}
Using \eqref{localrep}, the local representation of $\langle T^r_r\rangle{}^{[1]}-\langle T^r_{r}\rangle{}^{[1]}\rvert_{a=0}$ around each boundary of the entangling region is,
\begin{equation}
    \langle T^r_r\rangle{}^{[1]}-\langle T^r_{r}\rangle{}^{[1]}\rvert_{a=0}=\frac{\delta_{\Sigma}(\rho)\pi q}{Vol(M_{d-2})}\int_{M_q} d\rho d^{d-2}x\sqrt{h}\rho\mathcal{A}^{[1]}=N \delta_{\Sigma}(\rho).
\end{equation}
We have,
\begin{align}\label{regdepevendimcontr}
  \Delta r \frac{\d S_{\text{CFT}_d}}{\d \Delta r}\Bigg\rvert_{\text{reg dep}}=& \underset{q\to1}{lim}\left(\int_{M_q}d^dx\sqrt{g}  \left(\langle T^r_r\rangle{}^{[1]}-\langle T^r_{r}\rangle{}^{[1]}\rvert_{a=0}\right)\right)\nonumber\\
    =&\underset{q\to1}{lim}\Bigg[N\int_{M_q} \rho d\rho d\psi d^{d-2}\sqrt{h}\Big(\delta_{\Sigma}(\rho)\rvert_{\text{around }r_1}+\delta_{\Sigma}(\rho)\rvert_{\text{around }r_2}\Big)\Bigg]\nonumber\\
    =&2\pi \int_{M_q} d\rho d^{d-2}x\sqrt{h}\rho\mathcal{A}^{[1]}=2\pi Vol(M_{d-2})\int_{0}^\infty d\rho \rho\mathcal{A}^{[1]},
\end{align}
where in the last equality we have made use of the fact that we are dealing with a planar entangling region and hence $g_{\mu\nu}^{[1]}$ and $\mathcal{A}^{[1]}$ are independent of $y^i{}'s$.
\paragraph{2 dimensions:}See \eqref{2dimregdep}
\begin{align}\label{2dimregdep1}
  \mathcal{A}^{[1]}=-\frac{c}{24} R^{[1]}\implies \Delta r \frac{\d S_{\text{CFT}_2}}{\d \Delta r}\Bigg\rvert_{\text{reg dep}}=\frac{c}{6}.
\end{align}
\paragraph{4 dimensions: }
\begin{equation}
    \mathcal{A}^{(q)}[g_{\mu\nu}^{(q)}]=\chi_{\ \mu}^\mu[g_{\mu\nu}]^{d=4}=-\frac{a}{(4 \pi)^2} E^{(q)}_4+\frac{c}{(4 \pi)^2} I^{(q)}_4,
\end{equation}
where $I^{(q)}_4$ is the term involving weyl contraction $C^{(q)}_{\mu \nu \rho \sigma} C^{(q)}{}^{\mu \nu \rho \sigma}$.
\begin{align}
E_4^{(q)}&=R^{(q)}_{\alpha\beta\mu\nu}R^{\alpha\beta\mu\nu}{}^{(q)}-4R^{(q)}_{\mu\nu}R^{\mu\nu}{}^{(q)}+R^{(q)}{}^2,  \nonumber\\ I_4^{(q)}&=R^{(q)}_{\alpha\beta\mu\nu}R^{\alpha\beta\mu\nu}{}^{(q)}-2 R^{(q)}_{\mu\nu}R^{\mu\nu}{}^{(q)}+\frac{1}{3}R^{(q)}{}^2.
\end{align}
We make use of known integrals of Riemann contractions on replica metric i.e., \eqref{riemcontrsqcone}, \eqref{intEI} to get, 
\begin{align}\label{4dimlogterm}
  \Delta r \frac{\d S_{\text{CFT}_d}}{\d \Delta r}\Bigg\rvert_{\text{reg dep}}=&2\pi \int_{M_q} d\rho d^{d-2}x\sqrt{h}\rho\mathcal{A}^{[1]}\nonumber\\
  =&(a I^\Sigma_{a}-c I^\Sigma_{c}). 
\end{align}
 This is the universal logarithmic contribution to $\Sqft$. For a planar entangling region on flat space the Ricci scalar of the background, the extrinsic curvature of $\Sigma$ and projection of Ricci and Riemann tensors on $\Sigma$ is zero, therefore \eqref{4dimlogterm} is zero.

\paragraph{5 and 6 dimensions:}We can similarly compute the scaling of regulator-dependent contribution to EE in higher dimensions by using \eqref{regdepevendimcontr}. The replica trace anomaly is given by \eqref{evendanom} with $g_{\mu\nu}$ replaced by $g_{\mu\nu}^{(q)}$. The relevant replica integrals in 5 and 6 dimensions are listed in \cite{Fursaev:2013fta}.
\bigskip

The replica anomaly is independent of the choice of the CFT state. Therefore, the results in this section hold for any CFT state.

\paragraph{General deformation:}So far we have considered a particular kind of scaling i.e., we scale the entangling region along its finite length direction. Consider now \eqref{gensqftvar2} where we consider a general deformation of the background metric and the entangling surface geometry captured by $\delta_L g_{\mu\nu}$. For the CFT replica stress tensor given by \eqref{aindeprepstress} (i.e., all other coefficients except $C_d(q)$ is zero), it was shown in \cite{Rosenhaus:2014woa} that two terms in \eqref{general fol} contribute to logarithmic terms in $\delta_L \Sqft$ in 4-dimensions,
\begin{equation}
    \delta_L g_{ij}=x^a x^c R_{iacj}; \quad \delta_L g_{ab}=\frac{-1}{3}x^c x^d R_{acbd}.
\end{equation}
We have,
\begin{align}
    \delta_L \Sqft=\frac{c}{6\pi}\int d^2y(\delta^{ac}\delta^{bd}R_{abcd}+\delta^{ij}\delta^{ac}R_{iacj})\log(l/\epsilon).
\end{align}
here $l$ is the characteristic length scale of perturbations and $\epsilon$ is a UV cutoff. We have made use of \eqref{Cd1}. \eqref{CT} can be written as $C_T=40c/\pi^4$ where $c$ is the CFT central charge corresponding to the Weyl tensor contractions in the 4-dimensional trace anomaly. This should be compared with Solodukhin’s formula \cite{Solodukhin:2008dh} for the universal part of entanglement entropy in the case of a 4-dimensional CFT.

\subsection{General QFT}
The replica stress tensor vev $\langle T_{\mu\nu}\rangle^{[1]}$ for a planar entangling boundary in a general unitary $d$ dimensional field theory (free or interacting) was given in \cite{Smolkin:2014hba}
\begin{align}
 \left\langle T_{i j}\left(x_1, x_2, y\right) \right\rangle^{[1]}=&-\left\langle T_{i j}\left(x_1, x_2, y\right) K_0\right\rangle\nonumber\\
=&\frac{A_d \delta_{i j}}{(d-1)^2 \Gamma(d)} \int_0^{\infty} d \mu\left(c^{(0)}(\mu)-(d-1) c^{(2)}(\mu)\right) \mu^2 K_0(\mu \rho), \\
 \left\langle T_{ab}\left(x_1, x_2, y\right) \right\rangle^{[1]}=&-\left\langle T_{a b}\left(x_1, x_2, y\right) K_0\right\rangle\nonumber\\
 =&\frac{A_d}{(d-1)^2 \Gamma(d)} \int_0^{\infty} d \mu\left(c^{(0)}(\mu)+(d-1)(d-2) c^{(2)}(\mu)\right)\nonumber \\
 &\left(\mu^2 K_0(\mu \rho)\left(\delta_{a b}-\frac{x_a x_b}{\rho^2}\right)+\frac{\mu K_1(\mu \rho)}{\rho}\left(\delta_{a b}-\frac{2 x_a x_b}{\rho^2}\right)\right) .
\end{align}
where $x_a$ with $a=1,2$ are transverse to the entangling boundary and $y^i{}'s$ are along the entangling boundary. $\mu$ is the spectral parameter and $c^{(0)}(\mu)$ and $c^{(2)}(\mu)$ are spectral functions that encode the information of the exact QFT, and $A_d=\frac{\pi^{d/2}}{(d+1)2^{d-2}\Gamma(d/2)}$.
\bigskip

One could easily repeat the calculation for EE scaling that we just did for a CFT, for a $d-$dimensional field theory using the expressions above. We have,
\begin{align}\label{generalQFTEEscaling}
    &\Delta r \frac{\d S_{\text{$d$ dim, bulk}}}{\d \Delta r}\Bigg\rvert_{\text{reg indep}}\nonumber\\
    &=\underset{q\to1}{lim}\int_{\Sigma}d^{d-2}x \sqrt{h}\int_{0}^{2\pi q}d\psi\int_{\epsilon}^{\frac{r_2-r_1}{2}}d\rho \rho \langle T^{r}_r\rangle_{a=0}^{[1]}\nonumber\\
    &=\underset{q\to1}{lim}Vol(M_{d-2})\int_{0}^{2\pi q}d\psi\int_{\epsilon}^{\frac{r_2-r_1}{2}}d\rho \Bigg(\frac{A_d}{(d-1)^2 \Gamma(d)} \int_0^{\infty} d \mu C(\mu) \nonumber\\
& \left(\mu^2 \rho K_0(\mu \rho)\sin^2(\psi)-\mu K_1(\mu \rho)\cos(2\psi)\right)\Bigg)\nonumber\\
&=\frac{\pi A_d Vol(M_{d-2})}{(d-1)^2 \Gamma(d)}\int_{\epsilon}^{\frac{r_2-r_1}{2}}d\rho\Bigg( \int_0^{\infty} d \mu C(\mu) \mu^2 \rho K_0(\mu \rho)\Bigg)\nonumber\\
&=\frac{-\pi A_d Vol(M_{d-2})}{(d-1)^2 \Gamma(d)}\Bigg( \int_0^{\infty} d \mu C(\mu) (F(\mu,(r_2-r_1)/2)-F(\mu,\epsilon))\Bigg),
\end{align}
with $F(\mu,\rho)=\mu \rho K_1(\mu \rho)$, $C(\mu)=\left(c^{(0)}(\mu)+(d-1)(d-2) c^{(2)}(\mu)\right)$. In the CFT limit $c^{(0)}(\mu)\propto \mu^{d-2}\delta(\mu)$ and $c^{(2)}(\mu)=C_T\frac{d-1}{d}\mu^{d-3}$, this reproduces the CFT result we have in \eqref{aindeprepstress}, and \eqref{Sflatddimscaling1} with
\begin{equation}
    C_{d}^{[1]}= -\frac{C_T \pi^{d/2}\Gamma(d/2)}{\Gamma(d+2)},
\end{equation}
where $C_T$ is the charge that appears in the 2 point energy momentum correlation function in a $d-$dimensional CFT. 

\subsection{Entangling boundary with extrinsic curvature}
Consider a general CFT state on flat space on $M_1$ with a spherical or cylindrical entangling boundary i.e.,
\begin{equation}\label{sphfl}
    ds^2_{M_1,sph}=-dt^2+dr^2+r^2 h_{ij}(y) dy^i dy^j,
\end{equation}
with entangling region $r_1\leq r \leq r_2$ spanning the sphere directions $y^i$ on a constant $t$ slice. 

 or
\begin{equation}\label{cylfl}
    ds^2_{M_1,cyl}=-dt^2+dr^2+r^2 d\phi^2+dz^2,
\end{equation}
with entangling region $r_1\leq r \leq r_2$ spanning $\phi$ and $\frac{-L}{2}\leq z \leq \frac{L}{2}$ on a constant $t$ slice. 

The replica metric locally around the boundary of entangling region $r=r_0\in \{r_1, r_2\}$ is given by \eqref{sqcone} and \eqref{sqconecyl} i.e.,
\begin{align}\label{sqcone1}
     ds_q^2{}_{\text{ around } \Sigma,sph}&=U(\rho)d\rho^2+\rho^2 d\psi^2+(r_0+\rho^p c^{1-p}\cos(\psi))^2 h_{ij}(y) dy^i dy^j,\nonumber\\
     ds_q^2{}_{\text{ around } \Sigma,cyl}&=U(\rho)d\rho^2+\rho^2 d\psi^2+(r_0+\rho^p c^{1-p}\cos(\psi))^2 d\phi^2+dz^2.
\end{align}
Using \eqref{sqftvar2}, the scaling of $\Sqft$ in this case is given by,
\begin{equation}
    \alpha\Delta r \frac{\d S_{\text{$d$ dim}}}{\d \Delta r}=\underset{q\to1}{lim}\,\frac{1}{2}\int_{M_q}d^dx\sqrt{g}\,\delta_{\Delta r} g_{\mu\nu} \langle T^{\mu\nu}\rangle^{[1]}
\end{equation}
Since $\Sqft$ depends only on $\Delta r$ due to translation symmetry along $r$, we consider the scaling of $\Sqft$ wrt $\Delta r \to e^\alpha \Delta r$ or $r\to e^\alpha r$. On scaling the metric \eqref{sphfl} and \eqref{cylfl}, we get
\begin{align}
    \delta_{\Delta r}g_{\mu\nu,sph}&=2\alpha \eta_{\mu\nu,sph}, \quad \mu=\nu=r,\text{ and }  \mu,\nu \in \{i,j\},\nonumber\\
    \delta_{\Delta r}g_{\mu\nu,cyl}&=2\alpha \eta_{\mu\nu,cyl},\quad \mu=\nu=r \text{ or } \phi,  
\end{align}
therefore,
\begin{align}
     &\Delta r \frac{\d S_{\text{$d$ dim, sph}}}{\d \Delta r}=\underset{q\to1}{lim}\int_{M_q}d^dx\sqrt{g} \left(\langle T^{rr} \rangle^{[1]}+r^2 h_{ij}\langle T^{ij} \rangle^{[1]}\right),\\
     &\Delta r \frac{\d S_{\text{$d$ dim,cyl}}}{\d \Delta r}=\underset{q\to1}{lim}\int_{M_q}d^4x\sqrt{g} \left(\langle T^{rr} \rangle^{[1]}+r^2 \langle T^{\phi\phi} \rangle^{[1]}\right).
\end{align}

If we consider flat space and vacuum state, we have symmetry along $\tau$ and $-L/2 \leq z \leq L/2$, therefore $\Sqft$ is also independent of $\tau$ and $z$. We will restrict to vacuum state and scale $\tau$ and $z$ as well. We get
\begin{align}
    \delta_{\Delta r}g_{\mu\nu,sph}&=2\alpha \eta_{\mu\nu,sph},\nonumber\\
    \delta_{\Delta r}g_{\mu\nu,cyl}&=2\alpha \eta_{\mu\nu,cyl}.
\end{align}
Therefore, for the vacuum state, we have
\begin{equation}
    \Delta r \frac{\d S_{\text{$d$ dim}}}{\d \Delta r}=\underset{q\to1}{lim}\int_{M_q}d^dx\sqrt{g}  \,\eta_{\mu\nu}\langle T^{\mu \nu}\rangle^{[1]}. 
\end{equation}
Furthermore, for vacuum state we have $\langle T_{\mu\nu}\rangle^{[0]}=0$, which implies $\eta_{\mu\nu}\langle T^{\mu \nu}\rangle^{[1]}=\langle T^{\mu}_\mu\rangle^{[1]}$. This implies
\begin{equation}\label{extrinsicrepscal}
    \Delta r \frac{\d S_{\text{$d$ dim}}}{\d \Delta r}=\underset{q\to1}{lim}\int_{M_q}d^dx\sqrt{g} \langle T^{\mu}_\mu\rangle^{[1]} =\underset{q\to1}{lim}\int_{M_q}d^dx\sqrt{g} \mathcal{A}^{[1]}.
\end{equation}
As we see from the expressions above the only contribution to EE scaling for a spherical or cylindrical entangling region on flat background is from the regulator dependent replica anomaly $\mathcal{A}^{(q)}$ at $O(q-1)$. There are no regulator independent replica stress tensor contributions, that were present in the case of planar disconnected entangling boundaries. 

The replica anomaly contribution above is very similar to the regulator dependent planar contribution \eqref{evendanom}, except now the replica metric \eqref{sqcone1} and hence the replica anomaly can depend on the coordinate $\psi$ and have non zero extrinsic curvature for boundaries $r=r_0$. These integrals were performed in \cite{Fursaev:2013fta} and we have listed them in \eqref{riemcontrsqcone}, \eqref{Ksph}, \eqref{Kcyl}.\\ \\
\textbf{2 dimensions: } 
\begin{align}\label{2dimsphcyl}
  \mathcal{A}^{[1]}=-\frac{c}{24} R^{[1]}\implies \Delta r \frac{\d S_{\text{CFT}_2}}{\d \Delta r}\Bigg\rvert_{\text{reg dep}}=\frac{c}{6}.
\end{align}
\textbf{4 dimensions: }
\begin{equation}
    \mathcal{A}^{(q)}[g_{\mu\nu}^{(q)}]=\chi_{\ \mu}^\mu[g_{\mu\nu}]^{d=4}=-\frac{a}{(4 \pi)^2} E^{(q)}_4+\frac{c}{(4 \pi)^2} I^{(q)}_4.
\end{equation}
Using known integrals of Riemann contractions on replica metric with extrinsic curvature i.e., \eqref{riemcontrsqcone}, \eqref{intEI}, we have
\begin{align}
  \Delta r \frac{\d S_{\text{CFT}_4}}{\d \Delta r}=&2\pi \int_{M_q} d\rho d^{d-2}x\sqrt{h}\rho\mathcal{A}^{[1]}\nonumber\\
  =&(a I^\Sigma_{a}-c I^\Sigma_{c}). 
\end{align}
For a spherical and cylindrical $\Sigma$ on flat space we only have contributions from integrals of $K^2$ and $\text{Tr } K^2$. Using the value of these integrals in both cases given by \eqref{Ksph}, \eqref{Kcyl}, we have 
\begin{align}\label{4dimsphcyl}
  \Delta r \frac{\d S_{\text{CFT}_4}}{\d \Delta r}
  =& 4 a ; \quad\text{Spherical $\Sigma$},\nonumber\\
  =& \frac{L}{2 r_0} c;   \quad\text{Cylindrical $\Sigma$}.
\end{align}
at each entangling boundary. Therefore, the net contribution in our case would be the sum of the contributions from the two disconnected entangling boundaries at $r_0=\{r_1,r_2\}$. The spherical $\Sigma$ isolates central charge $a$ contribution and the cylindrical $\Sigma$ isolates central charge $c$ contribution.

\paragraph{5 and 6 dimensions:}We can similarly compute the scaling of EE in higher dimensions by using \eqref{extrinsicrepscal}. The replica trace anomaly is given by \eqref{evendanom} with $g_{\mu\nu}$ replaced by $g_{\mu\nu}^{(q)}$. The relevant replica integrals in 5 and 6 dimensions are listed in \cite{Fursaev:2013fta}.

\section{Entanglement entropy - non vacuum states} \label{sec:nine}
\subsection{Thermal state in odd dimensional CFTs}
 Odd dimensional CFTs do not have an anomaly, and more generally $\chi_{\mu\nu}^{[0]}=0=\chi_{\mu\nu}^{(q)}$. We consider this case with static metrics on $M_1$ satisfying $\Gamma^i_{k\mu}=0$, $\mu \in \{\rho,\psi\}$. This is true for planar entangling region in flat space on $M_1$
  \begin{equation}\label{planarodd}
     ds^2_{M_1}=d\tau^2 +dr^2+\delta_{ij}dx^i dx^j.
 \end{equation}
 It is also true for a CFT on a $d-$dimensional static metric with a finite volume along the $x^i$ directions 
 \begin{equation}\label{compactodd}
     ds^2_{M_1}=g_{\tau\tau}(r)d\tau^2 +g_{rr}(r)dr^2+\phi_0^2 (d\Omega_{d-2}^2=h_{ij}dx^i dx^j),
 \end{equation}
where $r \in [0,\infty)$, $\tau \in (-\infty, \infty)$, $M_{d-2}$ is compact.

We consider the usual entangling region $r \in [r_1,r_2]$ at $\tau=\tau_0$ spanning the $x^i$ directions. 

The replica metric in a local region around the boundary $r=r_0\, (r_1\text{ or }r_2)$ is given by
\begin{equation}
    ds^2_{M_q}\text{ around endpoint} = U(\rho,a)d\rho^2+\rho^2 d\psi^2+\phi_0^2 h_{ij}dx^i dx^j,
\end{equation}
where $(r-r_0)=\rho \cos\psi$, $(\tau-\tau_0)=\rho\sin\psi$, and the conical regulating function $U(\rho,a)\rvert_{\rho \to 0} \to q^2$, $U(\rho,a)\rvert_{\rho>>a} \to 1$. We could always use coordinate $x=\rho/a$ where $U(x)$ is independent of regulator a. We have $\Gamma^i_{i\rho}{}^{(q)}=0=\Gamma^i_{i\psi}{}^{(q)}$. 

 \subsubsection{Replica stress tensor}
The replica trace and conservation equations \eqref{trlessfull}-\eqref{psiconsfull}, \eqref{sigmaieqn} reduce to\\ \\
\textbf{$O(q-1)$ Traceless condition:}
\begin{equation}\label{trlessthd}
    t_{\rho\rho}^{[1]}+\frac{1}{\rho^2}t_{\psi\psi}^{[1]}+t^i_{\ i}{}^{[1]}=U^{[1]}(\rho,a)t_{\rho\rho}^{[0]}.
\end{equation}
\textbf{$O(q-1)$ $\rho$-component of Conservation:}(which includes trace condition)
\begin{equation}\label{rhoconsthd}
   \frac{1}{\rho^2}\d_\psi t^{[1]}_{\rho\psi}-\frac{1}{\rho^2}\d_\rho t^{[1]}_{\psi\psi}-\left(\d_\rho+\frac{1}{\rho}\right)t^i_{\ i}{}^{[1]}=0.
\end{equation}
\textbf{$O(q-1)$ $\psi$-component of Conservation:}
\begin{equation}\label{psiconsthd}
  \left(\d_\rho+\frac{1}{\rho}\right)t^{[1]}_{\rho\psi}+\frac{1}{\rho^2}\d_\psi t^{[1]}_{\psi\psi}=
   \left(\frac{1}{2}\d_\rho U^{[1]}+U^{[1]}\d_\rho+\frac{U^{[1]}}{\rho}\right)t^{[0]}_{\rho\psi}.
\end{equation}
\textbf{$O(q-1)$ $i$-component of Conservation:}
\begin{equation}\label{iconsthd}
  \text{For a fixed $i$: }       (\d_i+\sum_l \Gamma^l_{\ li})\langle T^i_{\ i}\rangle^{[1]}-\sum_{l} \Gamma^l_{\ li}\langle T^l_{\ l}\rangle^{[1]}=0.
\end{equation}
  
We will now calculate the scaling of the EE in this case. When the metric on $M_1$ is flat, scaling the metric using \eqref{scalingst}, we have
\begin{align}
    \delta_{\Delta r}g_{\mu\nu}&=2\alpha; \quad \mu=\nu=r,\\
    &=0;\quad \text{otherwise}.
\end{align}
Therefore using \eqref{sqftvar2}, the scaling of the EE is given by,
\begin{equation}\label{SqCFTodddimscaling}
    \Delta r \frac{\d S_{\text{CFT}_d}}{\d \Delta r}=\underset{q\to1}{lim}\int_{M_q}d^dx\sqrt{g}\, t^{rr}{}^{[1]}.
\end{equation}
Using the coordinate transformation \eqref{conversion} we have,
\begin{align}
    t^{rr}&=U^{-2}t_{\rho\rho}^{(q)}(\cos(\psi))^2-\frac{2U^{-1}}{\rho}t_{\rho\psi}^{(q)}\cos(\psi)\sin(\psi)+\frac{1}{\rho^2}t_{\psi\psi}^{(q)}(\sin(\psi))^2, 
\end{align}

\begin{align}
    t^{rr}{}^{[1]}\Big|_{\text{around }r_0,\tau_0}&=t_{\rho\rho}^{[1]}(\cos(\psi))^2-\frac{2}{\rho}t_{\rho\psi}^{[1]}\cos(\psi)\sin(\psi)+\frac{1}{\rho^2}t_{\psi\psi}^{[1]}(\sin(\psi))^2 \nonumber\\
    &\hspace{0.2cm}-2U^{[1]}t^{[0]}_{\rho\rho}(\cos(\psi))^2-\frac{2U^{[1]}}{\rho}t^{[0]}_{\rho\psi}\cos(\psi)\sin(\psi).
\end{align}
We will consider the CFT thermal stress tensor on a flat background (\eqref{thermalrhopsi} with $\epsilon = -(d+1)p$ and $g_{rr}=g_{\tau\tau}=1$,  $g_{r\tau}=0$, $g_{ij}=\phi_{0}^2 h_{ij}$) 
\begin{align}\label{thermalCFT}
   t_{\rho\psi}^{[0]}&=(-dp/2) \rho \sin(2\psi), \nonumber\\
   t_{\rho\rho}^{[0]}&=p[1- d(\sin(\psi))^2], \nonumber\\
   t_{\psi\psi}^{[0]}&=p\rho^2[1- d(\cos(\psi))^2] ,\nonumber\\
   t_{ij}^{[0]}&=p(g_{ij}=\phi_{0}^2 h_{ij} \text{ or } \delta_{ij}).
\end{align}

\paragraph{Homogenous solution:}Let us first consider the homogenous solution to the replica trace and conservation equations above. \eqref{iconsthd} has no sources and has the solution
\begin{equation}
    t^i_i{}^{[1]}=\frac{(d-2)C^{[1]}}{\rho^d}+ t^i_i{}^{[0]}.
\end{equation}
The $\frac{C^{[1]}}{\rho^d}$ term reflects the lack of $\rho$ translation symmetry due to the conical singularity on the replica manifold. Since the $i-$conservation equations on $M_1$ and $M_q$ are the same with $t^i_i{}^{[0]}\to t^i_i{}^{[1]}$, noting \eqref{homogenote} we additionally have the $t^i_i{}^{[0]}$ term.
\\

Noting that $t^i_i{}^{[0]}$ and $t^i_i{}^{[1]}$ are independent of $\psi$, \eqref{rhoconsthd} and \eqref{psiconsthd} implies,
\begin{equation}
    \left(1+3\rho\d_\rho+\rho^2 \d_\rho^2+\d_\psi^2\right)t^{[1]}_{\rho\psi}=0.
\end{equation}
 We use the ansatz $t^{[1]}_{\rho\psi}=A\rho^\Delta m(\psi)$, This gives
 \begin{align}
     &\d_\psi^2 m(\psi)=-k^2 m(\psi);\quad k^2=(\Delta+1)^2 \nonumber\\
     \implies&\Delta=(-k-1) \text{ or }(k-1),
 \end{align}
which implies
\begin{align}
     &m(\psi)=c_1 \cos(k\psi)+ c_2 \sin(k\psi).
 \end{align}
We have the following replica corrections to the replica stress tensor,
\begin{equation}\label{trs}
    t^{[1]}_{\rho\psi}=\sum_k(A_k \rho^{-(k+1)}+B_k\rho^{(k-1)})(c_1 \cos(k\psi)+ c_2 \sin(k\psi)).
\end{equation}
Noting our observations in \eqref{homogenote}, we must have $t^{[1]}_{\rho\psi}=t^{[0]}_{\rho\psi}$ which implies $B_2 c_2 = -dp/2$, $c_1=0$, $A_k=0$ $\forall\, k$ and $B_{k\neq 2}=0$. With this we have
\begin{equation}\label{trshomoge}
    t^{[1]}_{\rho\psi}= \frac{-dp}{2}\rho\sin(2\psi).
\end{equation}
\eqref{rhoconsthd} implies
\begin{align}
    t^{[1]}_{\psi\psi}&=\int d\rho\Bigg[ \d_\psi  t^{[1]}_{\rho\psi}-\rho^2\left(\d_\rho+\frac{1}{\rho}\right)t^i_{\ i}{}^{[1]}\Bigg]\nonumber\\
    &=\sum_k(B_k \rho^k-A_k\rho^{-k})(c_2 \cos(k\psi)-c_1 \sin(k\psi))+(1-d)\frac{C_{d}^{[1]}}{\rho^{d-2}}+\frac{(2-d)}{2}p\rho^2+c_3(\psi)\nonumber\\
    &=[t^{[0]}_{\psi\psi}=p\rho^2(1-d\cos^2(\psi))]+(1-d)\frac{C_{d}^{[1]}}{\rho^{d-2}},\label{tss}
\end{align}
where we have chosen $c_3(\psi)=0$ on dimensional grounds.
\eqref{trlessthd} implies,
\begin{align}
    t^{[1]}_{\rho\rho}&=-\frac{1}{\rho^2} t^{[1]}_{\psi\psi}-t^i_{\ i}{}^{[1]}\nonumber\\
    &=\sum_k(A\rho^{-k-2}-B \rho^{k-2})(c_2 \cos(k\psi)-c_1 \sin(k\psi))\nonumber\\
    &=[t^{[0]}_{\rho\rho}=p\rho^2(1-d\sin^2(\psi))]+\frac{C_{d}^{[1]}}{\rho^{d}}.\label{trr}
\end{align}

Notice that $\d_\psi$ is a killing vector on $M_1$ and $M_q$ but the thermal state breaks this symmetry. Considering this and the discussion in \eqref{homogenote}, the homogenous solution to $t^{[1]}_{\mu\nu}$ contains the vacuum part which has $\psi$ translation symmetry and the thermal homogenous part which does not have this symmetry i.e.,
\begin{equation}
    t^{[1]}_{\mu\nu}\rvert_{homogenous}=\frac{C^{[1]}}{\rho^d}((1-d)\rho^2,1,1,1,1,...,1)+ t^{[0]}_{\mu\nu}(\rho,\psi)\rvert_{thermal},
\end{equation}
where the basis is in order $\{\psi,\rho,x^i\}$.

\paragraph{Full solution:}Let us consider a general CFT state. Combining \eqref{rhoconsthd} and \eqref{psiconsthd} into a second order PDE in $t_{\rho\psi}^{[1]}$ only,
\begin{align}\label{flat3pde}
    &(\d_{\psi}^2 + \rho^2 \d_{\rho}^2 + 3\rho \d_{\rho}+1 )t_{\rho\psi}^{[1]}(\rho,\psi)
    =\Bigg(\left(1+2\rho \d_{\rho}+\frac{1}{2}\rho^2\d_{\rho}^2 \right)U^{[1]}(\rho,a)\nonumber\\
    &+\left(3\rho U^{[1]}(\rho,a)+\frac{3}{2}\rho^2\d_{\rho}U^{[1]}(\rho,a) \right)\d_{\rho}+\rho^2 U^{[1]}(\rho,a)\d_\rho^2\Bigg)t_{\rho\psi}^{[0]}(\rho,\psi),
\end{align}
We will now demand
\begin{align}\label{splitform}
    &t_{\rho\psi}^{[0]}=h(\rho) m(\psi),\\
    &\d_{\psi}^2 m(\psi)= -n^2 m(\psi) \text{ i.e., } m(\psi)= c_1 \cos(n\psi) + c_2 \sin(n\psi),\label{m3d}
\end{align}
for arbitrary constants $n$, $c_1, c_2$. 

The solution to \eqref{flat3pde} is 
\begin{equation}
    t_{\rho\psi}^{[1]}(\rho,\psi)=g(\rho) m(\psi),
\end{equation}
 where $g(\rho)$ satisifies the first-order ODE
\begin{align}
    &\rho  \left(2 \rho  g''(\rho )+6 g'(\rho )\right)-2 (n^2-1) g(\rho )\nonumber\\
    =&2 U^{[1]}(\rho,a ) \rho^2  h''(\rho )+3\left(\rho^2 U^{[1]}{'}(\rho,a ) +2\rho U^{[1]}(\rho,a )\right)h'(\rho)\nonumber\\
    &+h(\rho ) \left(\rho  \left(\rho  U^{[1]}{''}(\rho,a )+4 U^{[1]}{'}(\rho ,a)\right)+2 U^{[1]}(\rho,a
   )\right).
\end{align}
This can be solved for $f(\rho)$ but we will not write the solution here.

\eqref{splitform} is not true for a general CFT state, however,it is true  for the CFT thermal stress tensor on a flat background \eqref{thermalCFT}. Restricting further to this state i.e., $h(\rho)=-dp \rho/2$ and $m(\psi)=\sin(2\psi)$ (i.e., a particular solution to \eqref{m3d} with $n=2$), we have
\begin{align}
    t_{\rho\psi}^{[1]} = \Bigg\{\frac{d_1}{\rho^3}+d_2 \rho+\frac{d p}{16}a f(x)\Bigg\} \sin(2\psi),
\end{align}
with
\begin{align}
    f(x)=x\Bigg(2 U^{[1]}(x)+ x U^{[1]}{'}(x)
    -\int_1^x\left(\frac{8U^{[1]}(x')}{x'}+7 U^{[1]}{'}(x')+x'U^{[1]}{''}(x')\right)dx'\Bigg),
\end{align}
where $x=\rho/a$. $d_1, d_2$ are integration constants. Comparing with the homogenous solution \eqref{trshomoge} and noting \eqref{homogenote} we get
\begin{align}\label{d1d2}
    d_1=0; \quad
    d_2=\frac{-dp}{2},
\end{align}
i.e.,
\begin{equation}\label{trsth}
     t_{\rho\psi}^{[1]} = t_{\rho\psi}^{[0]} + \frac{d p}{16}a f(x) \sin(2\psi).
\end{equation}
\eqref{iconsthd} has no sources and has the same solution as the homogenous solution on $M_q$
\begin{equation}
    t^i_i{}^{[1]}=\frac{(d-2)C_d^{[1]}}{\rho^d}+ t^i_i{}^{[0]}.
\end{equation}
Using \eqref{rhoconsthd},
\begin{align}\label{tssth}
    t_{\psi\psi}^{[1]} &= \int d\rho\d_{\psi}t^{[1]}_{\rho\psi}-\int d\rho\rho \d_\rho(\rho t^{i}{}_i{}^{[1]})\nonumber\\
    &=\Bigg\{\frac{-d_1}{\rho^2}+d_2 \rho^2
    +\frac{d p}{8}a^2 \int dx f(x)\Bigg\} \cos(2\psi)+d_3(\psi)-\frac{\rho^2}{2}(t^{i}{}_i{}^{[1]}=t^{i}{}_i{}^{[0]}=(d-2)p)\nonumber\\
    &= (1-d)\frac{C_{d}^{[1]}}{\rho^{d-2}}+t_{\psi\psi}^{[0]}+\frac{d p}{8}a^2 \int dx f(x) \cos(2\psi)+d_3(\psi).
\end{align}
We again choose $d_3(\psi)=0$ on dimensional grounds. Using \eqref{trlessthd}
\begin{align}\label{troroth}
    t_{\rho\rho}^{[1]} =&-\frac{  t_{\psi\psi}^{[1]}}{\rho^2}-t^i_i{}^{[1]}+U^{[1]}t^{[0]}_{\rho\rho} \nonumber\\  
    =&\Bigg\{\frac{d_1}{\rho^4}-d_2
    -\frac{d p}{8 x^2} \int dx f(x)\Bigg\} \cos(2\psi)-\frac{t^i_i{}^{[1]}}{2}+ U^{[1]}t^{[0]}_{\rho\rho}\nonumber\\
    =& \frac{C_{d}^{[1]}}{\rho^d}+t^{[0]}_{\rho\rho}-\frac{d p}{8 x^2} \int dx f(x)\cos(2\psi)+ U^{[1]}t^{[0]}_{\rho\rho}.
\end{align}
We have,
\begin{align}\label{thermalt1rr}
    t^{rr}{}^{[1]}=&t_{\rho\rho}^{[1]}(\cos(\psi))^2-\frac{2}{\rho}t_{\rho\psi}^{[1]}\cos(\psi)\sin(\psi)+\frac{1}{\rho^2}t_{\psi\psi}^{[1]}(\sin(\psi))^2\nonumber\\ 
    &-2U^{[1]}t^{[0]}_{\rho\rho}(\cos(\psi))^2-\frac{2U^{[1]}}{\rho}t^{[0]}_{\rho\psi}\cos(\psi)\sin(\psi)\nonumber\\
    =&\frac{C_d^{[1]}}{\rho^d}(1-d\sin^2(\psi))+t_{rr}^{[0]}=\left(p\left(1-\frac{d}{2}\right)-d_2+d_1\frac{\cos(4\psi)}{\rho^4}\right)+\nonumber\\
    &I(x,\psi)=\Bigg[\left(U^{[1]}t^{[0]}_{\rho\rho} -\frac{d p}{8 x^2} \int dx f(x) \cos(2\psi) \right)\cos(\psi)^2-\left(\frac{d p}{16x} f(x) \sin^2(2\psi)\right)\nonumber\\
    &+\left(\frac{d p}{8x^2} \int dx f(x) \cos(2\psi)\right)\sin(\psi)^2-2U^{[1]}t^{[0]}_{\rho\rho}(\cos(\psi))^2-\frac{2U^{[1]}}{\rho}t^{[0]}_{\rho\psi}\cos(\psi)\sin(\psi)\Bigg].
\end{align}
The universal parts of $U^{[1]}(x)$ is determined by $\eqref{limits2}$. We have $U^{[1]}(x)\rvert_{x\to 0}=2$, $U^{[1]}(x)\rvert_{x>> 1}=0$. Notice that $t_{\mu\nu}^{[1]}$ with $\mu, \nu \in \{\rho,\psi\}$ depends on non-universal terms in the regulating function i.e., it seems to depend on the choice of the regulator. We will work with a specific regulating function \eqref{Uexplicit} i.e., $U(x,a)=\frac{x^2+ q^2}{x^2+1}$ with $U^{[1]}(x,a)=\frac{2 }{1+x^2}$. With this choice, we have,
\begin{align}
  f(x) &= 8x\left(\log\left(1+\frac{1}{x^2}\right)-\frac{1}{1+x^2}\right), \\
  \int f(x) dx&=8x^2\log\left(\sqrt{1+\frac{1}{x^2}}\right); \quad x=\frac{\rho}{a}.
\end{align}
\subsubsection{Entanglement Entropy}
The scaling of the EE in this case is given by,
\begin{align}\label{thermalEEscaling}
    &\Delta r \frac{\d S_{\text{CFT}_d}}{\d \Delta r}=\underset{q\to1}{lim}\int_{M_q}d^dx\sqrt{g} t_{rr}^{[1]}\nonumber\\
    &=\underset{q\to1,a\to 0}{lim}\int_{M_{d-2},\psi=0}^{2\pi q}\sqrt{h}d^{d-2}y^i d\psi\Bigg[\int_{\epsilon}^{\frac{r_2-r_1}{2}} d\rho\rho\Bigg(\frac{C_d^{[1]}}{\rho^d}[1-d\sin^2(\psi)]\nonumber\\
    &\quad+\left(t^{[0]}_{rr}=p\left(1-\frac{d}{2}\right)-d_2+d_1\frac{\cos(4\psi)}{\rho^4}\right)\Bigg)+a^2\int_\epsilon^\infty dx x I(x,\psi)\Bigg]\nonumber\\
    &=\underset{q\to1,a\to 0}{lim} \pi q Vol(M_{d-2})\Bigg[\int_\epsilon^\frac{r_2-r_1}{2} d\rho\rho\left((2-d) \frac{C_d^{[1]}}{\rho^{d-1}}+(2-d)p-2d_2\right)\nonumber\\
    &\quad+a^2 p\int_\epsilon^\infty dx x \left(\frac{2(d-1)}{1+x^2}-d\log\left(1+\frac{1}{x^2}\right)\right)\Bigg]\nonumber\\
    &=\underset{q\to1}{lim}\pi Vol(M_{d-2}) \left\{C_d^{[1]}\left(\frac{2^{d-2}}{(\Delta r)^{d-2}}-\frac{1}{\epsilon^{d-2}}\right)+\frac{(2-d)p-2d_2}{8}(\Delta r)^2\right\}\nonumber\\
    &=\pi Vol(M_{d-2}) \left\{C_d^{[1]}\left(\frac{2^{d-2}}{(\Delta r)^{d-2}}-\frac{1}{\epsilon^{d-2}}\right)+\frac{p}{4}(\Delta r)^2\right\}.
\end{align}
We have kept $d_1,d_2$ arbitrary until the last equality where we have used \eqref{d1d2}. As mentioned in \eqref{metstate}, we restrict the $\rho$ integral between $0$ to $\frac{r_2-r_1}{2}$ for the regulator-independent homogenous replica stress tensor contributions, and between $0$ to $\infty$ for the regulator dependent replica stress tensor contributions.

We have chosen a specific regulating function $U(x,a)$. But from the second equality above, we see that the $a$ dependence in the regulator-dependent integral goes as $a^2$, and hence the integral goes to zero as $a\to 0$ independent of the choice of $U(x,a)$. So the contribution to the scaling of EE is purely from the regulator independent homogenous part of the replica stress tensor. 

When solving the replica trace and conservation equations \eqref{trlessthd}-\eqref{iconsthd} in the case of a disconnected boundary $\Delta r=L=(r_2-r_1)$ distance apart, we can consider $L$ to be one more dimensionful parameter that the replica stress tensor $\langle T^\mu_{\, \nu}\rangle_q$ can in general depend on. If we follow an approach similar to the vacuum case we find that the stress tensor and EE change as follows
\begin{equation}\label{tiithwithL}
    t^i_i{}^{[1]}=\frac{(d-2)C_d^{[1]}}{\rho^d}K(\rho/L)+ t^i_i{}^{[0]}K_1(\rho/L),
\end{equation}
where $K(\rho/L)$ and $K_1(\rho/L)$ are dimensionless functions which tend to 1 as $\rho>>L$, which implies
\begin{equation}
    K(\rho,L)=\sum_{i\geq 0}A_i\left(\frac{L}{\rho}\right)^i;\quad K_1(\rho,L)=\sum_{i\geq 0}A^{(1)}_i\left(\frac{L}{\rho}\right)^i;\quad i\in \mathbb{Z}.
\end{equation}

\begin{fleqn}
\begin{equation}
\begin{split}
    t_{\rho\psi}^{[1]} = \Bigg\{\frac{d_1}{\rho^3}+d_2 \rho+\frac{d p}{16}a f(x)\Bigg\} \sin(2\psi),
\end{split}
\end{equation}
\end{fleqn}

\begin{fleqn}
\begin{equation}
\begin{split}
    &t_{\psi\psi}^{[1]} = \int d\rho\d_{\psi}t^{[1]}_{\rho\psi}-\int d\rho\rho \d_\rho(\rho t^{i}{}_i{}^{[1]})\\
    &=\Bigg\{\frac{-d_1}{\rho^2}+d_2 \rho^2
    +\frac{d p}{8}a^2 \int dx f(x)\Bigg\} \cos(2\psi)+d_3(\psi)-\frac{\rho^2}{2}(d-2)p K_1(\rho/L)\\ 
    &\quad-(d-2)p\int d\rho \frac{\rho^2}{2}  \d_\rho K_1(\rho/L)
   +\frac{(1-d)C_d^{[1]}}{\rho^{d-2}}K(\rho,L)+\int d\rho \d_\rho K(\rho,L) \frac{C_d^{[1]}}{\rho^{d-2}},
\end{split}
\end{equation}
\end{fleqn}

\begin{fleqn}
\begin{equation}
\begin{split}
   t_{\rho\rho}^{[1]} =&-\frac{  t_{\psi\psi}^{[1]}}{\rho^2}-t^i_i{}^{[1]}+U^{[1]}t^{[0]}_{\rho\rho} \\  
    =&\Bigg\{\frac{d_1}{\rho^4}-d_2
    -\frac{d p}{8 x^2} \int dx f(x)\Bigg\} \cos(2\psi)+ U^{[1]}t^{[0]}_{\rho\rho}-\frac{(d-2)}{2}pK_1(\rho/L) \\
    &+\frac{(d-2)p}{\rho^2}\int d\rho \frac{\rho^2}{2}  \d_\rho K_1(\rho/L)+\frac{C_d^{[1]}}{\rho^{d}}K(\rho,L)-\frac{1}{\rho^2}\int d\rho \d_\rho K(\rho,L) \frac{C_d^{[1]}}{\rho^{d-2}},  
\end{split}
\end{equation}
\end{fleqn}

\begin{fleqn}
\begin{equation}
\begin{split}
    t_{rr}^{[1]}=&t_{\rho\rho}^{[1]}(\cos(\psi))^2-\frac{2}{\rho}t_{\rho\psi}^{[1]}\cos(\psi)\sin(\psi)+\frac{1}{\rho^2}t_{\psi\psi}^{[1]}(\sin(\psi))^2\nonumber\\
    &-2U^{[1]}t^{[0]}_{\rho\rho}(\cos(\psi))^2-\frac{2U^{[1]}}{\rho}t^{[0]}_{\rho\psi}\cos(\psi)\sin(\psi)\\   
    =&\frac{C_d^{[1]}}{\rho^d}K(\rho/L)(1-d\sin^2(\psi))+p\left(1-\frac{d}{2}\right)K_1(\rho/L)-d_2+d_1\frac{\cos(4\psi)}{\rho^4}\\
    &+I(x,\psi)+\frac{\cos(2\psi)}{\rho^2}\int d\rho\left((d-2)p \frac{\rho^2}{2}  \d_\rho K_1(\rho/L)-\d_\rho K(\rho,L) \frac{C_d^{[1]}}{\rho^{d-2}}\right).
\end{split}
\end{equation}
\end{fleqn}
We have,
\begin{fleqn}
\begin{equation}
\begin{split}
 &\Delta r \frac{\d S_{\text{$d$ dim, bulk}}}{\d \Delta r}\Bigg\rvert_{\text{reg indep}}=\underset{q\to1}{lim}\int_{M_q}d^dx\sqrt{g} t_{rr}^{[1]}\\
&=\underset{q\to1}{lim}\int_{\Sigma}d^{d-2}x \sqrt{h}\int_{0}^{2\pi q}d\psi\int_{\epsilon}^{\alpha L}d\rho \rho \Bigg(\frac{C_d^{[1]}}{\rho^d}K\left(\frac{\rho}{L}\right)(1-d\sin^2(\psi))+p\left(1-\frac{d}{2}\right)K_1\left(\frac{\rho}{L}\right)-d_2\Bigg)\\
&=\pi Vol(M_{d-2})\Bigg[(2-d)\sum_{i\geq 0}\Bigg(\frac{C_d^{[1]}}{L^{d-2}} \frac{B_i}{\alpha^{d-2+i}}+p \frac{B^{(1)}_{i}}{\alpha^{i-2}} L^2-\frac{C_d^{[1]}B_i}{\epsilon^{d-2}}\frac{L^i}{\epsilon^i}-p B^{(1)}_{i} \frac{L^i}{\epsilon^{i-2}} \Bigg)-d_2 \alpha^2 L^2\Bigg],\nonumber
\end{split}
\end{equation}
\end{fleqn}
where $i\in \mathbb{Z}$ and $B_i=A_i/(2-d-i)$, $B_i^{(1)}=A_i^{(1)}/(2-i)$.

\section{Conclusions} \label{sec:ten}

We begin this section by summarising the main results of this paper. We have considered the properties of entanglement entropy on a generic static background \eqref{genrepentscaling}, \eqref{delr}. This analysis requires the replica stress tensor, which is determined using the replica stress tensor conservation \eqref{repdeqn} as well as other defining properties, such as the trace equation in a CFT. Note that the geometry of the entangling region is captured by the metric on the replica space. On static backgrounds we can uniquely and explicitly solve for the replica stress tensor expectation value using the replica stress tensor conservation equations \eqref{sigmaieqn}, \eqref{rhoconsfull}, \eqref{psiconsfull} and trace equation \eqref{trlessfull} in the case of a CFT. The $O(q-1)$ replica stress tensor expectation value completely determines the scaling of the entanglement entropy (see \eqref{sqftvar2}).  

Most computations of entanglement and R\'{e}nyi entropies have been limited to vacuum states on flat space in specific entangling subregions. The approach presented in this paper provides a way to calculate the scaling of R\'{e}nyi and entanglement entropies (and the quantities themselves up to a constant) for general states and subregions on arbitrary backgrounds. It also sets up the methodology required to calculate replica stress tensor correlations in any state given the corresponding stress tensor correlations on the base manifold. 

We have presented explicit computations for vacuum and quasithermal states for a subregion bounded by two planes in flat space using the replica metric corresponding to an unsquashed cone \eqref{repmet1}. We have also given explicit computations in spherical and cylindrical entangling boundaries in flat space. A natural extension would be to consider other states and subregions on static backgrounds. While our focus has been on specific scaling transformations, we could analyse in a similar way the behaviour of entanglement under generic variations, including deformations of the background metric, using \eqref{genrepentscaling}. 

Another straightforward extension would be to consider other states (non vacuum and non thermal) on flat backgrounds in higher dimensions. The replica conservation and trace equations that determine $\langle T_{\mu\nu}\rangle^{[1]}$ will now have sources $\langle T_{\mu\nu}\rangle^{[0]}$ and $\mathcal{A}^{[1]}[g_{\mu\nu}^{[q]}]$ (in even dimensions). We could also consider a more complicated entangling boundary geometry. The expansion of the metric around a general foliation is given by \eqref{general fol} and the corresponding replica metric is given in \eqref{repgen}. Using these in the replica conservation and trace equations gives the replica stress tensor expectation value, and using that one can obtain the scaling of entanglement entropy and also the R\'{e}nyi entropy for any entangling boundary of interest.

Further analysis of entanglement and its variations in non-flat static backgrounds would use the replica metric $\eqref{repmetstat}$ which has explicit dependence on the extrinsic curvature of the entangling boundary. In this generic context the replica conservation and trace equations are more complicated, due to the non-vanishing connection on the base space, although still straightforwardly solvable case by case.

To extend to stationary black holes, one would need to extend the computation of entanglement to the corresponding stationary backgrounds. One way of approaching this would be to make use of the generating boost or rotation transforms that relate stationary and static black holes. For example, the defining equations for the replica stress tensor in the stationary case will be related to those in the static case by the corresponding generating transformations. A more ambitious step would be to extend our analysis to entanglement in generic backgrounds, in which there is no time translation symmetry. Interesting first steps in this direction could follow from considering de Sitter type backgrounds.   

Carrollian and celestial CFTs have recently been extensively discussed in the context of holography in asymptotically flat spacetimes. Since our computation is a field theory computation using replica trick and does not rely on holography or on the nature of the asymptotics of the space, one could straightforwardly extend our methodology to compute entanglement entropy in Carrollian and celestial field theories. The trace and conservation equations will be modified, reflecting the underlying symmetry structure of these theories \cite{Baiguera:2022lsw, Armas:2023dcz, deBoer:2017ing, deBoer:2020xlc, deBoer:2023fnj, Have:2024dff}. It would be interesting to explore entanglement in such theories, and relate to the proposed holographic descriptions. More generally, explicit computations of $\Sqft$ in higher dimensions also facilitate exploration of the quantum focussing conjecture \cite{Bousso:2015mna}, with the possibility of refining bounds further.


\bigskip

Let us now turn to the connections between our analysis and islands of entanglement. 
The premise of islands follows from the semi-classical prescription for entropy, i.e. the quantum extremal surface approach. 
The quantum extremal surface (QES) prescription \cite{Engelhardt:2014gca} states that the generalised entropy associated with a subregion $A$ is obtained from extremising the contributions of the geometric area and the entanglement entropy of the quantum fields in subregion $A$:
\begin{equation}\label{QESpres}
    S_{\hbar^0}(A) = \textnormal{min}  \Biggr[ \underset{A}{\textnormal{ext}}\left(\Sgen=\Sqft(A)+\Sarea(\d A)\right) \Biggr].
\end{equation}
The quantum extremal surface is used in the island prescription \cite{Penington:2019npb, Almheiri:2019yqk} as a way to recover the Page curve and restore unitarity, by including quantum  $O(\hbar^0)$ corrections to the Hawking radiation from an evaporating black hole.

Specifically, if $S[\textbf{Rad}]$ is the total entropy and $\Sqft[A]$ is the renormalised entropy of the quantum fields (here radiation) associated with a subregion $A$, we have 
\begin{equation}\label{island extremum}
    S[\textbf{Rad}] = \textnormal{min}  \Biggr[ \underset{A}{\textnormal{ext}}\left(\Sgen=\Sqft(A)+\Sarea(\d A)\right) \Biggr].
\end{equation}
We vary the subregion $A$ until the generalised entropy is minimized. This minimum generalised entropy associated with a particular subregion gives the entropy of the radiation in the quantum description (to first order in the $\hbar$ expansion). It has been shown to recover the Page curve in certain setups. 

If the subregion $A_{min}$ is finite i.e., it does not share a boundary with asymptotic infinity, then the complementary region (which has the same entropy, provided we started with a pure state) will be disconnected and has a region whose boundaries are disconnected from the asymptotic infinity. The region which is disconnected is termed an island, and the presence of islands has been shown to be key in recovering the Page curve.

Now let us consider the extremisation of the generalised entropy in a static black hole background. Given the underlying symmetry of such a background, the extremisation corresponds to considering quantum fields in a slab region bounded by inner and outer radii, and varying the radius. This is precisely the variation that we have considered in this paper. The area term in the generalised entropy is purely geometric and typically behaves monotonically in the radius. The existence of islands, i.e. the quantum extremal surface having a component that is disconnected from asymptotic infinity, can thus be translated to conditions on the scaling of $S_{\rm QFT}$ with the radius. This analysis will be the focus of our upcoming work. 

It was shown in \cite{Geng:2021hlu, Geng:2023zhq, Geng:2020fxl, Geng:2020qvw}, that in theories with long-range gravity where Gauss law holds, islands which by definition are disconnected from the asymptotic region of spacetime do not constitute consistent entanglement wedges. This is because the energy of an excitation localized to the island can be detected from outside the island. It was argued that this can be resolved if islands appear in conjunction with a massive graviton. Resolutions retaining massless graviton, have been discussed in the context of the de Sitter information paradox\cite{Geng:2021wcq} and in the presence of black holes with baby universes\cite{Iizuka:2021tut}. It will be interesting to see how our conditions on the scaling of $\Sqft$ for the existence of islands, which in turn impose conditions on the QFT spectrum, compare with these.  



\section*{Acknowledgements}

MT acknowledges support from the Science and Technologies Facilities Council (STFC), via grants ST/T000775/1 and ST/X000583/1. AS is supported by the Research Studentship in Mathematical Science, University of Southampton, and scholarship from the Government of Karnataka, India. AS is grateful for general discussions with  Kostas Skenderis, Felix Haehl, Mritunjay Verma, Deepali Singh, Benjamin Withers, David Turton, Itamar Yaakov, Oscar Dias, Suvrat Raju, Hao Geng, Aninda Sinha, Prasad Hegde, Jyothsna Kormaragiri, Chethan Krishnan, Rekha, Ramya, Monisha Gopalan, Rifath Khan, Ioannis Matthaiakakis, Chandramouli Chowdhury, Ernesto Bianchi, Zezhuang Hao, Abdul Afzal, Hector, Pratyusha Chowdhury, Radu Moga, Filip Landgren, Rajnandhini Mukherjee, Somasundaram, Divya Krishna Kumar, Sujith, Shounak De, Rishabh Bhardwaj, Sai Joshi, Ayngaran Thavanesan, Tamanna Jain.

AS is thankful to Sangappa IAS, Nischith Victor Daniel, Nagambika Devi IAS, Siddalingaswamy for their support in securing funding during the Covid pandemic.
\\ \\
AS dedicates this work to his Mom Jeyabharathi and the loving memory of his Dad Shekar.

\newpage
\appendix
\section{Notations} \label{Notations}
Lorentzian Metric signature: $(-,+,+,+)$\\
Entanglement entropy of a QFT: $\Sqft$\\
Entanglement entropy of a CFT: $S_{CFT}$\\

\begin{tabular}{ |p{9cm}|p{2.8cm}|p{2.3cm}|  }
\hline
\multicolumn{3}{|c|}{Base vs Replica Manifold} \\
\hline
& Base/Original Manifold & Replica $q$-Manifold \\
\hline
Manifold & $M_1$ &$M_q$ \\
Metric & $g_{\mu \nu}$ or $g_{\mu \nu}^{[0]}$  & $g^{(q)}_{\mu \nu}$ (around endpoint) \\
Stress Tensor expectation& $\left< T^{\mu \nu} \right>$ or $\left<T^{\mu \nu} \right>^{[0]}$ & $\left< T^{\mu \nu} \right>_q$ \\
State part of stress tensor expectation    &$t^{\mu \nu}$ or $t^{\mu \nu[0]}$ & $t^{\mu \nu}{}^{(q)}$ \\
Geometric part of stress tensor expectation& $\chi_{\mu \nu}$ or $\chi_{\mu \nu}^{[0]}$ & $\chi_{\mu \nu}{}^{(q)}$ \\
Anomaly & $\mathcal{A}[g_{\mu\nu}]$ &$ \mathcal{A}[g_{\mu\nu}^{(q)}]$\\
Christoffel & $\Gamma^{\mu}_{\nu\rho}$ or $\Gamma^{\mu}_{\nu\rho}{}^{[0]}$&$\Gamma^{\mu}_{\nu\rho}{}^{(q)}$\\
Covariant derivative & $\nabla_\mu$ &$\nabla_\mu^{(q)}$\\
$O(q-1)$ correction to the stress tensor expectation & - &  $\left<T^{\mu \nu}\right>^{[1]}$   \\
$O(q-1)$ correction to the state part of stress tensor expectation & - &  $t^{\mu \nu[1]}$   \\
$O(q-1)$ correction to the Anomaly & - &$ \mathcal{A}^{[1]}[g_{\mu\nu}^{(q)}]$\\
\hline
\end{tabular}
\begin{align}
  \left< T^{\mu \nu} \right>_q &= \left< T^{\mu \nu} \right>+(q-1)\left< T^{\mu \nu} \right>^{[1]}+ \mathcal{O}( (q-1)^2 ),\\
  \left< T^{\mu \nu} \right>_q &=  t^{\mu \nu}{}^{(q)}+\chi^{\mu\nu}{}^{(q)}, \\
  t^{\mu \nu}{}^{(q)} &=t^{\mu \nu[0]}+(q-1)t^{\mu \nu[1]}+\mathcal{O}( (q-1)^2)).
\end{align}

\section{State and (reduced) density matrix in a QFT} \label{A}
In this section, we will review the literature on how to construct states and density matrices in the path integral formalism \cite{TomHartman}.
\paragraph{Path integrals and states:}Consider a generic set of fields denoted by $\phi_i(x)$. A Lorentzian Path integral defines a transition amplitude under unitary time evolution from a state $\rvert\phi_1(x)\rangle$ to a state $\langle \phi_2(x)\rvert$.  
\begin{align}
    \langle\phi_2(\Vec{x})\rvert e^{-i(t_2-t_1)H}\rvert\phi_1(\Vec{x})\rangle=\int_{ \phi(\Vec{x},t_1)=\phi_1(\Vec{x})}^{ \phi(\Vec{x},t_2)=\phi_2(\Vec{x})}D\phi e^{iS[\phi]}=\langle\phi_2(\Vec{x})\rvert\psi\rangle.
\end{align}
\begin{figure}[H]
\centering
\includegraphics[angle=0, width=0.4\linewidth]{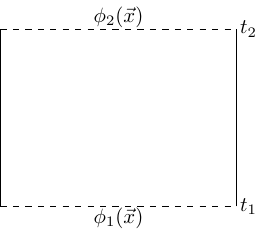}
\caption{Lorentzian Path integral}
\label{Lor}
\end{figure}
This can be interpreted as the overlap/ inner product of state $\rvert \psi \rangle = e^{-i(t_2-t_1)H}\rvert\phi_1(\Vec{x})\rangle$ with $\rvert\phi_2(\Vec{x})\rangle$. The state so defined satisfies the Schrodinger equation by construction.

We can Wick rotate to the Euclidean time ($t\to t'=-i\tau$), to get the Euclidean path integral
\begin{align}\label{Epath}
    \langle\phi_2(\Vec{x})\rvert e^{-(\tau_2-\tau_1)H}\rvert\phi_1(\Vec{x})\rangle=\int_{ \phi(\Vec{x},\tau_1)=\phi_1(\Vec{x})}^{ \phi(\Vec{x},\tau_2)=\phi_2(\Vec{x})}D\phi e^{-S_E[\phi]}=\langle\phi_2(\Vec{x})\rvert\psi\rangle.
\end{align}
\begin{figure}[H]
\centering
\includegraphics[angle=0, width=0.4\linewidth]{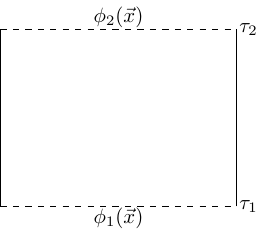}
\caption{Euclidean Path integral}
\label{Euc}
\end{figure}
This gives a transition amplitude as well, but the imaginary time evolution is not unitary i.e., $e^{-(\tau_2-\tau_1)H}$ is not a unitary operator. This can be viewed as an overlap of states. The transition amplitude can again be interpreted as the overlap with $\rvert\phi_2(\Vec{x})\rangle$ of the state 
\begin{equation}
    \rvert \psi \rangle = e^{-(\tau_2-\tau_1)H}\rvert\phi_1(\Vec{x})\rangle.
\end{equation}
If we now define a \textit{cut} as a spatial slice (codimension-1 surface) of the Euclidean manifold, then one can formally think of a Euclidean path integral with one set of boundary conditions and one open cut (i.e., cut with unspecified boundary condition) as a quantum state i.e.,
\begin{equation}\label{stateeqn}
   \rvert \psi \rangle = e^{-(\tau_2-\tau_1)H}\rvert\phi_1(\Vec{x})\rangle =  \int_{ \phi(\Vec{x},\tau_1)=\phi_1(\Vec{x})}^{}D\phi e^{-S_E[\phi]}.
\end{equation}
\begin{figure}[H]
\centering
\includegraphics[angle=0, width=0.5\linewidth]{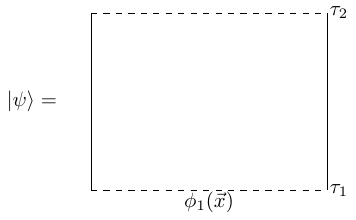}
\caption{Path integral with one open cut - state}
\label{state}
\end{figure}
The data on the top cut ($\tau=\tau_2$) is left unspecified. The functional $\rvert\psi\rangle$ along with the top boundary condition given by  $\langle \phi_2(\Vec{x})\rvert$ gives the complex numbers  $\langle \phi_2(\Vec{x})\rvert\psi\rangle$. States are defined on a spatial slice and do not care about Lorentzian vs Euclidean path integral. The state defined above by a Euclidean path integral is a state in the Hilbert space of the Lorentzian theory.

\paragraph{Density matrix using Path integrals:}A density matrix is an operator which takes a bra and a ket and produces a complex number. Thus any path integral with two identically shaped open cuts defines a density matrix (unnormalised).
\begin{figure}[H]
\centering
\includegraphics[angle=0, width=0.5\linewidth]{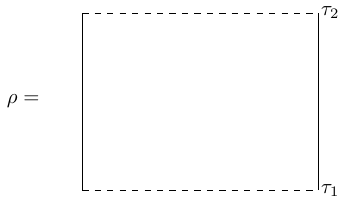}
\caption{An example of a path integral with two identical open cuts - density matrix}
\label{density}
\end{figure}
Consider the density matrix corresponding to the state defined by \eqref{stateeqn} i.e.,
\begin{equation}
    \rho =  \rvert\psi\rangle \langle\psi\rvert.
\end{equation}
\begin{figure}[H]
\centering
\includegraphics[scale=1.3]{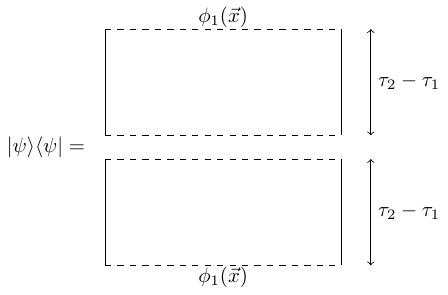}
\caption{Density matrix corresponding to a particular state}
\label{denstate}
\end{figure}
If $\phi_2(x)$ and $\phi_3(x)$ are the respective boundary conditions at the two open cuts, we have,
\begin{equation}
   \langle \phi_3(\Vec{x}) \rvert\rho\rvert \phi_2(\Vec{x})\rangle = \langle \phi_3(\Vec{x}) \rvert\psi\rangle \langle\psi\rvert \phi_2(\Vec{x})\rangle.
\end{equation}
This can be computed using the path integral expressions given by \eqref{Epath}.

The trace of the density matrix  
\begin{equation}
   Z= Tr \rho = \sum_{\phi(\Vec{x})}\langle \phi(\Vec{x}) \rvert\rho\rvert \phi(\Vec{x})\rangle = \sum_{\phi(\Vec{x})} \langle\psi\rvert \phi(\Vec{x})\rangle \langle \phi(\Vec{x}) \rvert\psi\rangle=\langle\psi\rvert\psi\rangle,
\end{equation}
is the path integral with the two open cuts having the same boundary condition $\phi(\Vec{x})$ and summed over all possible field configurations $\phi(\Vec{x})$ i.e., gluing the two open cuts.

\paragraph{Example 1:}The vacuum/ ground state:
Let us start with a state $\rvert X\rangle$. Expanding in the energy eigenstate $\rvert n\rangle$, we have,
\begin{equation}
    \rvert X\rangle=\sum_n x_n\rvert n \rangle; \quad H\rvert n\rangle=E_n\rvert n\rangle.
\end{equation}
We can define the (unnormalised) ground state by doing a path integral that extends to infinity in one direction i.e., by evolving over a long Euclidean time 
\begin{equation}
    e^{-\tau H}\rvert X\rangle \approx e^{-\tau E_0} x_0 \rvert 0\rangle;\quad \tau\to \infty.
\end{equation}
\begin{figure}[H]
\centering
\includegraphics[scale=1.3]{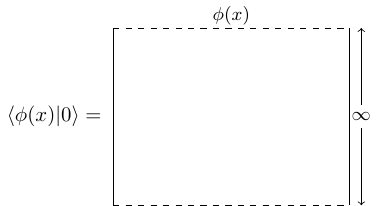}
\caption{Vacuum state}
\label{vac}
\end{figure}

The trace of the vacuum density matrix  
\begin{align}
   Z= Tr \rho &= \sum_{\phi(\Vec{x})}\langle \phi(\Vec{x}) \rvert\rho_{vacuum}\rvert \phi(\Vec{x})\rangle = \sum_{\phi(\Vec{x})} \langle\ 0 \rvert \phi(\Vec{x})\rangle \langle \phi(\Vec{x}) \rvert 0\rangle=\langle 0 \rvert 0 \rangle\nonumber\\
   &=\int D\phi e^{-S_E[\phi]}.
\end{align}
\begin{figure}[H]
\centering
\includegraphics[scale=1.3]{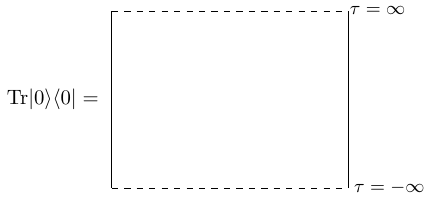}
\caption{Trace of the vacuum density matrix}
\label{Trvac}
\end{figure}

The path integral on the lower half plane produces $\rvert0\rangle$, the path integral on the upper half plane produces $\langle 0 \rvert$, and summing over all possible boundary conditions $\phi(\Vec{x})$ in the middle at $\tau=0$, glues the half-planes together giving the trace.
\paragraph{Example 2:}The thermal density matrix: Consider
\begin{equation}
    \rho_{thermal}=e^{-\beta H}.
\end{equation}
We have,
\begin{align}
   \langle \phi_2(\Vec{x}) \rvert \rho_{thermal}\rvert \phi_1(\Vec{x})\rangle& = \langle \phi_2(\Vec{x}) \rvert e^{-\beta H}\rvert \phi_1(\Vec{x})\rangle\nonumber\\
&=\int_{\phi(x,\tau_1)=\phi_1(\Vec{x})}^{\phi(x,\tau_2)=\phi_2(\Vec{x})}D[\phi]e^{-S_E}.
\end{align}
The trace 
\begin{align}
    Z(\beta)&=Tr e^{-\beta H} = \sum_{\phi(\Vec{x})}\langle\phi(\Vec{x})\rvert e^{-\beta H}\rvert\phi(\Vec{x})\rangle\nonumber \\
    &=\sum_{\phi(\Vec{x})} \int_{\phi(x,\tau_1)=\phi(\Vec{x})}^{\phi(x,\tau_2)=\phi(\Vec{x})}D[\phi]e^{-S_E}=\int_{S_1^{\tau}}D[\phi]e^{-S_E},
\end{align}
\begin{figure}[H]
\centering
\includegraphics[scale=1]{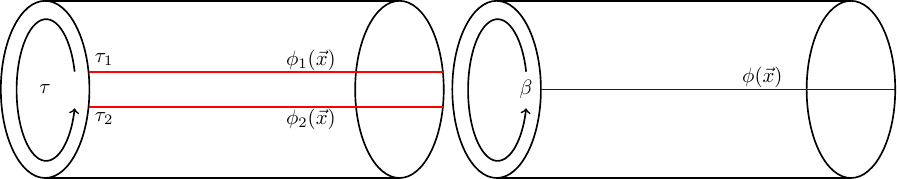}
\caption{Thermal density matrix and its trace}
\label{thermal}
\end{figure}
is an Euclidean path integral over a cylinder of time length $\beta$\cite{Calabrese:2009qy}. The boundary conditions above and the sum over all possible field configuration $\phi(\Vec{x})$ in the path integral, has the effect of sewing together edges along $\tau_1$ and $\tau_2$ forming a cylinder of circumference $\tau_2-\tau_1=\beta$. \footnote{$\beta$ is a placeholder for the coordinate length $(\tau_2-\tau_1)$. The physical length depends on the metric.}

\paragraph{Reduced density matrix:}The reduced density matrix $\rho_A=Tr_{\Bar{A}}\rho$, corresponding to a spatial interval $A$ on a fixed $\tau$ slice is the path integral with the two open cuts having the same boundary condition $\phi(\Vec{x})$ and summed over all possible field configurations $\phi(\Vec{x})$ only along the complement region $\Bar{A}$ i.e., gluing the two open cuts in $\rho$ only along the complement region $\Bar{A}$. $\langle \phi_2(\Vec{x})\rvert\rho_A\rvert\phi_1(\Vec{x})\rangle$ has the following path integral field boundary conditions along the open cuts
\begin{equation}
\begin{split}
&\phi(\text{cut 1})=\phi_1(\Vec{x}) \text{ $\forall$ x },\\
&\phi(\text{cut 2})=\phi_2(\Vec{x}) \text{ if $x \in A$ },\\
&\phi(\text{cut 2})=\phi_1(\Vec{x}) \text{ if $x \notin A$ },
\end{split}
\end{equation}
and summed over all possible $\phi_1(\Vec{x})$.

\begin{figure}[H]
\centering
\includegraphics[scale=1.8]{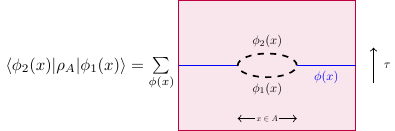}
\caption{Reduced density matrix corresponding to a region $A$}
\label{Redden}
\end{figure}

\bibliographystyle{JHEP}
\bibliography{bib}

\end{document}